\documentclass[useAMS,usenatbib,onecolumn]{mn2e}

\usepackage{epsfig}
\usepackage{amssymb}
\usepackage{graphics}
\usepackage{amsmath}
\usepackage{multirow}
\usepackage{widetext}
\usepackage{color}
\title[RMHD with ADER-DG]{Solving the relativistic magnetohydrodynamics equations with ADER discontinuous Galerkin methods, a posteriori subcell limiting and adaptive mesh refinement}

\author[O. Zanotti, F. Fambri, M. Dumbser] 
{O. Zanotti$^{1}$\thanks{E-mail:
olindo.zanotti@unitn.it}, 
F. Fambri$^{1}$, M. Dumbser$^{1}$ \\
$^{1}$Laboratory of Applied Mathematics, University of
  Trento, Via Mesiano 77, I-38123 Trento, Italy}
\begin{document}

\date{}

\pagerange{\pageref{firstpage}--\pageref{lastpage}} \pubyear{2009}

\maketitle

\newcommand{\be}{\begin{equation}}
\newcommand{\ee}{\end{equation}}
\newcommand{\bdm}{\begin{displaymath}}
\newcommand{\edm}{\end{displaymath}}
\newcommand{\bea}{\begin{eqnarray}}
\newcommand{\eea}{\end{eqnarray}}
\newcommand{\PNM}{P_NP_M}
\newcommand{\halb}{\frac{1}{2}}
\newcommand{\FQi}{\tens{\mathbf{F}}\left(\Qi\right)}
\newcommand{\FQj}{\tens{\mathbf{F}}\left(\Qj\right)}
\newcommand{\FQjj}{\tens{\mathbf{F}}\left(\Qjj\right)}
\newcommand{\nj}{\vec n_j}
\newcommand{\FORCE}{\textnormal{FORCE}}
\newcommand{\GFORCE}{\textnormal{GFORCEN}}
\newcommand{\LF}{\textnormal{LF}'}
\newcommand{\LW}{\textnormal{LW}'}
\newcommand{\WL}{\mathcal{W}_h^-}
\newcommand{\WR}{\mathcal{W}_h^+}
\newcommand{\nur}{\boldsymbol{\nu}^\textbf{r} }
\newcommand{\nuf}{\boldsymbol{\nu}^{\boldsymbol{\phi}} }
\newcommand{\nut}{\boldsymbol{\nu}^{\boldsymbol{\theta}} }
\newcommand{\ar}{\phi_1\rho_1}
\newcommand{\arr}{\phi_2\rho_2}
\newcommand{\ur}{u_1^r}
\newcommand{\uf}{u_1^{\phi}}
\newcommand{\ut}{u_1^{\theta}}
\newcommand{\urr}{u_2^r}
\newcommand{\uff}{u_2^{\phi}}
\newcommand{\utt}{u_2^{\theta}}
\newcommand{\ub}{\textbf{u}_\textbf{1}}
\newcommand{\ubb}{\textbf{u}_\textbf{2}}
\newcommand{\RoeMat}{{\tilde A}_{\Path}^G} 

\newcommand{\x}{\mathbf{x}}
\newcommand{\Q}{\mathbf{Q}}
\renewcommand{\u}{\mathbf{u}}
\newcommand{\w}{\mathbf{w}}
\newcommand{\q}{\mathbf{q}}
\newcommand{\F}{\mathbf{F}}
\newcommand{\f}{\mathbf{f}}
\newcommand{\g}{\mathbf{g}}
\newcommand{\h}{\mathbf{h}}
\renewcommand{\v}{\mathbf{v}}
\newcommand{\B}{\mathbf{B}}
\newcommand{\emm}{m}

\newcommand{\apriori}{\textit{a priori} }
\newcommand{\aposteriori}{\textit{a posteriori} }
\newcommand{\oz}[1]{ \textcolor{red}   {\texttt{\textbf{OZ: #1}}} }

\newfont{\numerikEleven}{ecrm1000}
\newfont{\numerikTen}{cmss10}
\newfont{\numerikNine}{cmss9}
\newfont{\numerikEight}{cmss8}

\label{firstpage}

\begin{abstract}
We present a new numerical tool for solving the special relativistic ideal MHD equations that is based on the combination of the following three 
key features: 
(i) a one-step ADER discontinuous Galerkin (DG) scheme that allows for an \textit{arbitrary} order of accuracy in both space and time, 
(ii) an \aposteriori subcell finite volume limiter that is activated to avoid spurious oscillations at discontinuities without destroying the
natural subcell resolution capabilities of the DG finite element framework 
and finally (iii) a space-time adaptive mesh refinement (AMR) framework with time-accurate local time-stepping.

The divergence-free character of the magnetic field is instead taken into account through the so-called "divergence-cleaning" approach.
The convergence of the new scheme is verified up to $5^{{\rm th}}$ order in space and time and the results for a set of significant 
numerical tests including shock tube problems, the RMHD rotor and blast wave problems, as well as the Orszag-Tang vortex system are shown. 
We also consider a simple case of the relativistic Kelvin--Helmholtz instability with a magnetic field, emphasizing the potential of the 
new method for studying turbulent RMHD flows. 
We discuss the advantages of our new approach when the equations of relativistic MHD need to be solved with high accuracy within various 
astrophysical systems. 
\end{abstract}


\section{INTRODUCTION}
\label{Introduction}

Special relativistic magnetohydrodynamics (RMHD) is supposed to provide a sufficiently accurate description of the dynamics 
of those astrophysical plasma that move close to the speed of light and which are
subject to electromagnetic forces dominating over gravitational forces.
This is the case of  high energy astrophysical phenomena like
extragalactic jets~\citep{Begelman1984}, gamma-ray bursts~\citep{Kouveliotou1993} and magnetospheres of neutron stars~\citep{Michel1991}. 
In all these physical systems, in fact, leaving aside the problem of the
origin of relativistic jets, which clearly involves the role of the accretion disc and of the corresponding central compact object, general relativistic effects can be fairly neglected. 
The degree of complexity related to magnetohydrodynamics can of course vary notably. For example, as a first approximation we can neglect dissipation due to resistivity, or we can consider the fluid as a single-component one, although we know for sure that both magnetic reconnection and multi-fluids
effects can  become important under specific physical conditions.

The numerical solution of the special relativistic magnetohydrodynamics equations has been particularly fostered by the introduction of Godunov methods based on Riemann solvers, which had already been successfully applied to relativistic hydrodynamics. This was the approach
followed in the pioneering works by \cite{Komissarov1999} and \cite{BalsaraRMHD}, who implemented for the first time
second order Total Variation Diminishing (TVD) schemes with a specific interest towards astrophysical applications. Since then, relativistic magnetohydrodynamics
has developed along different directions with impressive results. From one side several approximate Riemann solvers have been introduced 
\citep{Mignone2006,HLLCRMHD,Mignone2009,Kim2014}.
From another side, relativistic magnetohydrodynamics has been extended to the general relativistic regime 
\citep{Duez05MHD0,Baumgarte2003,Anton06,DelZanna2007,Giacomazzo:2007ti}, and it is currently used to study a variety of high energy physical processes.
An additional direction of research has been represented by the inclusion of dissipative effects, namely non-ideal resistive magnetohydrodynamics, 
with encouraging results \citep{Komissarov2007,Palenzuela:2008sf,DumbserZanotti,Zenitani2010,Takamoto2011b,Bucciantini2012a}. 
Moreover, high order numerical schemes have also been pursued \citep{DelZanna2003,Anderson2006a}, while simulations of 
multi-fluids in RMHD are emerging as a new frontier \citep{Zenitani2009,Barkov2014}.
Finally, Adaptive Mesh Refinement (AMR) within RMHD codes has been also considered \citep{Balsara2001b,Neilsen2006,Etienne:2010ui,Mignone2012,Keppens2012,Zanotti2015}
and it is an active field of research.
In most of the approaches mentioned so far the evolution in time is performed through the method of lines, resulting in 
multistep Runge--Kutta
schemes, either explicit or implicit. 
A valuable alternative  is provided by ADER schemes, which were introduced by \citet{titarevtoro,Toro2002} and became 
popular after the modern reformulation by \cite{DumbserEnauxToro,Dumbser2008,Balsara2013934}.
In a nutshell, ADER schemes are high order numerical schemes with a single step for the time update and
they have been already applied to the equations of relativistic MHD, both in the ideal case \citep{Dumbser2008,Zanotti2015}
and in the resistive case \citep{DumbserZanotti}.
Another common choice that is typically adopted in the majority of modern RMHD codes is that of using
finite difference or finite volume conservative schemes. Although rather successful, these schemes require larger and larger stencils when the
order of accuracy is increased, a fact that can give rise to substantial overhead when they are parallelized.
Discontinuous Galerkin (DG) schemes \citep{cbs1,cbs2,cbs3,cbs4}, on the contrary, do not need any spatial reconstruction and they allow for an arbitrary order of 
accuracy. DG schemes are still relatively unknown in high energy astrophysics, and only a few investigations have been performed so far in the relativistic regime 
\citep{zumbusch_2009_fed,Radice2011,Zanotti2011b}. 
Unfortunately, DG schemes suffer from a serious problem, which has negatively affected their popularity. Namely, since they are linear in the sense of Godunov's theorem, they produce oscillations as
soon as a discontinuity appears in the solution, even though they exploit the conservative formulation of the equations and even though Riemann solvers are used for the computation of the fluxes. The procedures that have been adopted to overcome this difficulty can be roughly divided in two classes. From one side, it is possible to introduce additional numerical dissipation, either in the form of 
artificial viscosity \citep{Hartman_02,Persson_06,Feistauer4}, or by means of filtering \citep{Radice2011}. 
From another side, it is possible to isolate the so-called \emph{troubled cells}, namely those affected by spurious oscillations, and adopt for them some sort of nonlinear
finite-volume-type slope-limiting procedure 
\citep{cbs4,QiuShu1,Qiu_2004,balsara2007,Zhu_2008,Zhu_13,Luo_2007,Krivodonova_2007}, either based on nonlinear WENO/HWENO reconstruction or by 
applying a TVB limiter to the higher order moments of the discrete solution. The drawback of this strategy is that in most cases the subcell 
resolution properties of the DG scheme are immediately lost.

Very recently, a promising alternative has been proposed by \cite{Dumbser2014}, which is based on a previous idea of \cite{CDL1} and 
\cite{CDL2} called MOOD (multi-dimensional optimal order detection), and which adopts an \aposteriori approach 
to the problem of limiting of high order schemes in the finite volume framework. 
In a few words, the novel \aposteriori DG limiter method of \cite{Dumbser2014} consists of (i) computing the solution by means of an \textit{unlimited} ADER-DG scheme, (ii) 
detecting \aposteriori the troubled cells by applying a simple discrete maximum principle (DMP) and positivity of density and pressure on the discrete solution, 
(iii) creating a local sub-grid within these troubled DG cells, and (iv) \textit{recomputing} the discrete solution at the sub-grid level via a more robust Total Variation Diminishing 
(TVD) or Weighted Essentially Non Oscillatory (WENO) finite volume scheme. The final non-oscillatory DG solution on the main grid is then recovered from the subcell
averages by means of a finite-volume reconstruction operator that acts on the cell averages of the subgrid. 
In the present paper we apply this idea for the first time to solve the RMHD equations in combination with space-time adaptive mesh refinement and time-accurate local time stepping, 
extending a similar work proposed for classical fluid dynamics by \cite{Zanotti2015c}. For alternative work on DG subcell limiters see also \citep{CasoniHuerta1,Sonntag,Fechter1}. 

The plan of the paper is the following. In
Section~\ref{sec:Mathematical_formulation} we report the RMHD equations
and the basic physical
assumptions, while Section~\ref{sec:Numerical_method} is
devoted to the presentation of the numerical
method, which is validated in Section~\ref{sec:Tests}.
Section~\ref{sec:KH} contains a first simple analysis of the turbulence induced by
the Kelvin--Helmholtz instability,
while in Section~\ref{sec:Conclusions} we conclude the work.
We have considered a flat spacetime in pseudo-Cartesian
coordinates, namely the metric 
$\eta_{\mu\nu}={\rm diag}(-1,1,1,1)$, where 
Greek letters run from 0 to 3 and Latin letters $i,j,k,\ldots$ run from 1 to 3.
The speed of light is set to $c=1$ and we make use of the
Lorentz-Heaviside notation for the electromagnetic
quantities, such that all $\sqrt{4\pi}$ factors disappear.  
Finally, we use Einstein summation convention over
repeated indices. 

\section{Mathematical formulation and physical assumptions}
\label{sec:Mathematical_formulation}

The energy-momentum tensor of a single-component
plasma with infinite conductivity is given
by~\citep{Anile_book}
\be
T^{\mu\nu}=(\rho h +b^2) u^\mu
u^\nu+(p + b^2/2) \eta^{\mu\nu}- b^\mu b^\nu \,,
\label{eq:T}
\ee
where $u^\mu$ is the four velocity of the fluid, $b^\mu$ is the four vector
magnetic field, $b^2=b_\mu b^\mu$, while
$h$, $\rho$ and $p$ are  the specific enthalpy, the rest
mass density and the thermal pressure, each of them 
measured in the co-moving frame of the fluid. The metric of the spacetime is the Minkowski one, namely $\eta^{\mu\nu}=\eta_{\mu\nu}={\rm diag}(-1,1,1,1)$.
We recall the in ideal MHD the electric field in the comoving frame of the fluid vanishes.
If we instead select a static laboratory observer defined by the
four-velocity vector $n^\mu=(-1,0,0,0)$, then the electric field $E^\mu$ and $B^\mu$ measured in such a frame are related to the 
electromagnetic tensor $F^{\mu\nu}$, and to its dual
$F^{\ast\mu\nu}$, by
\bea
\label{emtensor1}
F^{\mu\nu} &=& n^{\,\mu}E^{\nu} - E^{\mu}n^{\nu} +
\epsilon^{\,\mu\nu\lambda\kappa}B_{\lambda}n_{\kappa} \\
\label{emtensor2}
F^{\ast\mu\nu} &=& n^{\,\mu}B^{\nu} - B^{\mu}n^{\nu} -
\epsilon^{\,\mu\nu\lambda\kappa}E_{\lambda}n_{\kappa} \,,
\eea
where $\epsilon^{\,\mu\nu\lambda\kappa}$
is the completely antisymmetric spacetime Levi-Civita
tensor, with the convention that $\epsilon^{\, 0 1 2 3}=1$.
Note that the four vectors of the
electric and of the magnetic field are purely spatial, i.e.
$E^0=B^0=0$, $E^i=E_i$, $B^i=B_i$.
Moreover, the fluid four velocity
$u^\mu$ and the standard three velocity in the laboratory
frame are related as $v^i=u^i/W$, where
$W=({1-v^{\, 2}})^{-1/2}$ is the 
Lorentz factor of the fluid. We stress that the electric field does not need to be evolved in time through the Maxwell equations, since 
within the ideal MHD assumption it can always be computed a posteriori as $\vec E = -\vec v \times \vec B$.
In the rest of the paper we also assume that the fluid obeys the
ideal gas equation of state, namely
\be
p=\rho\epsilon(\gamma-1)\,,
\ee
where $\epsilon$ is the specific internal energy, which is a function of the temperature only, and $\gamma$ is the adiabatic index.
The equations of ideal RMHD, which in covariant form are
\bea
\label{0eq:mass}
&&\nabla_{\alpha} (\rho u^{\,\alpha})=0, \\
\label{0eq:momentum}
&&\nabla_{\alpha}T^{\alpha\beta}=0,\\
\label{0eq:maxwell}
&&\nabla_{\alpha}F^{\ast\alpha\beta}=0\,,
\eea
for numerical purposes are better expressed   
in conservative form as \citep{Komissarov1999,BalsaraRMHD} 
\be
\partial_t{\bf u} + \partial_i{\bf f}^i=0\,,
\label{eq:UFS}
\ee
where the conserved variables and the corresponding
fluxes in the $i$ direction are given by\footnote{Although 
formally written in conservative form,
the evolution of the magnetic field is based on Stokes' theorem rather than on Gauss' theorem. See \citet{Londrillo2000} for a careful discussion about these aspects.}
\be
{\bf u}=\left[\begin{array}{c}
D \\ S_j \\ U \\ B^j 
\end{array}\right],~~~
{\bf f}^i=\left[\begin{array}{c}
 v^i D \\
 W^i_j \\
 S^i \\
\epsilon^{jik}E^k 
\end{array}\right]\,.
\label{eq:fluxes}
\ee
The conserved variables $(D,S_j,U,B^j)$ 
are related to the rest-mass density $\rho$, to the thermal
pressure $p$, to the fluid velocity $v_i$ and to the magnetic field $B^i$ 
by
\bea
\label{eq:cons1}
&&D   = \rho W ,\\
\label{eq:cons2}
&&S_i = \rho h W^2 v_i + \epsilon_{ijk}E_j B_k, \\
\label{eq:cons3}
&&U   = \rho h W^2 - p + \frac{1}{2}(E^2 + B^2)\,,
\eea
where $\epsilon_{ijk}$ is  the spatial Levi--Civita tensor and $\delta_{ij}$ is the Kronecker symbol.       
The spatial tensor $W^i_j$ in (\ref{eq:fluxes}), representing the momentum  flux density, is 
\be
W_{ij} \equiv \rho h W^2 v_i v_j - E_i E_j - B_i B_j + \left[p +\frac{1}{2}(E^2+B^2)\right]\delta_{ij}\,, \\
\label{eq:W} 
\ee
where $\delta_{ij}$ is the Kronecker delta. 
Eqs.~\eqref{0eq:maxwell} above include the divergence free condition $\vec \nabla\cdot\vec B=0$. Although the Maxwell equations guarantee that 
such a constraint is mathematically fulfilled for all times if it is satisfied in the initial conditions, from a numerical point of view specific actions must be taken in order  to preserve the divergence-free property of the magnetic field during the evolution of the system. Several strategies have been proposed over the years to
solve this problem [see \cite{Toth2000} for a review]. In this paper we have 
adopted the so called 
{\em divergence-cleaning
approach} presented in \cite{Dedneretal}, which amounts to
augmenting
the system (\ref{eq:UFS}) with an additional equation for a scalar 
field $\Phi$, in order to propagate away the deviations from $\vec \nabla\cdot\vec B=0$.
Hence, we must solve 
\be
\label{eq:divB}
\partial_t \Phi + \partial_i B^i = -\kappa \Phi\,,
\ee
while the fluxes for the evolution of the magnetic field are also modified, namely ${\bf f}^i(B^j)\rightarrow \epsilon^{jik}E^k + \Phi \delta^{ij}$.
The damping coefficient $\kappa$ in Eq.~\eqref{eq:divB} drives the solution towards  $\vec \nabla\cdot\vec B=0$ over a time scale $1/\kappa$.
In our calculations we have typically used $\kappa\in[1;10]$. More details about this approach can be found in 
 \citet{Komissarov2007}, \citet{Palenzuela:2008sf}, \citet{Dionysopoulou:2012pp}.

As well known, in the relativistic framework the conversion from the conserved variables $(D,S_i,U,B_i)$ to the primitive 
variables $(p,\rho,v_i,B_i)$, which are needed for the computation of the fluxes,  is not 
analytic, and a numerical root-finding 
approach is therefore needed.  In our numerical code we adopted the third method reported in Sect. 3.2 of 
\cite{DelZanna2007}. A full account about alternative methods to invert the system (\ref{eq:cons1})--(\ref{eq:cons3}) was given in
\cite{NGMD2006}.
Additional information about the mathematical properties of the RMHD equations can be found in \citet{BalsaraSpicer1999,Komissarov1999,Anton06,DelZanna2007,Anton2010}. The latter, in particular, contains a detailed discussion about the renormalization of the eigenvectors
of the associated Jacobian.

\section{Numerical method}
\label{sec:Numerical_method}
The numerical scheme that we adopt results from the combination of a few 
different steps, which in principle could be used separately. Here we provide a brief but 
self-consistent presentation of the scheme, while addressing to \cite{Dumbser2014} and to \cite{Zanotti2015c} for additional  discussion.
%
\subsection{Basic mathematical definitions}
We use spatial Cartesian coordinates over a domain $\Omega$ which is composed by  
elements $T_i$ as
\begin{equation}
\label{eqn.tetdef}
 \Omega = \bigcup \limits_{i=1}^{N_E} T_i\,,
\end{equation}
where the index $i$ ranges from 1 to the total number of elements $N_E$.
At the generic time $t^n$, the numerical solution of Eq.~\eqref{eq:UFS}
is represented within each cell $T_i$ 
by polynomials of maximum degree $N \geq 0$, namely
\begin{equation}
\label{eqn.ansatz.uh}
  \u_h(\x,t^n) = \sum_{l=0}^{N}\Phi_l(\x) \hat{\u}^n_l= \Phi_l(\x) \hat{\u}^n_l \quad \x \in T_i\,,
\end{equation}
where $\u_h$ is referred to as the \emph{discrete representation} of the solution, 
while the coefficients
$\hat{\u}^n_l$ are the \emph{degrees of freedom}.\footnote{Here we slightly abuse of  the Einstein summation convention, 
which is adopted even if the mute indices over which the summation is performed do not refer to co-variant and contra-variant vectors.}
In one spatial dimension, the basis functions $\Phi_l(x)$ are given by 
the Lagrange interpolation polynomials, all of degree $N$, 
which pass through the $(N+1)$ Gauss-Legendre 
quadrature points \citep{Solin2006}. The resulting basis is therefore a nodal basis, with the property that $\Phi_l(x_k)=\delta_{lk}$, where $x_k$ are the
coordinates of the Gauss-Legendre nodal points. In multiple space dimensions, 
the basis functions $\Phi_l(\x)$ are the dyadic products of the one-dimensional basis.

%
\subsection{The Discontinuous Galerkin scheme}
The system of equations \eqref{eq:UFS} is in conservative form and, therefore, numerical schemes derived from it are guaranteed to converge to the weak solution~\citep{Lax60}, even if this contains a discontinuity. The vast majority of numerical schemes for the solution of the RMHD equations use either conservative finite difference or finite volume schemes, which incorporate in a natural way Riemann solvers, thus assuring the upwind property of the method. An effective  alternative to these approaches is represented by Discontinuous Galerkin methods~\citep{cbs0,cbs1,cbs3,cbs4}, which still exploit the conservative form of the equations and the usage of Riemann solvers, but which evolve in time the degrees of freedom with respect to the chosen basis, rather then the point values or the cell averages  of the solution.

To illustrate the method, we first 
multiply the governing equations \eqref{eq:UFS} by a test 
function $\Phi_k \in \mathcal{U}_h$, identical to the spatial basis functions of Eq.~\eqref{eqn.ansatz.uh}. After that,
we integrate over the space-time control volume $T_i \times [t^n;t^{n+1}]$. 
If we integrate by parts in space
the flux divergence term, we obtain
\begin{equation}
\label{eqn.pde.nc.gw1}
\int \limits_{t^n}^{t^{n+1}} \int \limits_{T_i} \Phi_k \frac{\partial \u_h}{\partial t} d\x dt
+\int \limits_{t^n}^{t^{n+1}} \int \limits_{\partial T_i} \Phi_k \, \f \left(\u_h \right)\cdot\mathbf{n} \, dS dt 
-\int \limits_{t^n}^{t^{n+1}} \int \limits_{T_i} \nabla \Phi_k \cdot \f \left(\u_h \right) d\x dt 
= 0, 
\end{equation}
where $\mathbf{n}$ is the outward pointing unit normal vector on the surface $\partial T_i$ of element $T_i$. 
The second term of Eq.~\eqref{eqn.pde.nc.gw1} contains a surface integration, which is conveniently performed through the solution of a Riemann problem at
the element boundary, like in traditional conservative finite volume schemes. This guarantees that the final method is upwind. 
The time integration of Eq.~\eqref{eqn.pde.nc.gw1} can be performed through Runge--Kutta schemes, leading to RKDG schemes~\citep{cbs0,cbs1,cbs2,cbs3,cbs4,Zhu_hadap_2013}
but it can also be obtained through the ADER philosophy, see \citep{dumbser_jsc,QiuDumbserShu}.
 More precisely, we want to devise a one-step time integration scheme for \eqref{eqn.pde.nc.gw1}, while 
preserving high order of accuracy both in space and in time. This can be done, provided an approximate \textit{predictor} state $\q_h$ is available 
at any intermediate time between $t^n$ and $t^{n+1}$ and with the same spatial accuracy of the initial DG polynomial. We have denoted this spacetime 
predictor solution $\q_h$ with a separate symbol, to distinguish it from the discrete solution of the DG scheme $\u_h$.  
After inserting $\u_h$, as given by \eqref{eqn.ansatz.uh}, in the first term of \eqref{eqn.pde.nc.gw1} and by using the spacetime predictor $\q_h$
in the other terms, we find the following one-step ADER discontinuous Galerkin scheme: 
\begin{widetext}
\begin{equation}
\label{eqn.pde.nc.gw2}
\left( \int \limits_{T_i} \Phi_k \Phi_l d\x \right) \left( \hat{\u}_l^{n+1} -  \hat{\u}_l^{n} \right) +
\int\limits_{t^n}^{t^{n+1}} \int_{\partial T_i} \Phi_k \, \mathcal{G}\left(\q_h^-, \q_h^+ \right)\cdot\mathbf{n} \, dS dt 
-\int\limits_{t^n}^{t^{n+1}} \int_{T_i} \nabla \Phi_k \cdot \f\left(\q_h \right) d\x dt  = 0\,.
\end{equation}
\end{widetext}
In the equation above $\mathcal{G}$ is a numerical flux function, which in practice is given by a Riemann solver, while 
$\q_h^-$ and $\q_h^+$  are the corresponding left and right states of the spacetime predictor solution that is typical of ADER schemes. 
%
%
This will allow us to compute the spacetime integrals of the second and of the third terms of Eq.~\eqref{eqn.pde.nc.gw2} to the desired order of accuracy.
The strategy for obtaining the predictor $\q_h$ from the DG polynomials $\u_h$ is explained in Sect.~\ref{sec:STDG}. 
Concerning the choice of the Riemann solver, in this paper we have used 
the simple Rusanov flux and the HLL solver \citep{toro-book}. 

\subsection{The spacetime discontinuous Galerkin predictor}
\label{sec:STDG}
In the original ADER approach by \cite{toro3} and \cite{titarevtoro}, the time evolution $\q_h$ of the data $\u_h$ available at time $t^n$ is 
obtained by means of the so-called Cauchy-Kowalevski procedure, which implies a Taylor expansion in time, and a subsequent replacement of time
derivatives with spatial derivatives through the governing system of PDEs. As simple as it is in principle, this approach becomes prohibitively 
complex for highly non-linear systems of equations. It has been successfully implemented for the classical Euler equations \citep{Dumbser2007} but it has never 
been extended to the relativistic regime. In the modern ADER version proposed by \cite{DumbserEnauxToro}, the time evolution is instead performed 
trough a spacetime discontinuous Galerkin predictor, which operates locally for each cell.
To illustrate the method, we first transform the PDE system of  Eq.~\eqref{eq:UFS} into a space-time reference coordinate system 
$(\xi,\eta,\zeta, \tau)$. Hence, the 
space-time control volume 
$\mathcal{C}_{ijkn}=[x_{i-\halb};x_{i+\halb}] \times [y_{j-\halb};y_{j+\halb}] \times [z_{k-\halb};z_{k+\halb}] \times [t^n;t^{n+1}]$ 
is mapped into the space-time reference element $T_E\times[0;1]$ as
\begin{equation}
\label{eq:xi}
x = x_{i-\halb} + \xi   \Delta x_i, \quad 
y = y_{j-\halb} + \eta  \Delta y_j, \quad 
z = z_{k-\halb} + \zeta \Delta z_k, \quad
t = t^n + \tau \Delta t\,,
\end{equation} 
where $T_E=[0;1]^d$ denotes the spatial reference element in $d$ spatial dimensions.
In these reference coordinates, Eq.~\eqref{eq:UFS} rephrases into
\begin{equation}
\label{eqn.pde.nc.2d}
 \frac{\partial \u}{\partial \tau}
    + \nabla_\xi \cdot \f^* \left( \u \right) = 0\,,
\end{equation}
where
\begin{equation}
  \f^* := \Delta t \left( \partial \boldsymbol{\xi} / \partial \x  \right)^{T} \cdot \f(\u)\,,
\end{equation}
with $\boldsymbol{\xi}=(\xi,\eta,\zeta)$ and
$\nabla_\xi = \partial \boldsymbol{\xi} / \partial \x \cdot \nabla$.
We now 
multiply \eqref{eqn.pde.nc.2d} by a space-time test function 
$\theta_k=\theta_k(\boldsymbol{\xi},\tau)$ and integrate over 
the space-time reference control volume $T_E \times [0;1]$, to obtain
\begin{equation}
\label{eqn.pde.nc.weak1}
 \int \limits_{0}^{1} \int \limits_{T_E} \theta_k \frac{\partial \u}{\partial \tau} \, d \boldsymbol{\xi} \, d \tau\,
    + \int \limits_{0}^{1} \int \limits_{T_E}\theta_k \nabla_\xi \cdot \f_h^* \left(\u\right)\, d \boldsymbol{\xi} \, d \tau\,
    = 0\,.
\end{equation}
The \emph{discrete spacetime solution} of equation \eqref{eqn.pde.nc.weak1} is the $\q_h$ that we have mentioned above. In analogy to 
Eq.~(\ref{eqn.ansatz.uh}), we expand it as
\begin{equation}
\label{eqn.st.state}
 \q_h = \q_h(\boldsymbol{\xi},\tau) =
 \theta_l \hat{\q}_l\,.
\end{equation}
Something similar is done for the fluxes, which are represented as
\begin{equation}
\label{eqn.st.flux}
 \f^*_h = \f^*_h(\boldsymbol{\xi},\tau) =
 \theta_l \hat{\f}^*_l\,.
\end{equation}
Both the space-time test function $\theta_k$ in Eq.~\eqref{eqn.pde.nc.weak1}
and the basis 
functions $\theta_l$ are chosen as dyadic products of Lagrange interpolation polynomials passing through the Gauss-Legendre quadrature points. 
As a result, 
the degrees of freedom for the fluxes can be computed as  the point--wise evaluation of the physical fluxes, i.e.
\begin{equation}
  \hat{\f}^*_l = \f^*(\hat{\q}_l)\,.
\end{equation}
Since the time evolution of the discrete solution $\q_h$ is now hidden in the basis functions, we 
can integrate the first term by parts in time in \eqref{eqn.pde.nc.weak1}, which allows us to introduce the DG 
solution $\u_h(\x,t^n)$ as initial condition at time $t^n$ in a weak form. We thus obtain
\begin{widetext}
\begin{equation}
\label{eqn.pde.nc.dg1}
  \int \limits_{T_E}  
	\theta_k (\boldsymbol{\xi},1) \q_h \, d \boldsymbol{\xi}  - 
	\int \limits_{T_E}  
	\theta_k (\boldsymbol{\xi},0) \u_h \, d \boldsymbol{\xi} 
	   - \int \limits_{0}^{1} \int \limits_{T_E}
		\frac{\partial \theta_k}{\partial \tau}  \q_h  \, d \boldsymbol{\xi} \, d \tau
    + \int \limits_{0}^{1} \int \limits_{T_E} \theta_k \nabla_\xi \cdot \f^*_h \, d \boldsymbol{\xi} \, d \tau
    = 0.
\end{equation}
\end{widetext}
Inserting (\ref{eqn.st.state}) and (\ref{eqn.st.flux}) into Eq.~(\ref{eqn.pde.nc.dg1}) provides \citep{Dumbser2008,HidalgoDumbser,DumbserZanotti}
\begin{widetext}
\begin{equation}
\label{eqn.pde.nc.dg2}
   \left( \int \limits_{T_E}  
	\theta_k (\boldsymbol{\xi},1) \theta_l (\boldsymbol{\xi},1)\, d \boldsymbol{\xi}
	- \int \limits_{0}^{1} \int \limits_{T_E}
		\frac{\partial \theta_k}{\partial \tau}   \theta_l  \, d \boldsymbol{\xi} \, d \tau
    \right)
    \hat{\q}_l 
    =   \left( \int \limits_{T_E}  
	\theta_k (\boldsymbol{\xi},0) \Phi_l \, d \boldsymbol{\xi}  \right)\hat \u_l^n
		 - \left(\int \limits_{0}^{1} \int \limits_{T_E} \theta_k \nabla_\xi \theta_l \, d \boldsymbol{\xi} \, d \tau\right) \f^*(\hat{\q}_l)\,,
\end{equation}
\end{widetext}
which is a nonlinear system to be solved in the unknown expansion 
coefficients $\hat{\q}_l$. A few comments should be given at this stage. The first one is that, being local in space, the
spacetime discontinuous Galerkin predictor does not require the solution of any Riemann problem, which is instead invoked in the global scheme 
\eqref{eqn.pde.nc.gw2}. The second comment is that the discontinuous Galerkin predictor just described can be used also in combination with more traditional
finite volume schemes, which has been done for the RMHD equations for instance in \cite{Zanotti2015}. 
The third comment is that this approach, unlike the original ADER approach, remains valid  even in the presence of stiff source terms, as it has been done
for various physical systems by 
\cite{DumbserZanotti,HidalgoDumbser,Zanotti2011,Dumbser-Uuriintsetseg2013}.
Finally, we emphasize that for DG schemes the 
timestep must be restricted as \citep{krivodonova2013analysis}
\begin{equation}
\Delta t < \frac{1}{d} \frac{1}{(2N+1)} \frac{h}{|\lambda_{\max}|}\,,
\label{eqn.cfl} 
\end{equation}
where $h$ and $|\lambda_{\max}|$ are a characteristic mesh size and the maximum signal velocity, respectively. 

\subsection{An \aposteriori subcell limiter}
\label{sec:limiter}
Should we implement the ADER-DG scheme as it is described in the two previous Sections, we would obtain a numerical scheme capable of 
resolving smooth solutions with an order of accuracy equal to $N+1$, where $N$ is the degree of the chosen polynomials,
but totally inadequate for discontinuous solutions, for which the 
Gibbs phenomenon would quickly lead to spurious oscillations and even to a breakdown of the scheme. A novel idea for an \aposteriori limiter has been recently proposed by \cite{Dumbser2014} and 
it works as follows.
\begin{itemize}
\item The \emph{unlimited} ADER-DG scheme \eqref{eqn.pde.nc.gw2} is first used to evolve the solution from time $t^n$ to $t^{n+1}$, producing a so-called
candidate solution $\u_h^*(\x,t^{n+1})$ inside each cell.
\item The candidate solution $\u_h^*(\x,t^{n+1})$ is then checked against two different criteria to verify its validity, namely
\begin{enumerate}
\item \emph{Physical admissibility detection:} if the conversion from conservative to primitive variables fails, or if either the pressure or the rest mass density drops below a threshold value, 
or if we encounter superluminal velocities, then the cell is flagged as \emph{troubled}. 
\item \emph{Numerical admissibility detection:} if the polynomial representing the candidate solution does not lie between the minimum and the maximum 
of the polynomials representing the solution at the previous time step in the set ${\cal{V}}_i$, then the cell is flagged as \emph{troubled}. 
The set ${\cal{V}}_i$ contains the cell $T_i$ and all its Voronoi neighbor cells that share a common node with $T_i$. This second detection criterion is 
specifically designed to remove spurious Gibbs oscillations from the solution. 
\end{enumerate} 
\item 
As soon as a cell is flagged as 
troubled at the future time $t^{n+1}$, it generates a local sub-grid formed by $N_s= 2N+1$ cells per space dimension, each of which is assigned a subcell average $\v_h(\x,t^n)$ by means of  
a $L_2$ projection obtained from the DG polynomial at the \textit{previous time level} $t^n$, i.e. 
\begin{equation} \label{vh}
	v_{i,j}^n=\frac{1}{|S_{i,j}|}\int_{S_{i,j}}{\textbf{u}_h(\textbf{x},t^n)\,d\textbf{x}}=
	\frac{1}{|S_{i,j}|}\int_{S_{i,j}}{\hat{\textbf{u}}_l^n\phi_l(\textbf{x})\,d\textbf{x}}, \qquad \forall S_{i,j}\in {\cal{S}}_i\,,
\end{equation}
where ${\cal{S}}_i=\bigcup_j S_{i,j}$ is the set of the sub-grid cells. In this way the high accuracy of the DG polynomial is transferred 
to the subgrid level \textit{before} the spurious oscillations arise. We have chosen $N_s= 2N+1$ in order to guarantee that 
the maximum timestep of the ADER-DG scheme on the main grid (c.f. Eq.~\eqref{eqn.cfl}) matches the maximum 
possible time step of the ADER finite volume scheme on the sub-grid. 

\item The alternative data representation, provided by Eq.~\eqref{vh}, is now used as initial condition to evolve the discrete solution 
with a more robust finite volume scheme on the sub-grid. This is done by resorting to either an ADER-WENO finite volume scheme, or to 
an even more robust second order TVD shock capturing scheme. 
For details about the implementation of WENO within our ADER framework we refer to \cite{Dumbser2012b,Zanotti2015}.
In practice, on the sub-grid a new evolution from time $t^n$ to $t^{n+1}$ is performed combining a third order 
WENO finite volume scheme with the spacetime discontinuous Galerkin predictor described in Sect.~\ref{sec:STDG}.
Only for particularly challenging problems we sacrifice WENO in favor of a simpler second order TVD scheme. 
We emphasize that both the DG scheme on the main grid as well as the WENO finite volume scheme on the sub-grid are \emph{one-step} ADER schemes.

\item The last step requires that the new solution at time $t^{n+1}$ over the sub-grid is projected  back to the main grid. This is done 
imposing that
\begin{equation}
\label{eqn.subcell.r} 
  \int \limits_{S_{i,j}} \u_h(\x,t^{n+1}) d\x = \int \limits_{S_{i,j}} \v_h(\x,t^{n+1}) d\x, \qquad \forall S_{i,j} \in \mathcal{S}_i\,.
\end{equation} 
which is a standard reconstruction problem in high order finite volume methods \citep{barthlsq,Titarev2004,titarevtoro,Dumbser2012b} and 
spectral finite volume schemes \citep{spectralfv2d,spectralfv3d}. 

\end{itemize}
In all the numerical simulations described in Sect.~\ref{sec:Tests}, the DG scheme over the main grid has been implemented with $N\in[2;5]$, hence up to the sixth
order of accuracy both in space and in time, while the WENO scheme on the sub-grid is always at the third order.
Moreover, in the Figs.~\ref{fig:Balsara3D} and \ref{fig:RMHDrotor} below we have represented 
in blue the unlimited cells, namely those that  have been successfully evolved through the standard ADER-DG scheme,
while we have represented in red the troubled cells, which required the activation of the subcell limiter.
%
\subsection{Adaptive Mesh Refinement}
The whole scheme described so far can be implemented over adaptively refined meshes (AMR), together with time-accurate local time-stepping (LTS). 
In \cite{Dumbser2012b}, \cite{Zanotti2015c} and \cite{Zanotti2015} we have already described all details of our AMR strategy, hence, in the following 
we recall the most important aspects only. 

Our AMR approach can be referred to as a ''cell-by-cell'' refinement \citep{Khokhlov1998}, according to which every cell $T_i$ 
is individually refined with no creation of grid patches. As customary, 
the refinement criterion involves up to the second order derivative of a suitable \emph{indicator function} $\Phi$, in terms of which a suitable refinement function is build~\citep{Loehner1987}, 
\begin{equation} 
\label{eq:chi_m}
	\chi_\emm (\Phi) =  \sqrt{ \frac{\sum_{k,l}{\left(\left. \partial ^2 \Phi \middle/ \partial x_k \partial x_l \right. \right)^2}}{ \sum_{k,l}{ \left[ \left. \Big( \left| \left. \partial \Phi \middle/ \partial x_k\right.\right|_{i+1} + \left| \left.\partial \Phi \middle/ \partial x_k \right. \right|_i \Big) \middle/ \Delta x_l \right. + \epsilon \left|\frac{\partial^2 }{\partial x_k\partial x_l} \right|\left| \Phi \right|    \right]^2}}  }\,.
\end{equation}
The refinement function $\chi_\emm$ is checked for each cell $T_i$, and 
if $\chi_i > \chi_\text{ref}$, the cell is refined, while  it is recoarsened if $\chi_i < \chi_\text{rec}$. In most of our simulations, except 
for Sect.~\ref{sec:conv} where we have used  $\Phi=B_y$, the indicator function has been assumed to be the relativistic mass density, i.e. $\Phi=D=W \rho$.
The implementation of the AMR infrastructure is based on the following general rules:
\begin{itemize}
\item A maximum level of refinement $\ell_\text{max}$ is chosen, such that $0\le\ell\le\ell_\text{max}$, where $\ell$ indicates the actual refinement level. 
\item When a \emph{mother cell} $T_i$ is refined, it generates $\mathfrak{r}^d$ \emph{children cells}, where $d$ is the spatial dimension, while $\mathfrak{r}$
is the refinement factor, typically chosen between  $2$ and $4$.
\item
Each cell $T_i$, at any level of refinement, is given a specific 
\textit{status}, denoted by $\sigma$ for convenience, with the following meaning
\begin{enumerate}
\item \textit{active cell} ($\sigma=0$), updated through the standard ADER-DG scheme;
\item \textit{virtual child cell} ($\sigma=1$), updated according to 
standard $L_2$ projection of the high order polynomial of the mother cell at the $(\ell-1)$-th level;
\item \textit{virtual mother cell} ($\sigma=-1$), updated by recursively  averaging over all children cells from higher refinement levels.  
\end{enumerate}
A virtual child cell has always an active mother cell, while a virtual mother cell has 
children with status $\sigma \leq 0$.
\item 
Only active cells ($\sigma=0$) can be refined. Hence, if a virtual cell needs to be refined, it must be first activated.
\item 
The levels of refinement of two cells that are Voronoi neighbors\footnote{The \textit{Voronoi neighbors} $\mathcal{V}_i$ of a cell $T_i$ are cells which 
share common nodes.} of each other can only 
differ by at most unity. 
Moreover, every cell has  Voronoi neighbors, which can be either active or virtual, at the same level of refinement. 
\end{itemize} 
The rules above are quite general, and they would still hold even if a pure finite volume scheme was adopted. In addition to them, a few more 
instructions are needed when the AMR framework is combined with the presence of the limiter for the DG scheme. Namely,
\begin{itemize}
\item The virtual children cells inherit the limiter status of their active mother cell.
\item If at least one active child is flagged as troubled, then the (virtual) mother is also flagged as troubled.
\item Cells which need the subcell limiting cannot be recoarsened.
\end{itemize}
A proper description of the AMR-projection and of the AMR-averaging at the sub-grid level involving different levels of refinement can be found in \cite{Zanotti2015c}.

\section{Numerical tests} 
\label{sec:Tests} 

 \begin{table*}
 \centering
 \numerikNine
 \begin{tabular}{|c|c||ccc|ccc|c|}
   \hline
   \multicolumn{9}{|c|}{\textbf{2D circularly polarized Alfven Wave problem --- ADER-DG-$\mathbb{P}_N$ + WENO3 SCL}} \\
   \hline
    & $N_x$ & $L_1$ error & $L_2$ error & $L_\infty$ error & $L_1$ order & $L_2$ order & $L_\infty$ order &
   Theor. \\
   \hline
   \hline
   \multirow{4}{*}{\rotatebox{90}{{DG-$\mathbb{P}_2$}}}
& 30	& 2.9861E-3 & 7.2314E-4	& 4.2388E-4	&---	&---	&--- & \multirow{4}{*}{3}\\
& 60	& 2.9229E-4	& 8.0346E-5	& 7.0230E-5	& 3.35	& 3.17	& 2.59 &\\
& 90	& 8.8069E-5	& 2.5059E-5	& 2.3319E-5	& 2.95	& 2.87	& 2.72 &\\
& 120	& 3.6687E-5	& 1.0900E-5	& 1.0948E-5	& 3.04	& 2.89	& 2.63 &\\
   \cline{2-8}
   \hline
	 \multirow{4}{*}{\rotatebox{90}{{DG-$\mathbb{P}_3$}}}
& 15	& 1.2671E-4	& 2.5939E-5	& 1.1433E-5	&---	&---	&--- & \multirow{4}{*}{4}\\
& 20	& 3.1455E-5	& 6.5949E-6	& 2.9456E-6	& 4.48	& 4.76	& 4.71 &\\
& 25	& 1.1743E-5 & 2.5410E-6	& 1.4527E-6	& 4.41	& 4.27	& 3.17 &\\
& 30	& 5.7046E-6 & 1.2767E-6	& 7.5875E-7	& 3.96	& 3.77	& 3.56 &\\
   \cline{2-8}
   \hline
	 \multirow{4}{*}{\rotatebox{90}{{DG-$\mathbb{P}_4$}}}
& 10	& 6.6600E-5	& 1.4648E-5	& 7.5420E-6	&---	  &---	 &--- & \multirow{4}{*}{5}\\
& 15	& 7.8640E-6	& 1.9384E-6	& 1.2828E-6	& 5.26	& 4.98 & 4.36 & \\
& 20	& 1.8748E-6	& 4.9562E-7	& 3.6520E-7	& 4.98	& 4.74 & 4.36 &\\
& 25	& 6.1631E-7 & 1.6408E-7	& 1.3283E-7	& 4.98	& 4.95 & 4.53 &\\
   \cline{2-8}
   \hline	
 \end{tabular}
 \caption{ \label{tab:CPA} $L_1, L_2$ and $L_\infty$ errors and convergence rates for the 
   2D circularly polarized Alfven wave problem for the ADER-DG-$\mathbb{P}_N$ scheme with subcell limiter and adaptive mesh refinement. 
	Two levels of refinement have been used with a refinement factor $\mathfrak{r}=3$. The errors have been computed for the variable $B^y$.}
 \end{table*}

\subsection{Convergence test}
\label{sec:conv}
We have tested the convergence of our new numerical scheme by  considering the propagation of a circularly polarized Alfven wave, for which an analytic solution
is known \citep{Komissarov1997,DelZanna2007}. Choosing $x$ as the direction of propagation, and $\eta$ as the amplitude of the wave, the magnetic field is given by
\begin{eqnarray}
B_x&=&B_0 \\
B_y&=&\eta B_0\cos[k(x-v_A t)]\\
B_z&=&\eta B_0\sin[k(x-v_A t)]\,,
\end{eqnarray}
where $B_0$ is the uniform magnetic field along $x$, $k$ is the wave number, while $v_A$ is the Alfven speed at which the wave propagates
(see \cite{DelZanna2007} for its analytic form).
%
%
The vector tips of the transverse velocity field
describe circles in the $yz$ plane normal to $\vec{B}_0$, according to 
\begin{equation}
v_y=-v_A B_y/B_0,~~~v_z=-v_A B_z/B_0\,.
\end{equation}
We have used $\rho=p=B_0=\eta=1$, and since the wave is incompressible, the background values of $\rho$ and $p$ are not affected.
The test has been performed in two spatial dimensions, using periodic boundary conditions, over the computational domain $\Omega=[0; 2\pi]\times[0; 2\pi]$. 
We compare the numerical solution with the analytic one
after one period $T=L/v_A=2\pi/v_A$.
The results of this analysis are reported in Tab.~\ref{tab:CPA}, which report 
the $L_1$, $L_2$ and $L_\infty$ norms of the error of $B^y$. 
The Rusanov flux has been adopted, with $\ell_{\rm max}=2$ and a Courant factor
$\rm{CFL}=0.8$. 
We emphasize that, due to the smoothness of the solution, the subcell limiter is never activated.
As it is apparent from the table,  the nominal order of convergence is essentially confirmed.

\subsection{Riemann problems} 
%
\begin{table} 
  \numerikNine  
\begin{center} 
\begin{tabular}{|c|c||cccccccc|c|c|} 
\hline
Problem    && $\rho$ &$(v_x$&$v_y$&$v_z)$ & $p$ & $(B_x$&$B_y$&$B_z)$ & $t_{\text{final}}$ & $\gamma$ \\
\hline
\hline
\multirow{1}{*}{\rotatebox{0}{\textbf{RP1}} } 
                              &$x > 0$  & 0.125   &  0.0 & 0.0 &0.0 & 0.1  & 0.5&-1.0&0.0 & \multirow{2}{*}{0.4}&\multirow{2}{*}{2.0}\\ 
(Test 1 in \cite{BalsaraRMHD})&$x \leq 0$  & 1.0  &    0.0 & 0.0 &0.0 & 1.0 & 0.5& 1.0&0.0 &   &\\ 
\hline
\hline
\multirow{1}{*}{\rotatebox{0}{\textbf{RP2}}}
                                 &$x > 0$ & 1.0   &  -0.45 & -0.2 & 0.2 & 1.0  & 2.0&-0.7&0.5 &\multirow{2}{*}{0.55}&\multirow{2}{*}{ $\left.5\middle/ 3\right.$}\\ 
(Test 5 in \cite{BalsaraRMHD})&$x \leq 0$ & 1.08  &    0.4 & 0.3 & 0.2 & 0.95 & 2.0& 0.3&0.3 &  &\\ 
\hline
\end{tabular} 
\caption{
\label{tab:RP1D}
Initial conditions for the one--dimensional Riemann problems.} 
\end{center}
\end{table} 
%
\begin{figure}
  \begin{center} 
  \begin{tabular}{c} 
   \includegraphics[width=0.75\textwidth]{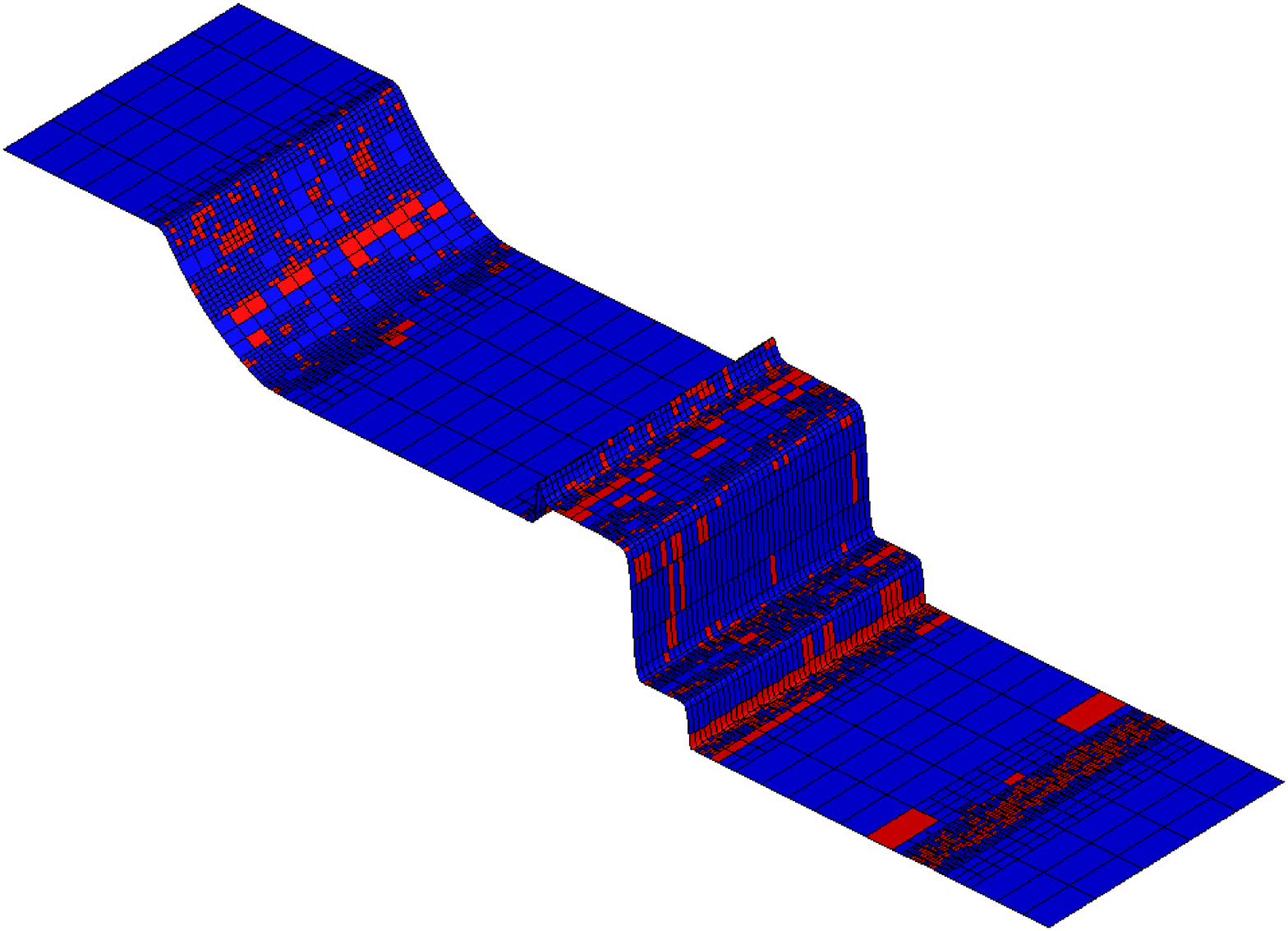} \\
    \includegraphics[width=0.75\textwidth]{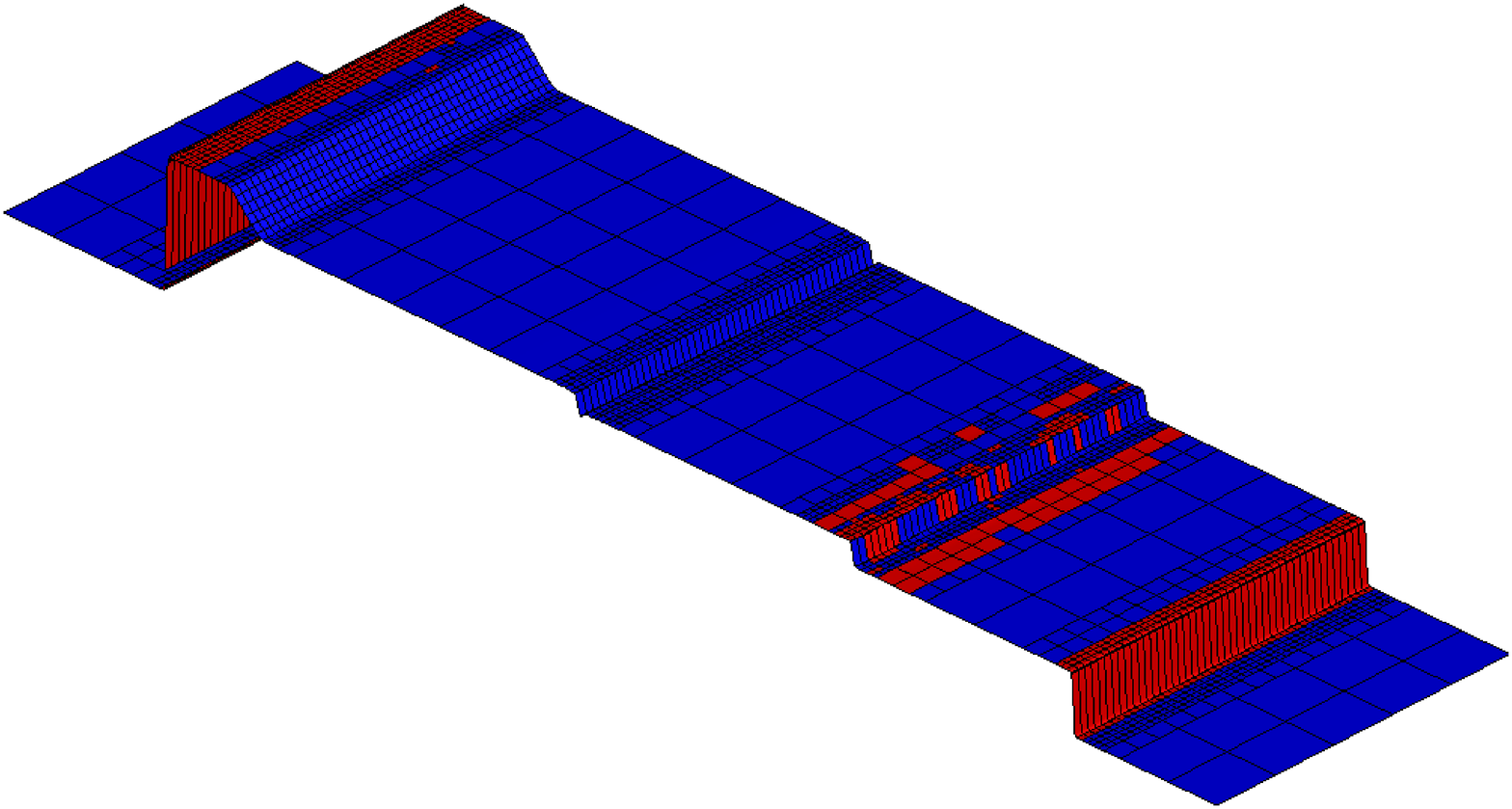}
  \end{tabular}
   \caption{ \label{fig:Balsara3D}
			3D view of the density variable and the corresponding AMR grid. Top panel: RP1 at $t_{\text{final}}=0.4$ (coarsest grid of $40\times5$ elements).  Bottom panel: 
      RP2 at $t_{\text{final}}=0.55$ (coarsest grid of $25\times5$ elements). The limited cells, using the subcell ADER-WENO3 finite volume scheme, 
			are highlighted in red, while unlimited DG-$\mathbb{P}_3$ cells are highlighted in blue.
     }
  \end{center}
\end{figure}

\begin{figure}
  \begin{center} 
  \begin{tabular}{cc} 
    \includegraphics[width=0.47\textwidth]{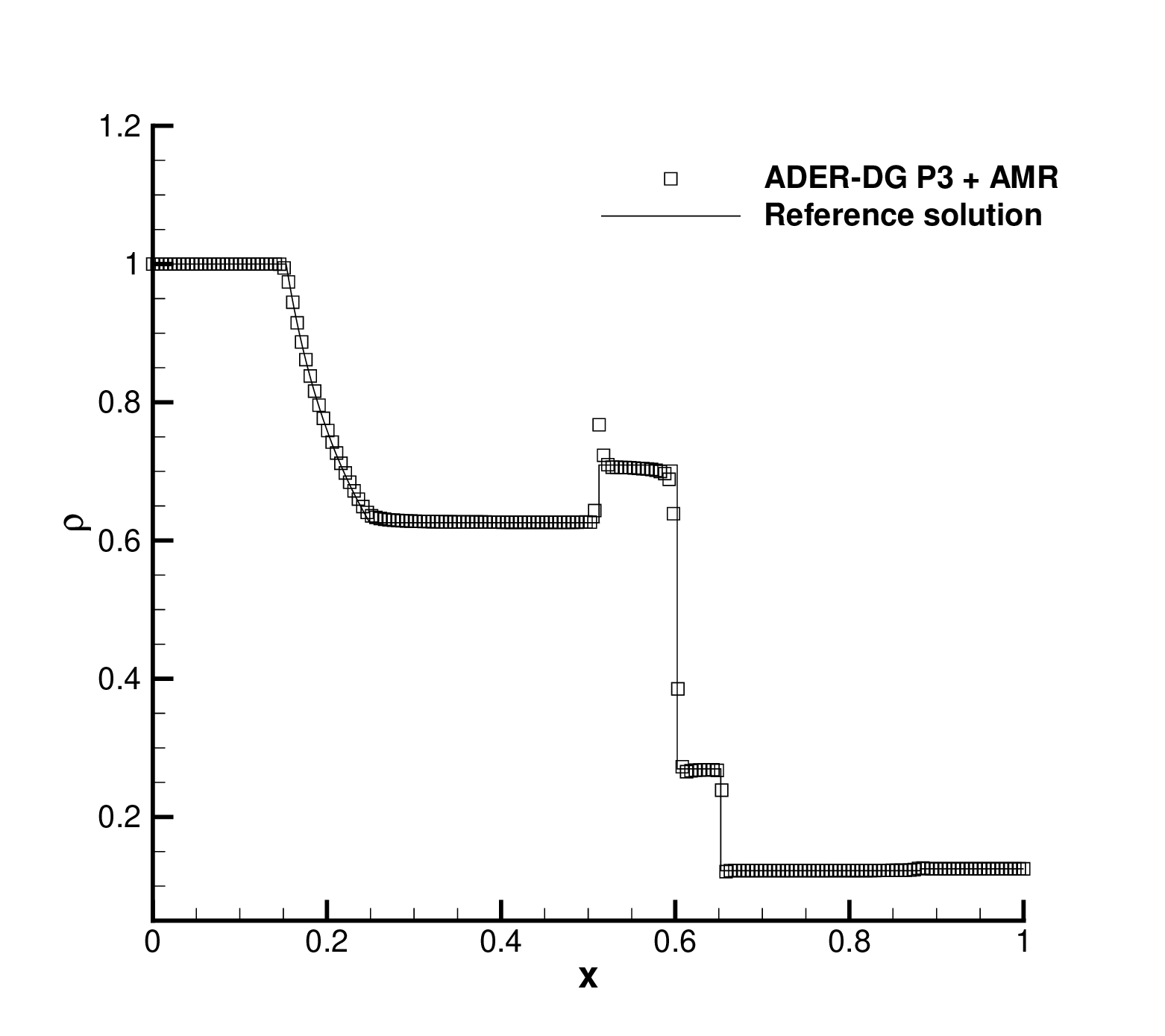}&
    \includegraphics[width=0.47\textwidth]{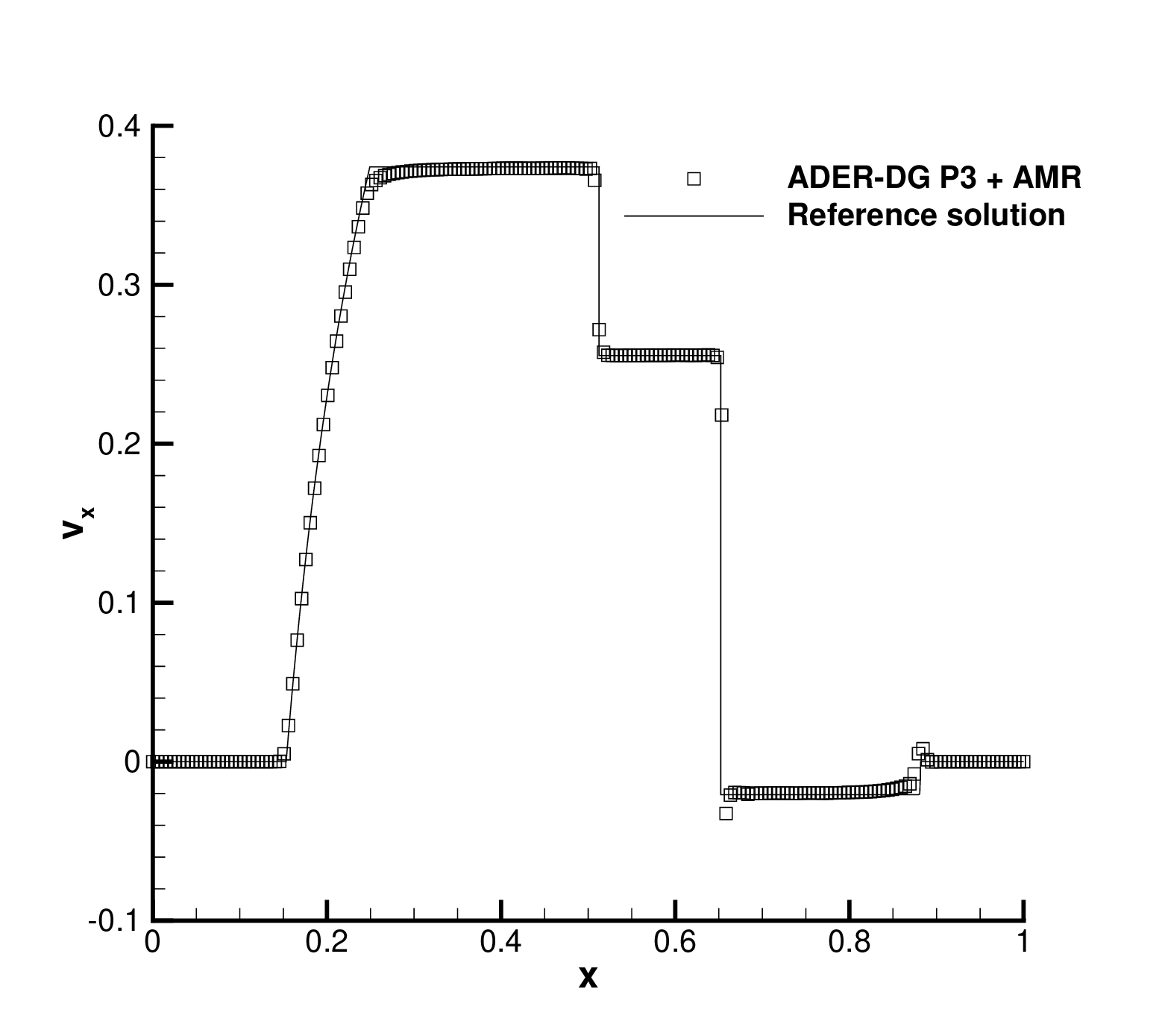}
		\\
    \includegraphics[width=0.47\textwidth]{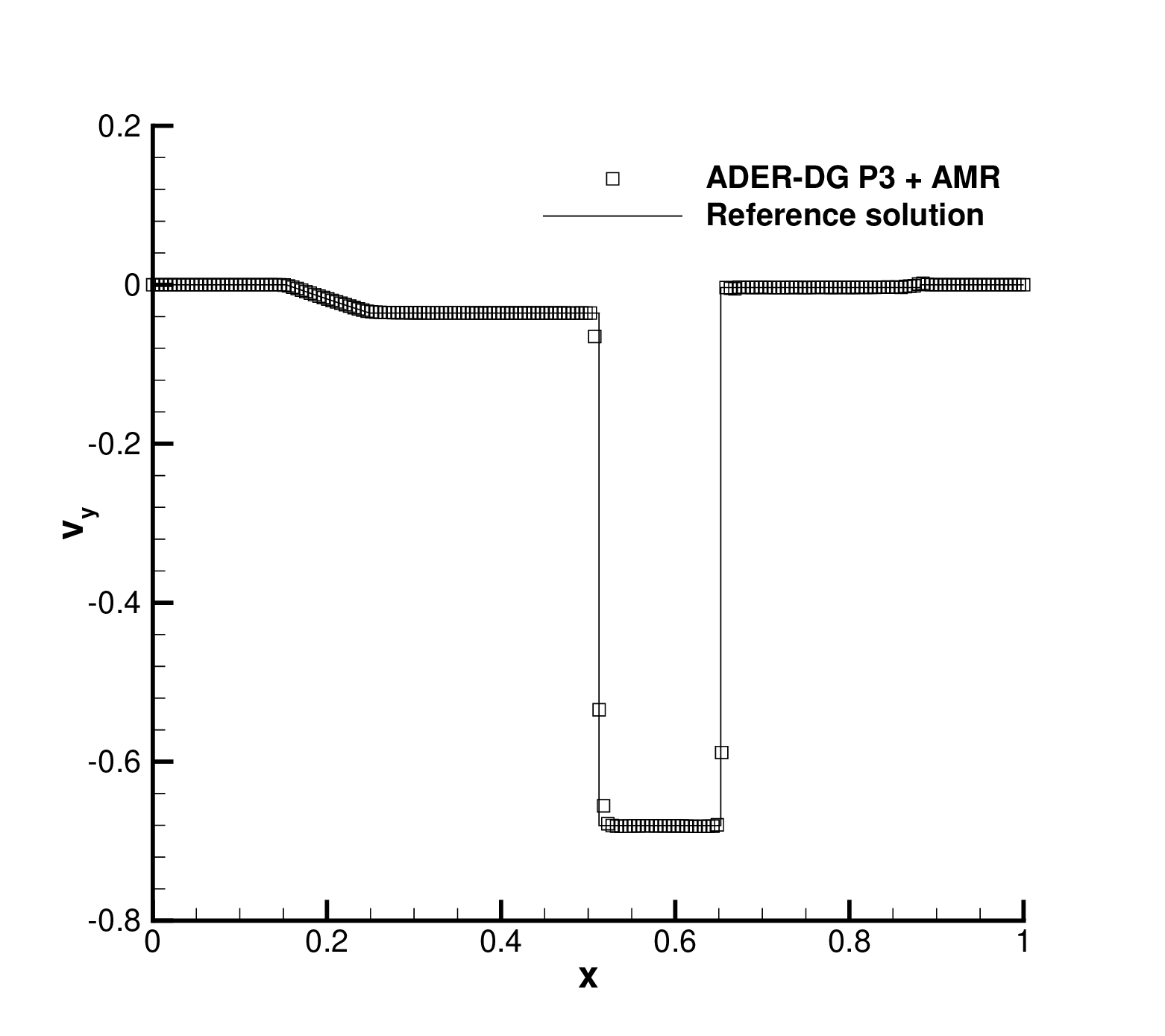} &
    \includegraphics[width=0.47\textwidth]{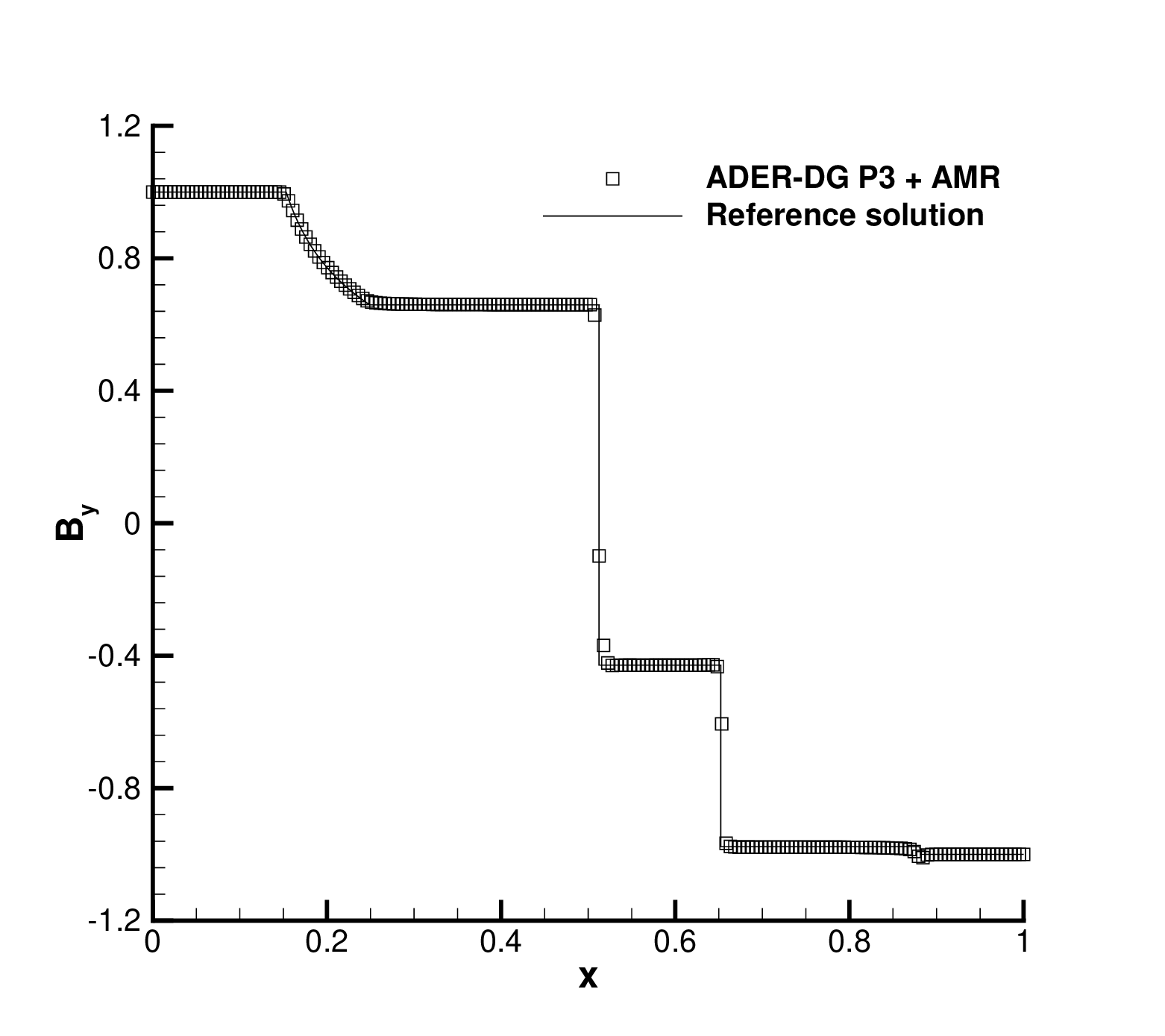} 
  \end{tabular}
   \caption{ \label{fig:Balsara1}
			RP1: physical variables interpolated along a 1D cut on $200$ equidistant points at $t_{\text{final}}=0.4$, starting 
			from a coarsest grid of $40\times5$ elements by using the ADER-DG-$\mathbb{P}_3$ scheme supplemented with the \aposteriori ADER-TVD subcell limiter. 
     }
  \end{center}
\end{figure}
\begin{figure}
  \begin{center} 
  \begin{tabular}{cc} 
    \includegraphics[width=0.47\textwidth]{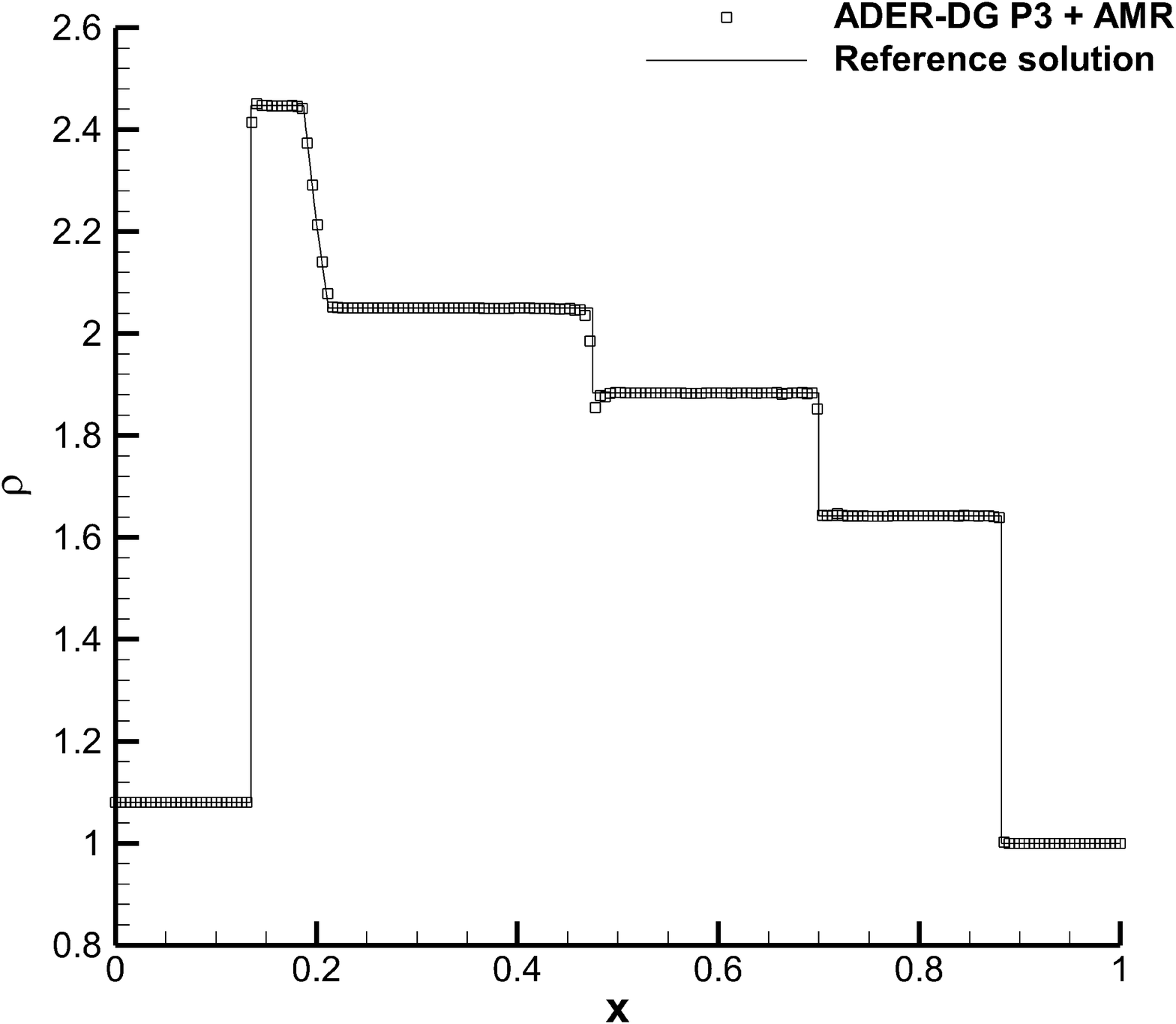}&
    \includegraphics[width=0.47\textwidth]{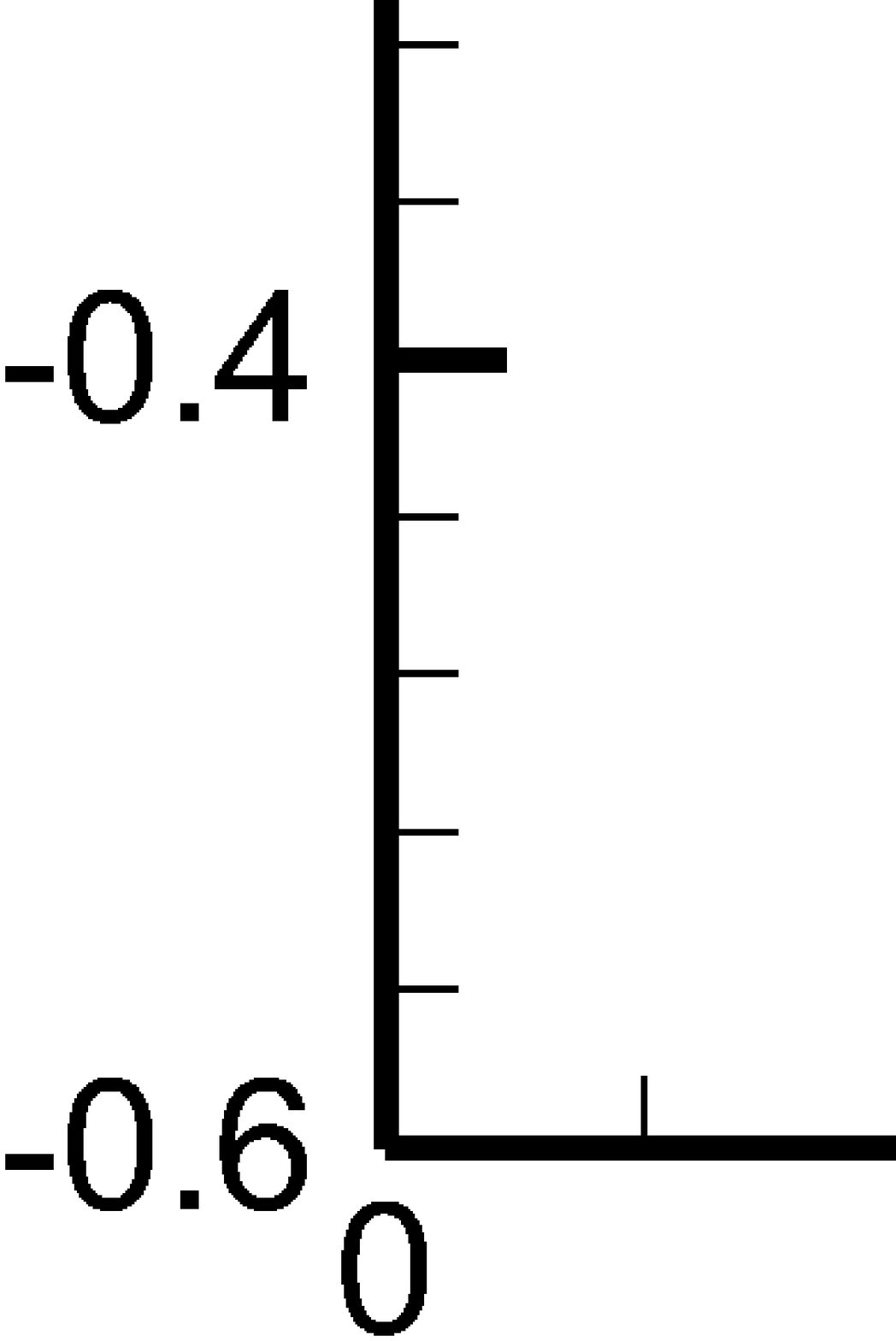}
		\\
    \includegraphics[width=0.47\textwidth]{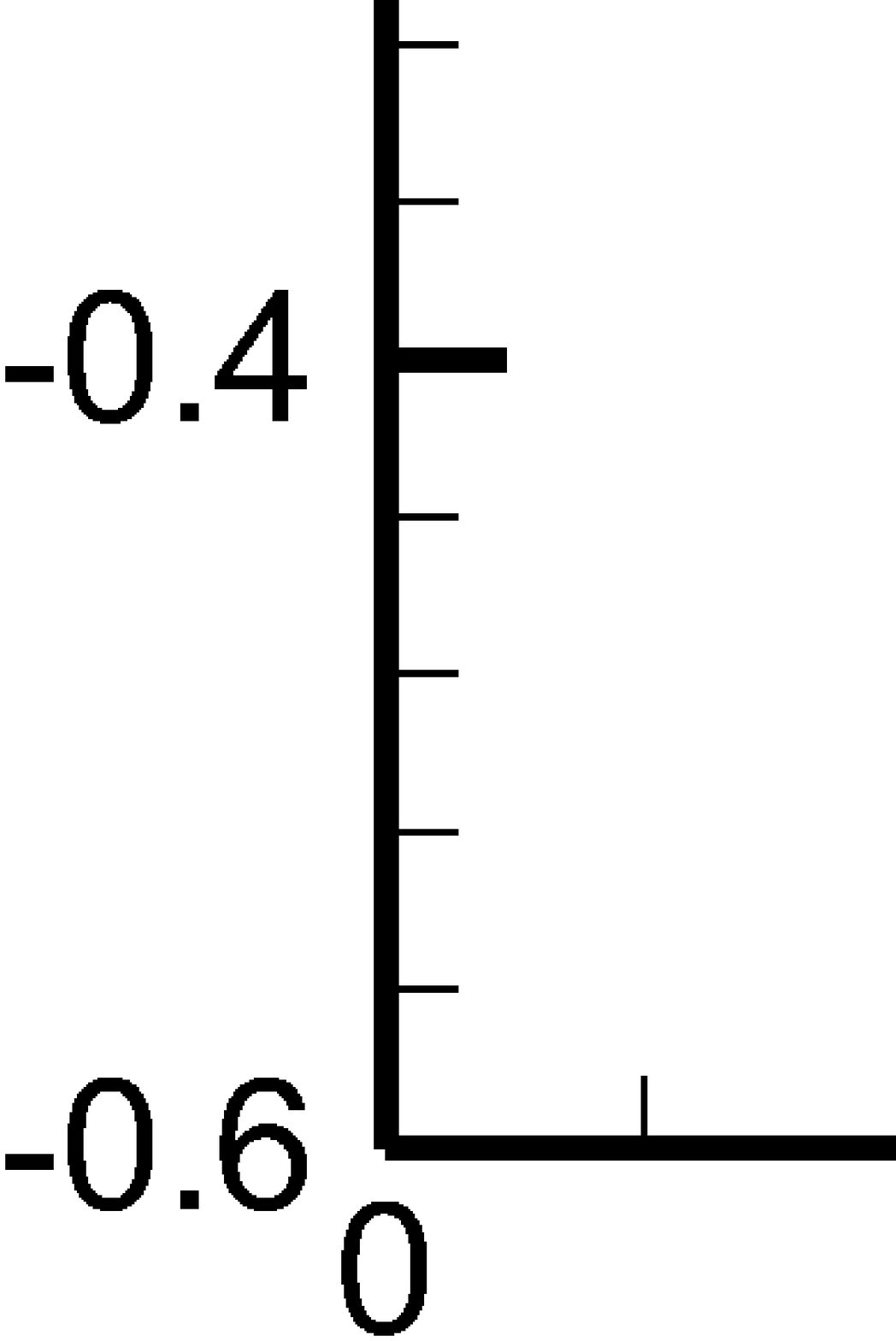}&
    \includegraphics[width=0.47\textwidth]{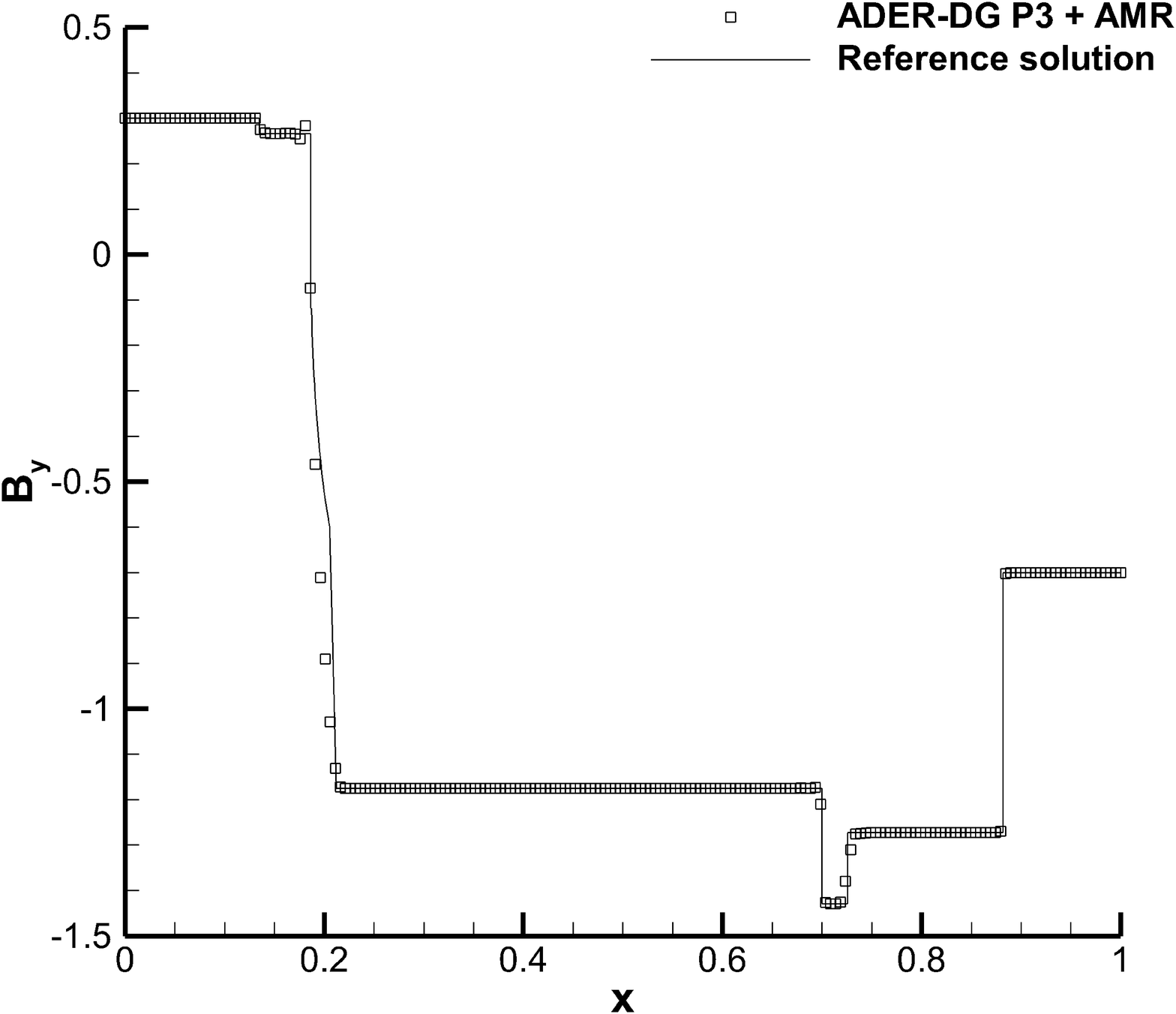} 
  \end{tabular}
   \caption{ \label{fig:Balsara5}
			RP2: physical variables interpolated along a 1D cut on $200$ equidistant points at $t_{\text{final}}=0.55$, starting 
			from a coarsest grid of $25\times5$ elements by using the ADER-DG-$\mathbb{P}_3$ scheme supplemented with the \aposteriori ADER-WENO3 subcell limiter. 
     }
  \end{center}
\end{figure}

Once the convergence properties have been verified, we  consider a few relevant shock-tube problems
to test the new ADER-DG-AMR method. Specifically, we concentrate on 
two classical Riemann problems for RMHD, already proposed by \cite{vanPutten1993} and 
classified as Test 1 and Test 5 in Table 1 of \cite{BalsaraRMHD}. 
The corresponding initial conditions, referred to as RP1 and RP2 in the following, are given in Tab.~\ref{tab:RP1D}, which reports also
the final times and the adiabatic indices.\footnote{
The adiabatic index $\gamma$ of RP1 in unphysical, as it violates Taub's inequality based on kinetic theory \citep{Taub1948,Mignone2007}
but it is fixed equal to $2$ anyway to ease comparison with \cite{vanPutten1993} and \cite{BalsaraRMHD}.}

The two chosen Riemann problems are solved along two coarse grids of $40\times5$ and $25\times5$ elements, respectively. Then, the intial grid is adaptively refined in space and time according to $\mathfrak{r}=3$ and $\ell_{\rm max}=2$. The computational domain is only formally two-dimensional, 
since the second direction $y$ acts as a passive one. 
Both tests have been solved using the ADER-DG-$\mathbb{P}_3$ scheme, 
but they differ in the subcell limiter, which is the second order TVD finite volume scheme for RP1, while it is 
the third order ADER-WENO finite volume scheme for RP2.
The HLL solver has been used for both RP1 and RP2 with a Courant factor $\rm{CFL}=0.5$. 
The damping factor for the divergence-cleaning procedure is set to $\kappa=10$.

Figure~\ref{fig:Balsara3D} shows the three-dimensional plot of the solution for the rest mass density and the corresponding AMR grid, by plotting the real DG polynomials (highlighted in blue) for every single unlimited cell and the piecewise linear interpolation of the ADER-WENO limiter along the subcell averages 
(highlighted in red) for the limited cells.
A reference solution for these Riemann problems is computed with the exact Riemann solver proposed by \cite{Giacomazzo:2005jy}. 
Figures~\ref{fig:Balsara1} and \ref{fig:Balsara5} show the comparison with the reference solution by plotting
the rest mass density, the $x-$ and the $y-$ velocity components and the $y-$ component of the magnetic field, interpolated over a one-dimensional cut composed of $200$ equidistant points at the final state.
A remarkable agreement between the numerical and the reference solution is obtained. All the  waves are well captured, five for RP1 and seven for RP2.
More specifically, RP1 has
a left-going and a right-going fast rarefaction wave, 
a left-going compound wave, a central contact discontinuity, and a right-going slow shock.
RP2 has instead 
a left-going and a right-going fast shock, a left-going and a right going Alf\'en wave, a left-going rarefaction wave, a central contact discontinuity and a right-going slow shock. 
Due to the combined action of the subcell-limiter and of AMR, all discontinuities are resolved within just one cell or two cells at most.
We note that, while the compound wave is absent by construction in the exact solution,
its width in the numerical solution
is rather small and its amplitude is also comparatively smaller with respect to that obtained with other numerical schemes,  indicating that this might really be a numerical artifact. However, see also the discussion in \cite{Mignone2009}.

The small asymmetries visible in Fig.~\ref{fig:Balsara3D} for RP1 along the passive $y$ direction  are attributable to the joint interaction between: (1) 
the lack of reconstruction in characteristic variables, which could typically help in these cases;
(2) some residual \emph{post-shock} oscillations, that in  \citep{Balsara98} 
were suppressed by means of artificial viscosity.
In spite of these small defects, these results show the capabilities of the
new scheme, which does not resort to any artificial viscosity,
in resolving the strongly non-linear waves of RMHD equations, for which an unlimited DG schemes would catastrophically fail.

%
\subsection{The rotor problem} 
%
\begin{figure*}
\begin{center}
\begin{tabular}{lr}
\includegraphics[width=0.45\textwidth]{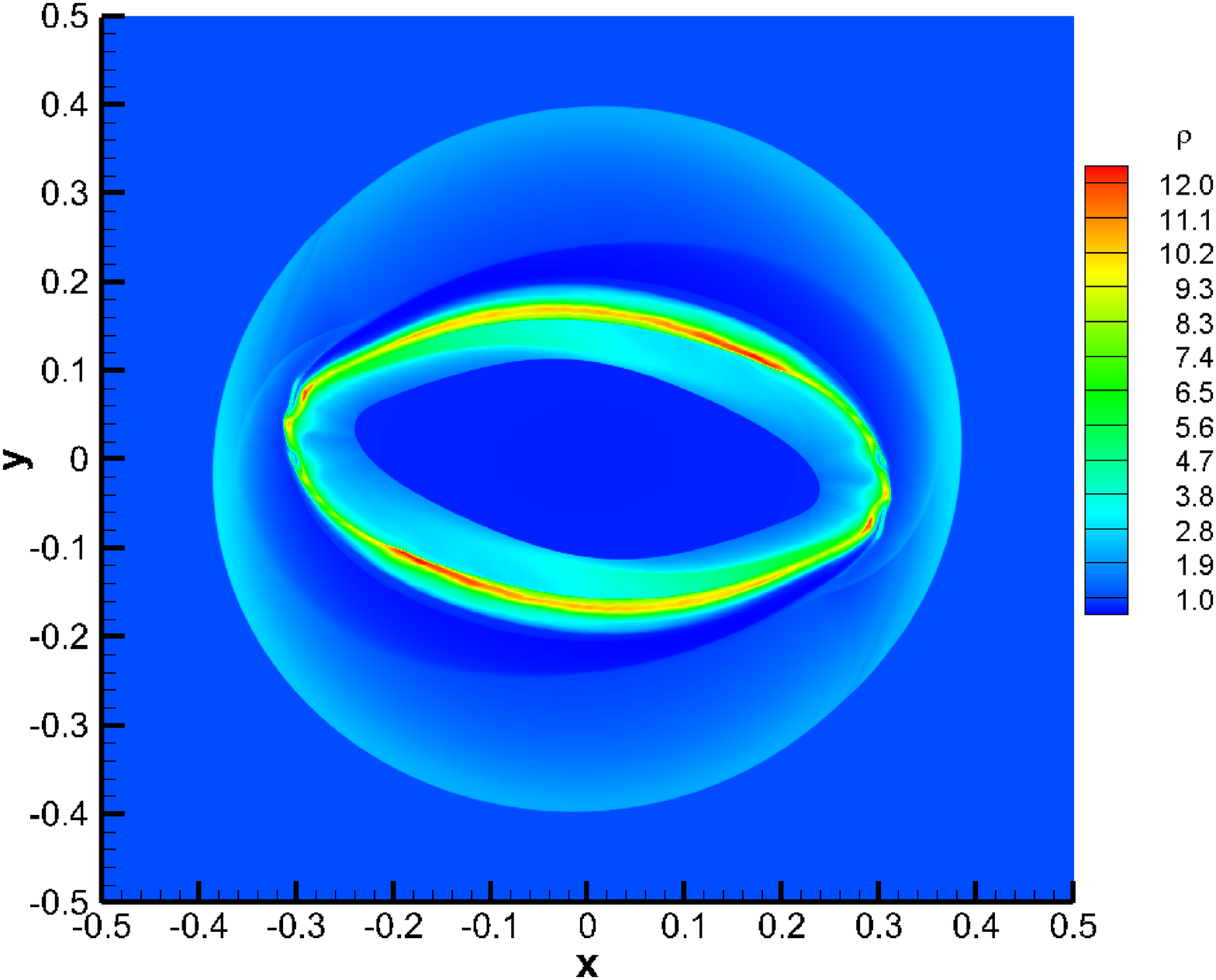}   &  
\includegraphics[width=0.45\textwidth]{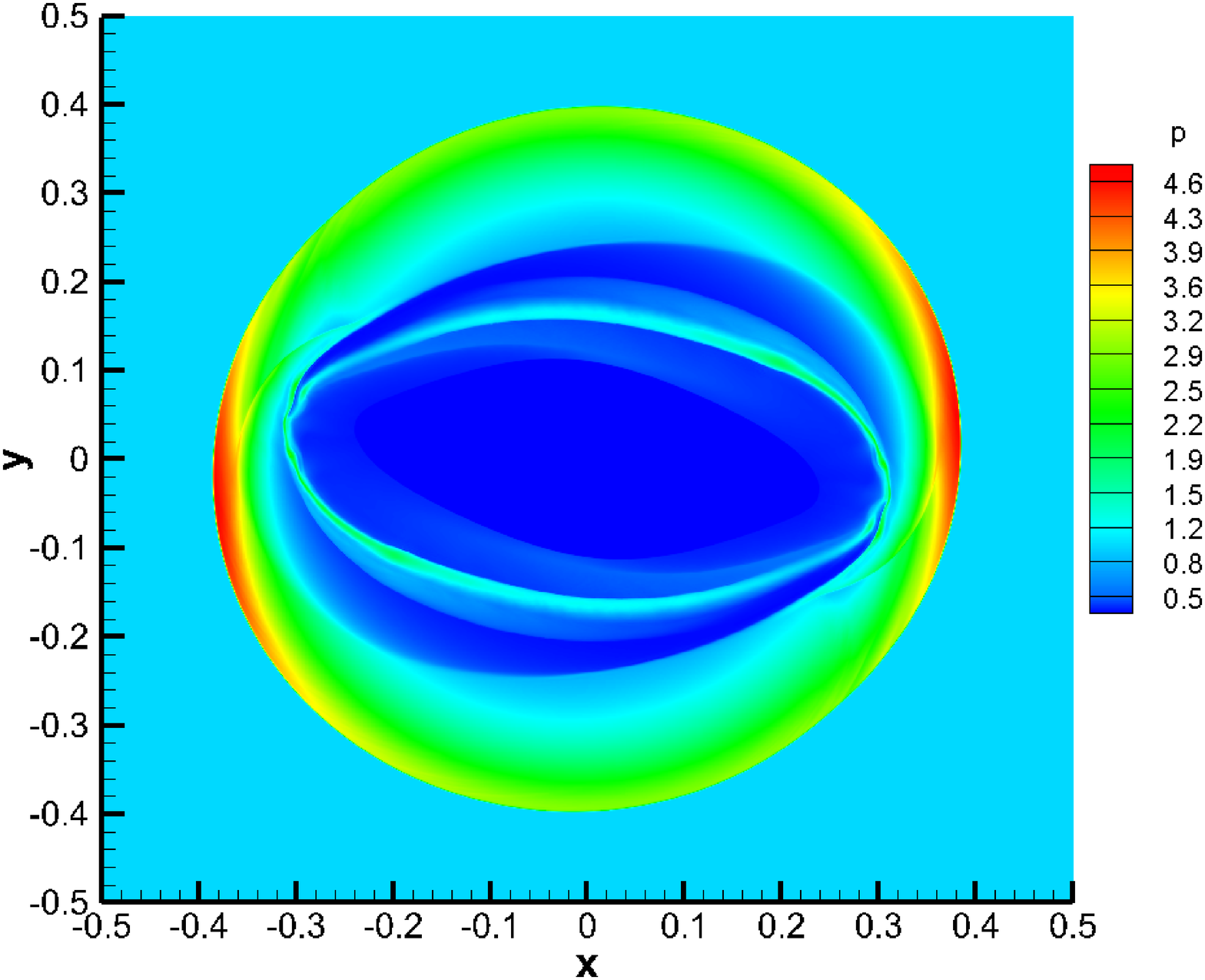}   \\    
\includegraphics[width=0.45\textwidth]{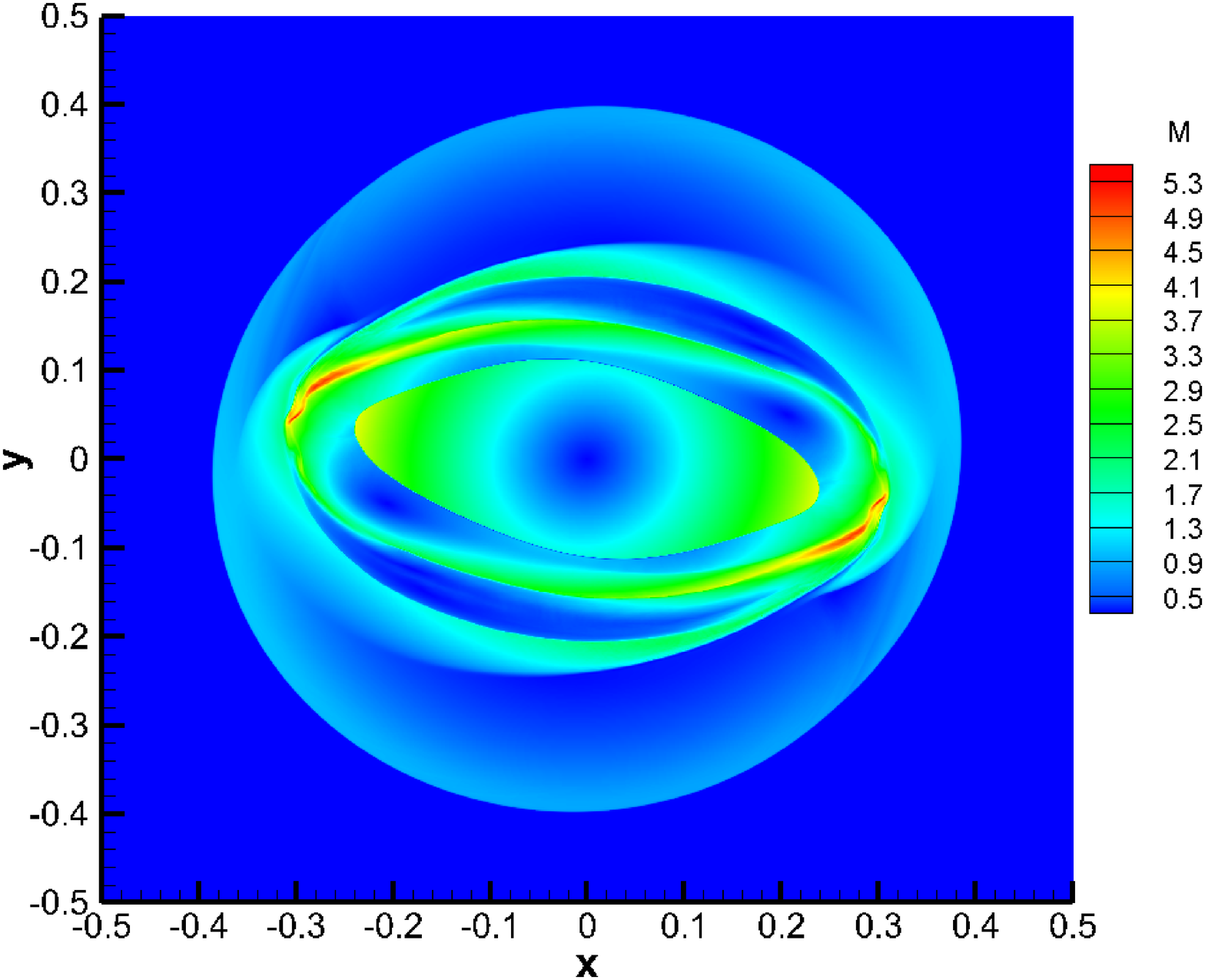} & 
\includegraphics[width=0.45\textwidth]{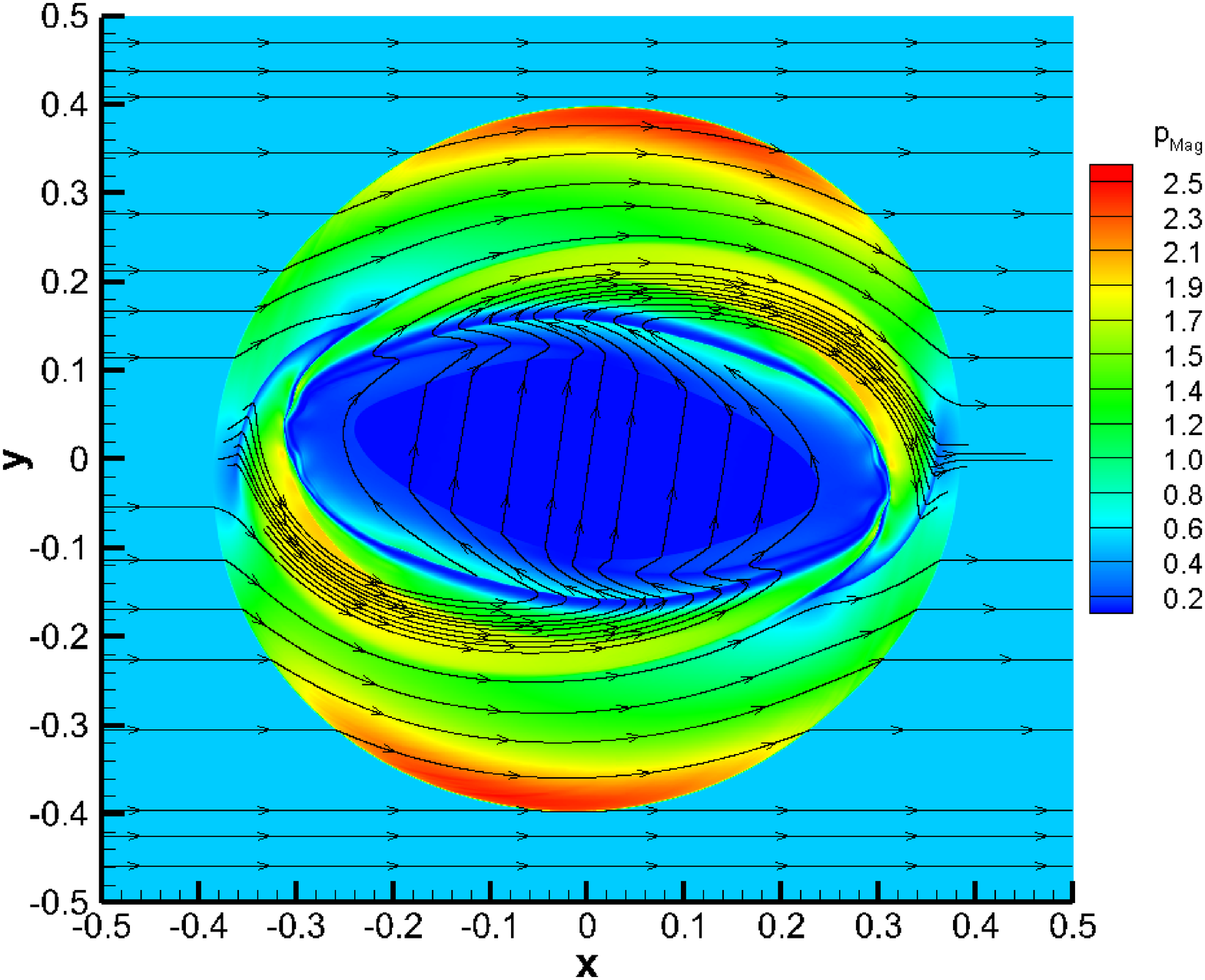} \\ 
\includegraphics[width=0.43\textwidth]{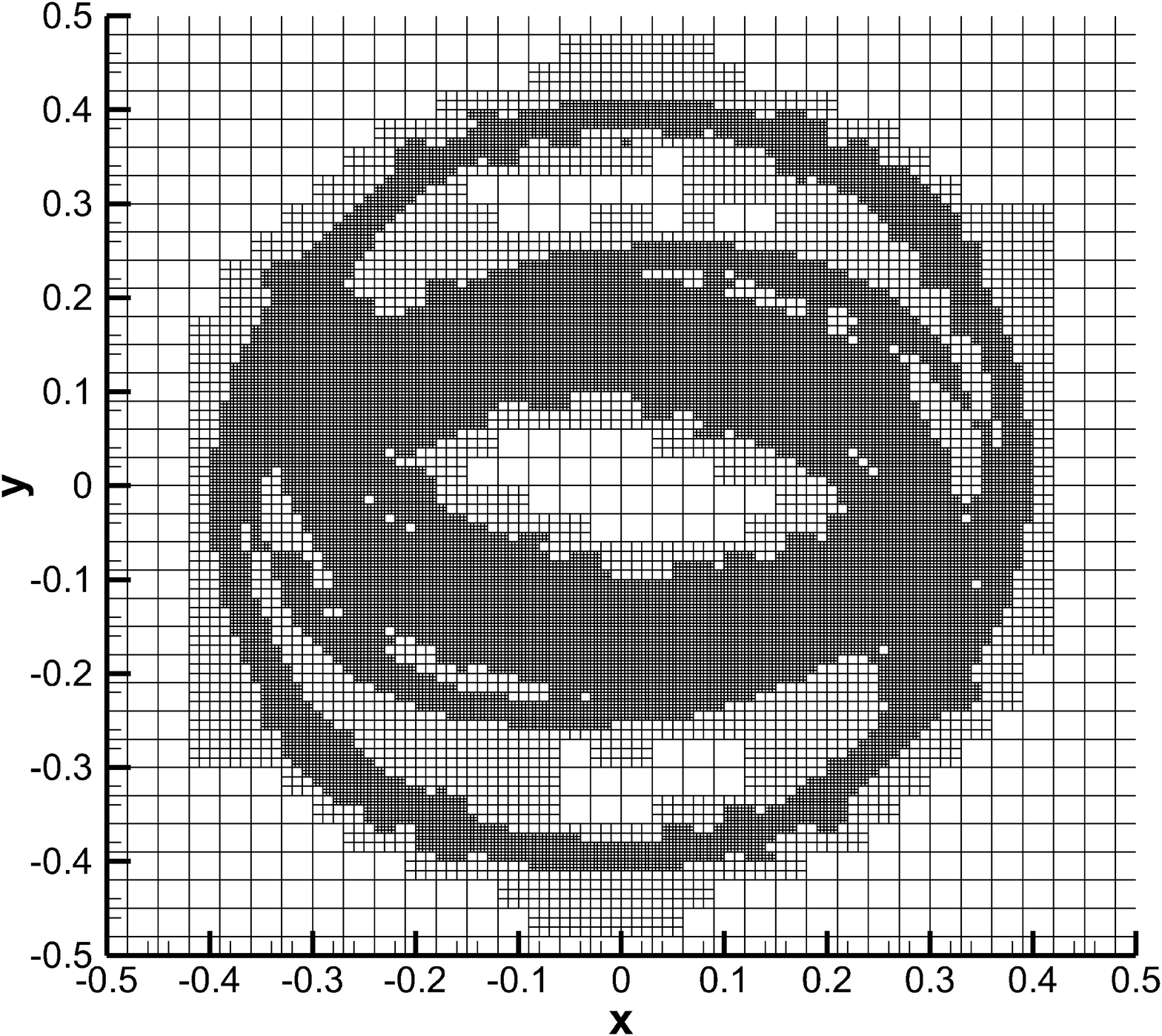}     & 
\includegraphics[width=0.43\textwidth]{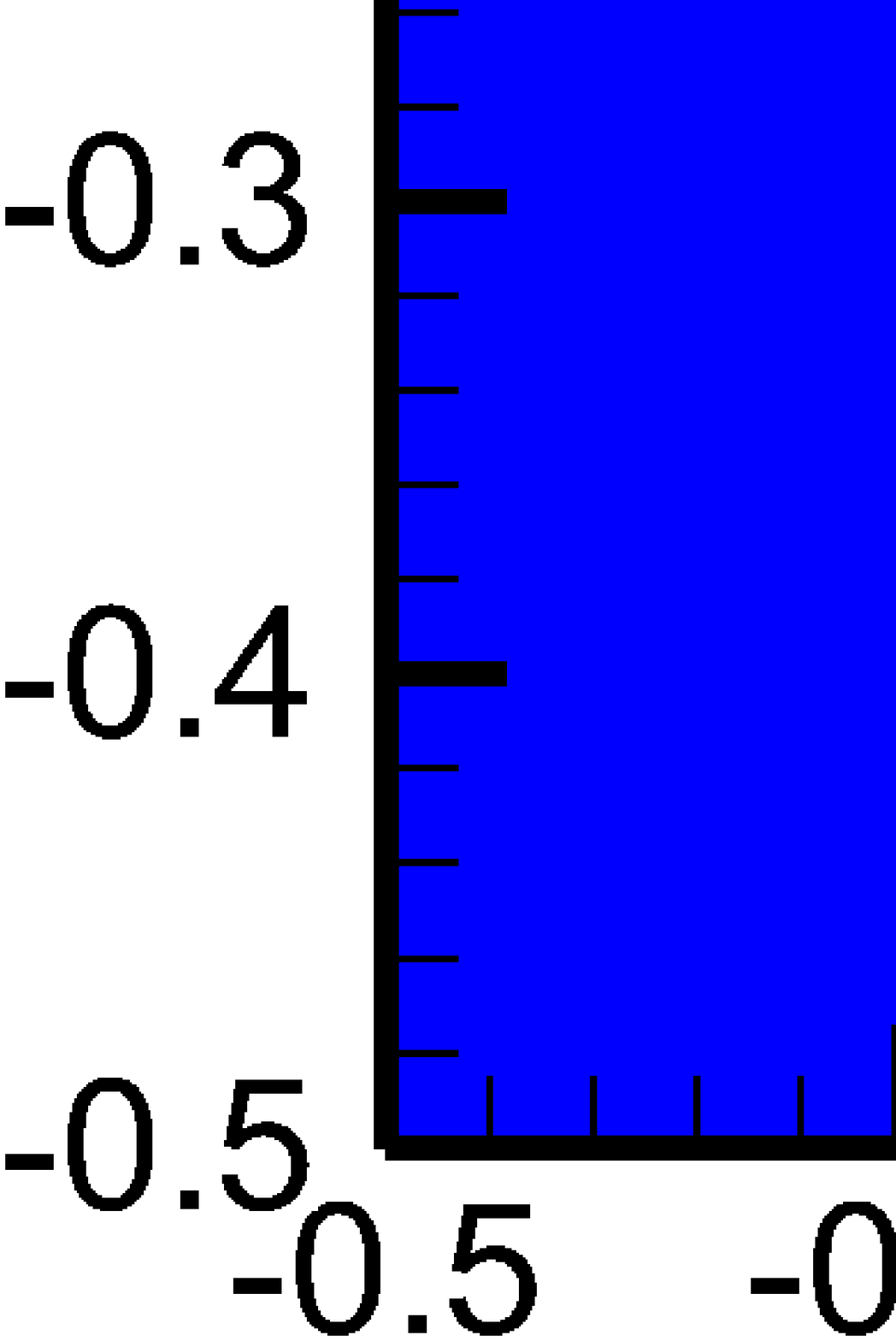} 
\end{tabular} 
\caption{Solution of the RMHD rotor problem at time $t=0.4$, obtained with the ADER-DG $\mathbb{P}_5$ scheme 
supplemented with the \aposteriori second order TVD subcell limiter. Top panels: rest-mass density (left) and thermal pressure (right). Central panels: Mach number (left) and magnetic pressure (right).
Bottom panels: AMR grid (left) and limiter map (right) with troubled cells marked in red and regular unlimited cells marked in blue. 
}
\label{fig:RMHDrotor}
\end{center}
\end{figure*}
As a first genuinely two dimensional test we consider the relativistic version of the MHD rotor problem, originally proposed 
by \cite{BalsaraSpicer1999}, and solved by a number of authors over the years, including \cite{DelZanna2003}, \cite{DumbserZanotti}, 
\cite{ADER_MOOD_14} and \cite{Kim2014}. The computational domain is chosen to be $\Omega = [-0.6,0.6]\times[-0.6,0.6]$, 
discretized on a coarse initial grid formed by $40\times40$ elements. The AMR 
framework is activated with a refinement factor $\mathfrak{r}=3$ and a number of refinement levels $\ell_\text{max}=2$. In this problem a cylinder of a high density fluid is rotating rapidly with angular velocity $\omega$, surrounded by a low density fluid at rest. 
The initial conditions are in fact given by
\begin{equation}
\rho=\left\{\begin{array}{cl}
10 & \text{for}\;\; 0\le r\le 0.1; \\ 1 & \text{otherwise};
\end{array}\right.,~~~
\omega=\left\{\begin{array}{cl}
9.95 & \text{for}\;\; 0\le r\le 0.1; \\ 0 & \text{otherwise};
\end{array}\right.,~~~
{\B} = \left(\begin{array}{c}
1.0  \\ 0 \\ 0
\end{array}\right),~~~
p = 1\,,
\label{eq:MHDrotor_ic}
\end{equation}
which imply an initial maximum Lorentz factor $W_{\rm max}\approx 10$ at $r=0.1$.
Transmissive boundary  conditions are applied at the borders.
The spinning of the rotor produces torsional Alfv\'en waves that are launched outside the cylinder, transferring amounts of its initial angular momentum into the external medium. 
The simulation is performed without any linear taper, that means the physical variables between the internal rotor and the fluid at rest are really discontinuous. The adiabatic index is $\gamma=4/3$. For this test, the $\mathbb{P}_5$ version of our ADER-DG scheme was used, combined with the Rusanov Riemann solver. 
Due to the challenging nature of the problem, a  robust second-order TVD scheme, rather then the standard WENO scheme, has been used on the subgrid where the limiter is activated.

Fig.~\ref{fig:RMHDrotor} shows the rest-mass density, the thermal pressure, the relativistic Mach number $M$ and 
the magnetic pressure $p_{\text{Mag}}$ at time $t=0.4$. The latter are computed according to
\begin{equation}
M = \frac{W v }{W_s v_s}, \;\;\;\;\;
p_{\text{Mag}} = \frac{1}{2} b^2 =\frac{B^2/W + (\mathbf{v}\cdot\mathbf{B})^2}{2}\,,
\label{eq:MachMPressure}
\end{equation}
where $v_s$ is the speed of sound and $W_s=(1-v_s^2)^{-1/2}$ is the corresponding Lorentz factor. 
Although an analytic solution is not available for this test, the results shown are in 
very good qualitative agreement with those already reported in the literature. 
In particular, the maximum Lorentz factor of the rotor, which is considerably slowed down by magnetic braking, is 
$W_{\rm max}\approx 2.1$.
Moreover, the 
adopted divergence-cleaning approach works accurately as expected, with no appreciable spurious oscillations 
generated in the rest mass density or in the magnetic field.
Lastly, the behaviour of the space-time AMR  and of the \aposteriori limiter is depicted in the two bottom panels of 
Fig.~\ref{fig:RMHDrotor}: 
the final mesh is shown in the left panel, whereas in the right the troubled zones are represented in red.
Clearly, the activation of the limiter becomes necessary only in a limited number of cells, and precisely where discontinuities are stronger.

%
\subsection{Cylindrical blast wave} 
\begin{figure*}
\begin{center}
\begin{tabular}{lr}
\includegraphics[width=0.45\textwidth]{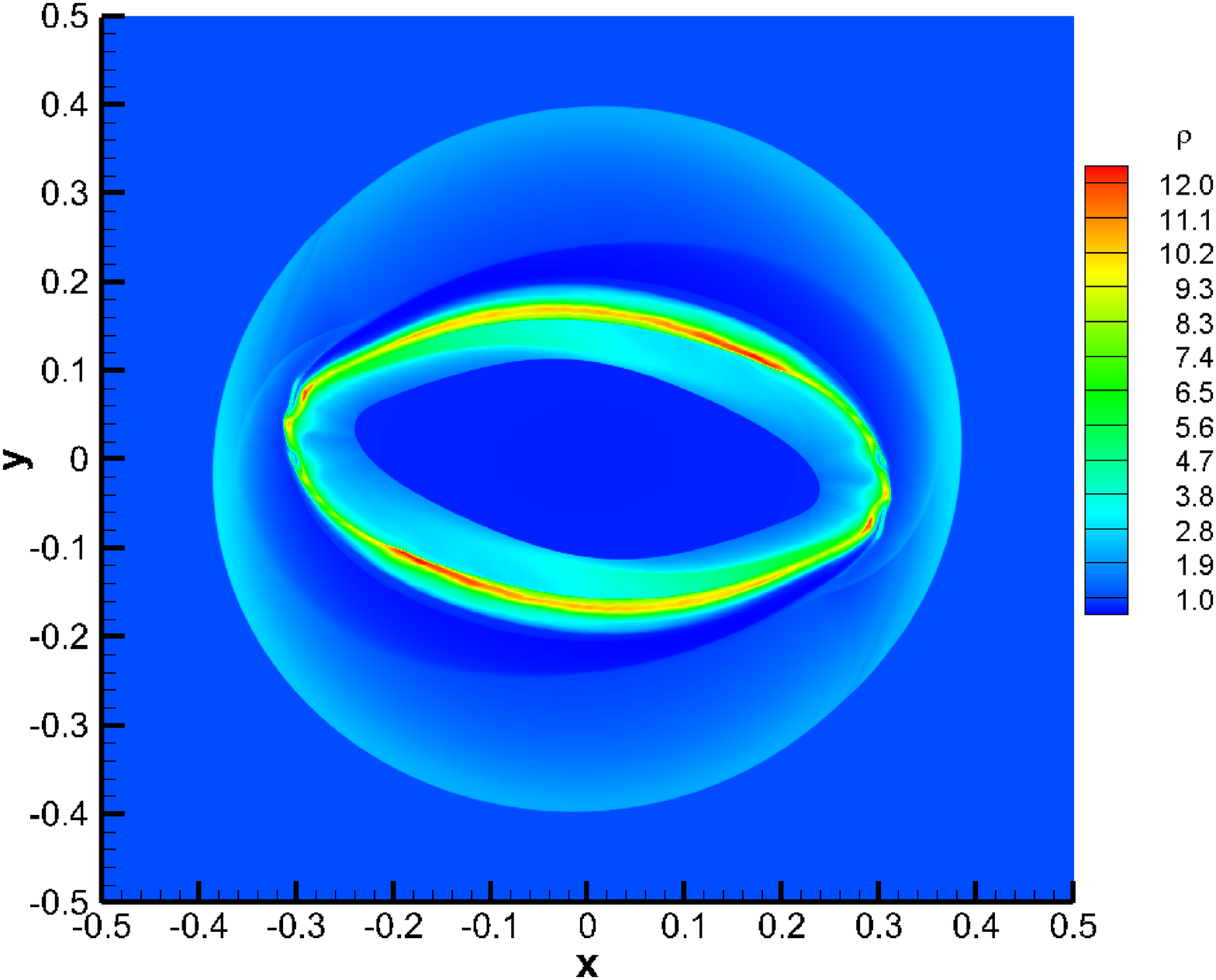}   &  
\includegraphics[width=0.45\textwidth]{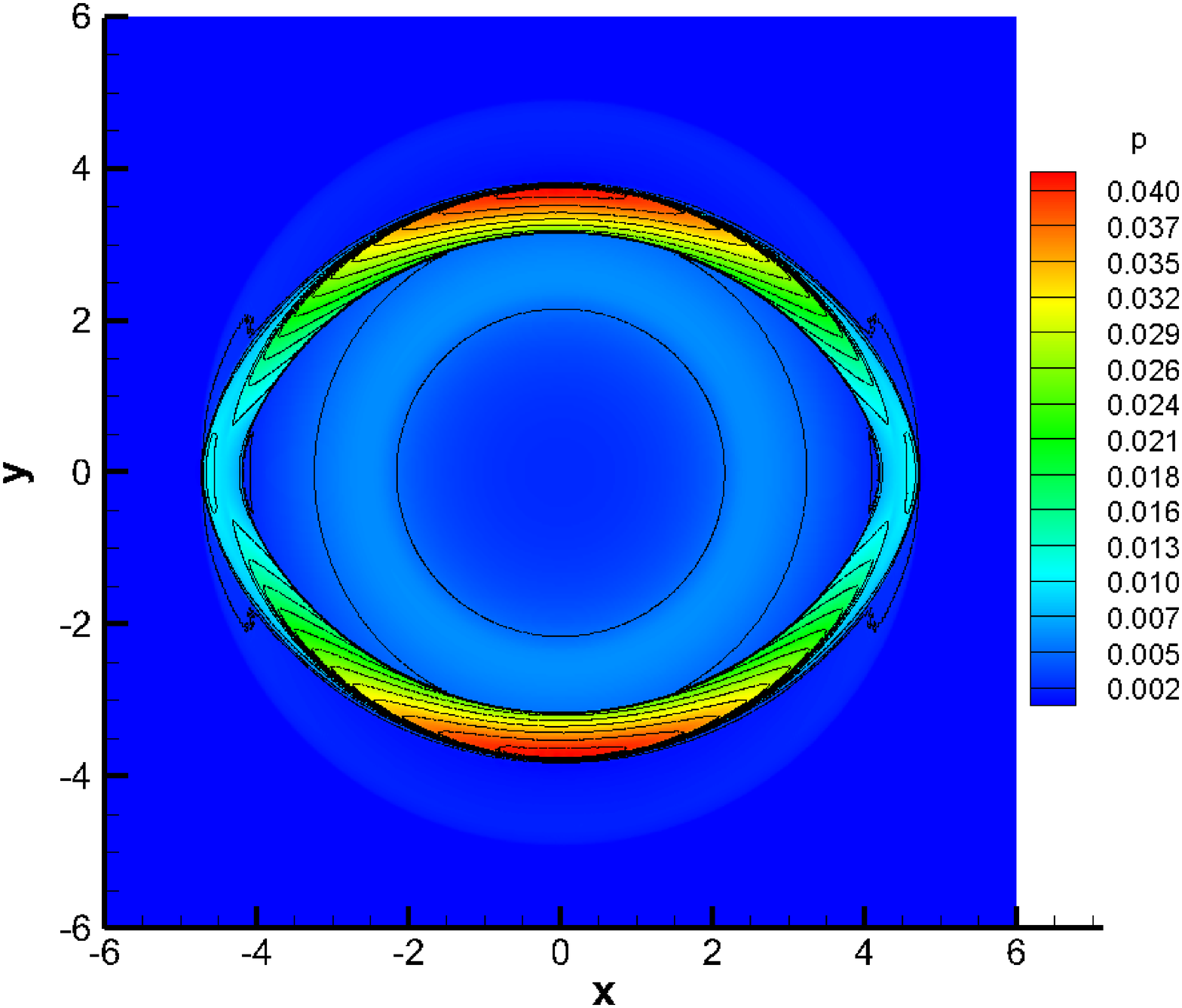}   \\  
\includegraphics[width=0.45\textwidth]{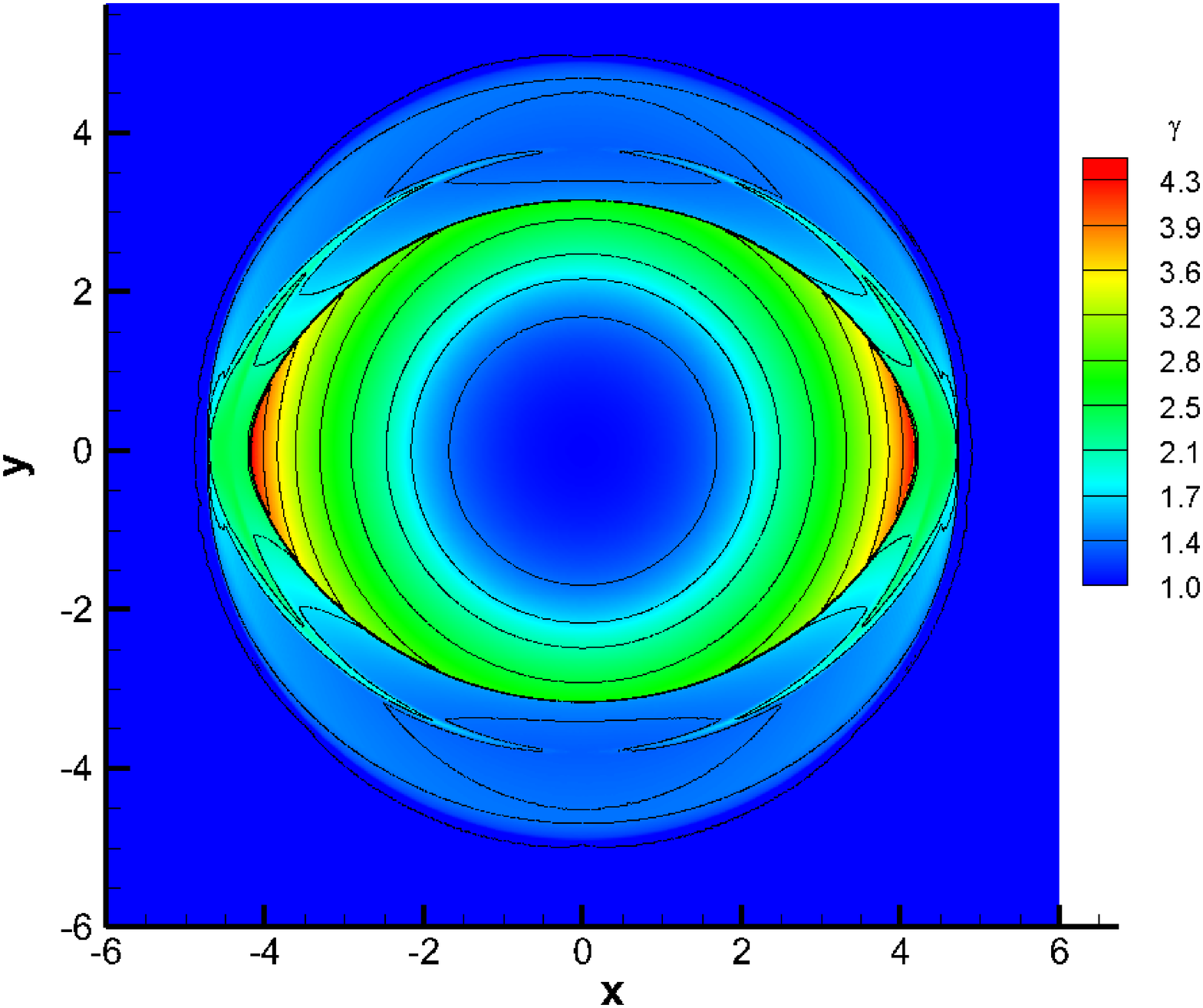} & 
\includegraphics[width=0.45\textwidth]{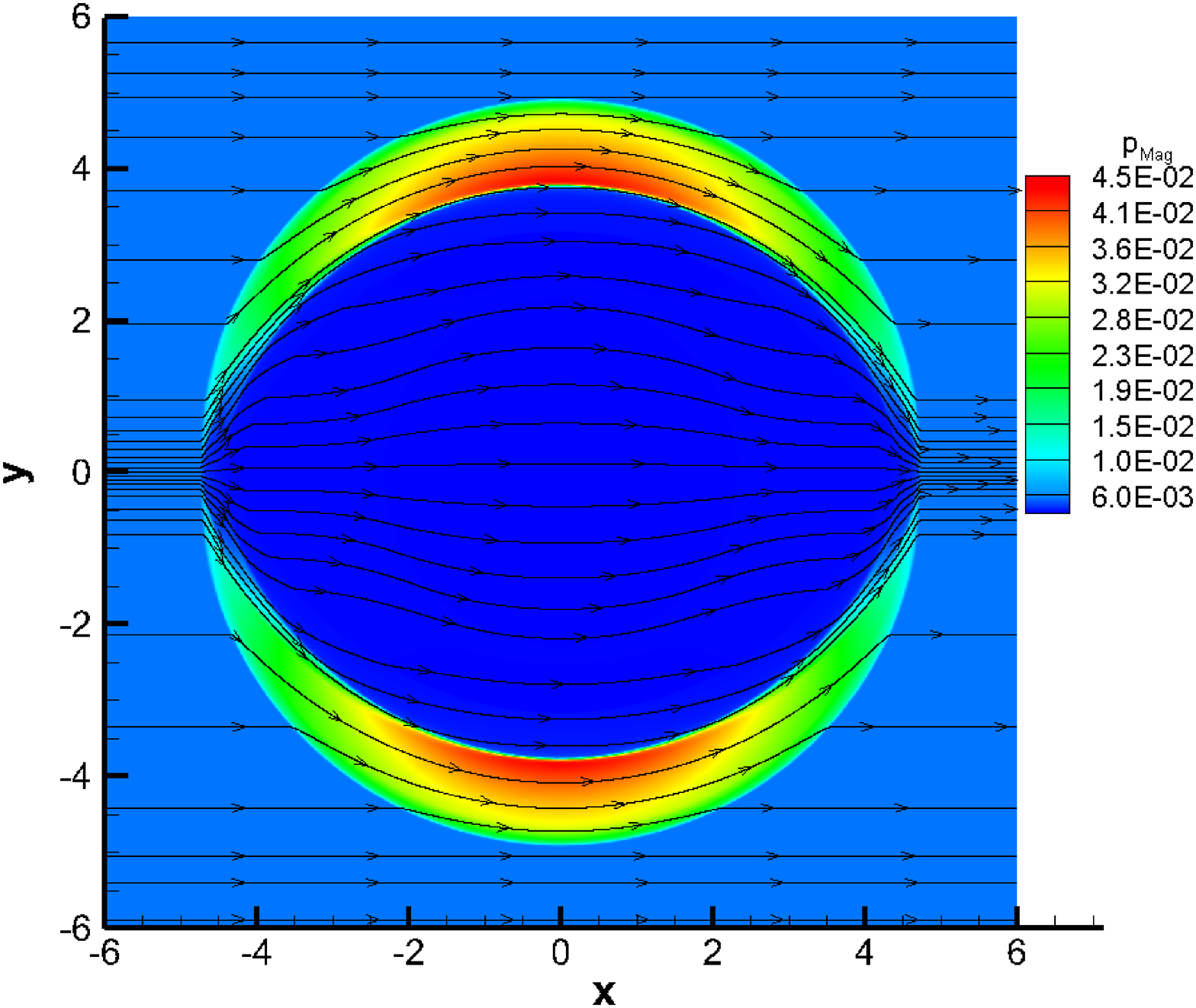} \\ 
\includegraphics[width=0.43\textwidth]{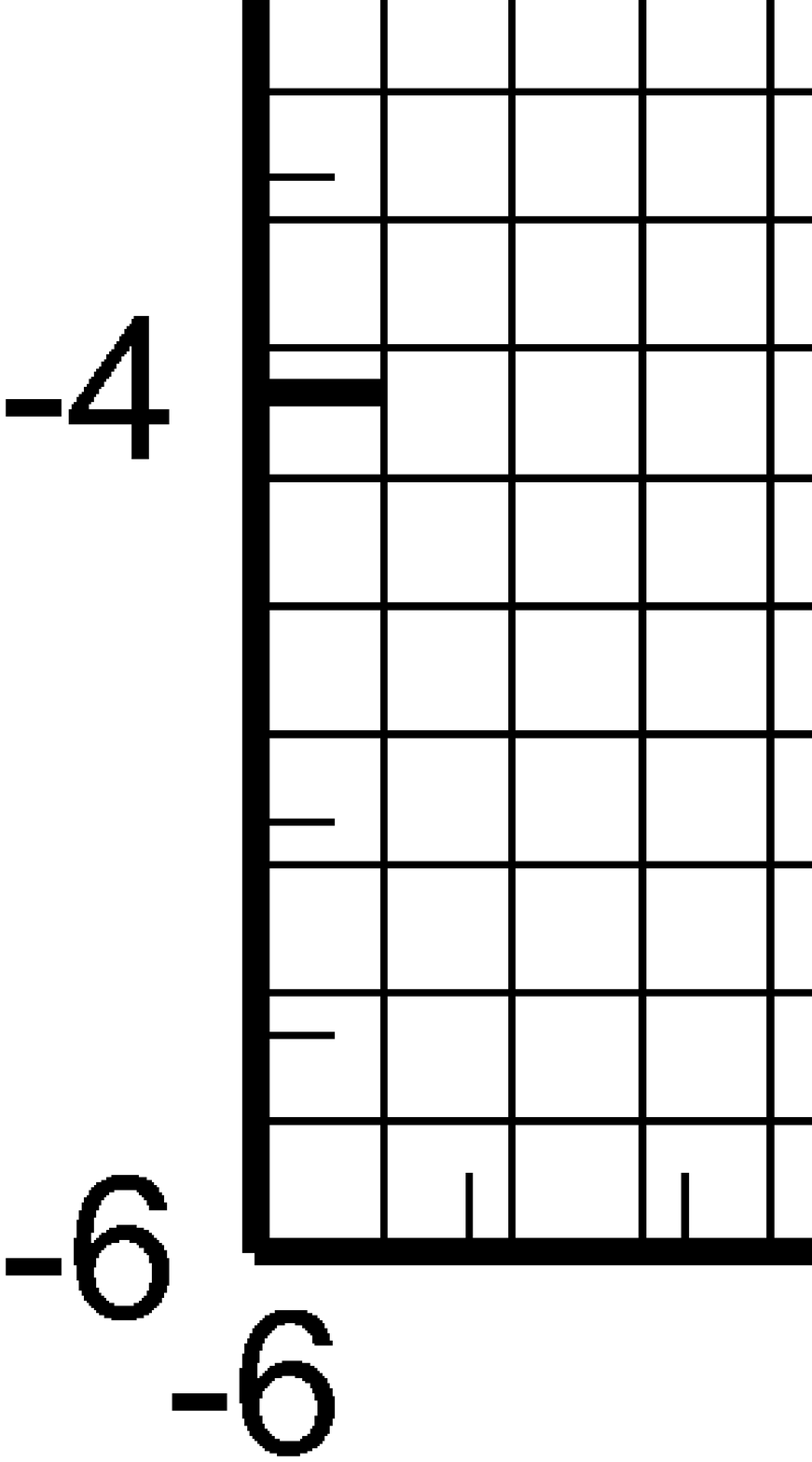}     &  
\includegraphics[width=0.43\textwidth]{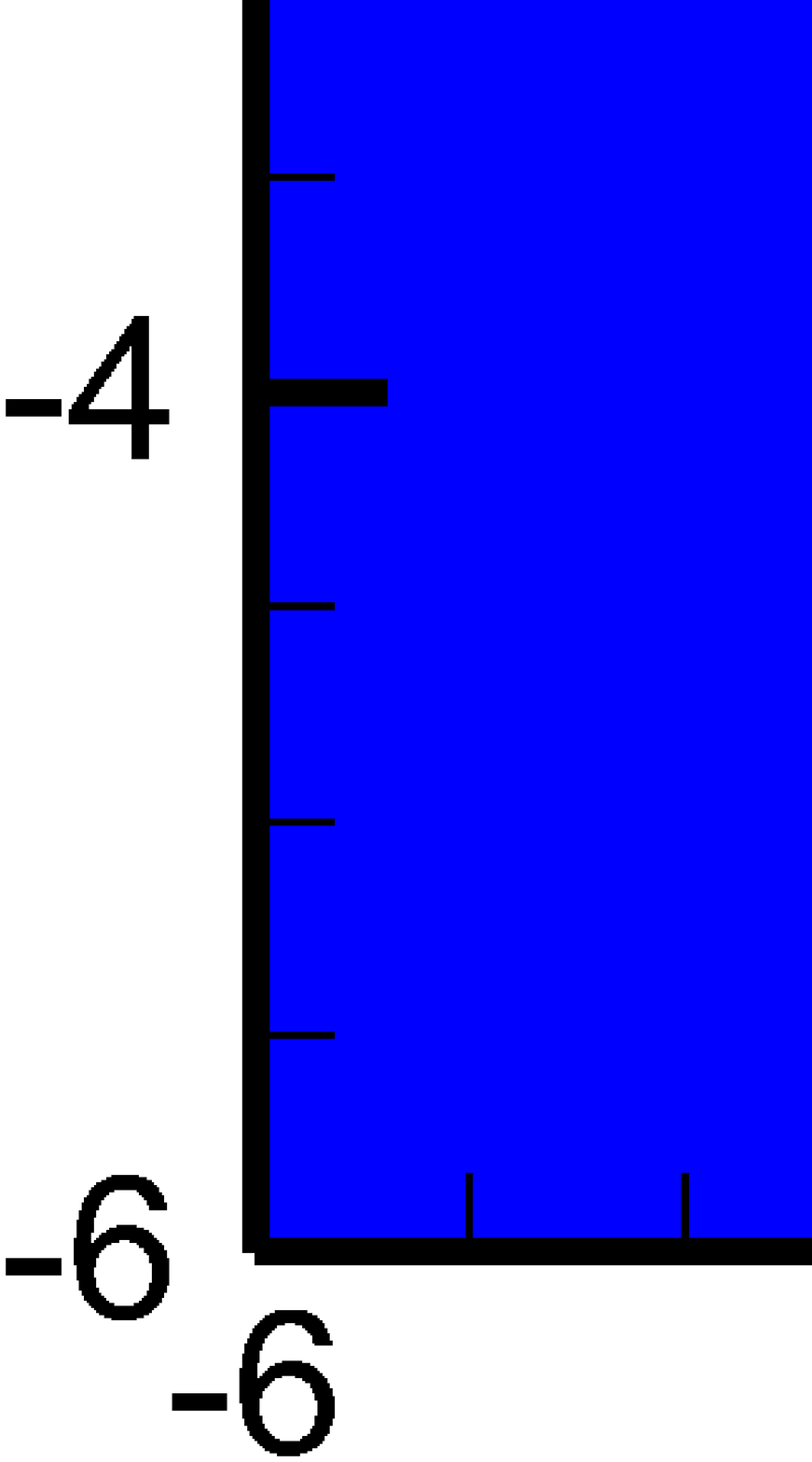} 
\end{tabular} 
\caption{Solution of the RMHD blast wave with $B_x=0.1$ at time $t=4.0$, obtained with the ADER-DG $\mathbb{P}_3$ scheme 
supplemented with the \aposteriori second order TVD subcell limiter. 
Top panels: rest-mass density (left) and thermal pressure (right). Central panels: Lorentz factor (left) and magnetic pressure (right), with magnetic field lines reported.
Bottom panels: AMR grid (left) and limiter map (right) with troubled cells marked in red and regular unlimited cells marked in blue. 
}
\label{fig:RMHDBlast-B01}
\end{center}
\end{figure*}
\begin{figure*}
\begin{center}
\begin{tabular}{lr}
\includegraphics[width=0.45\textwidth]{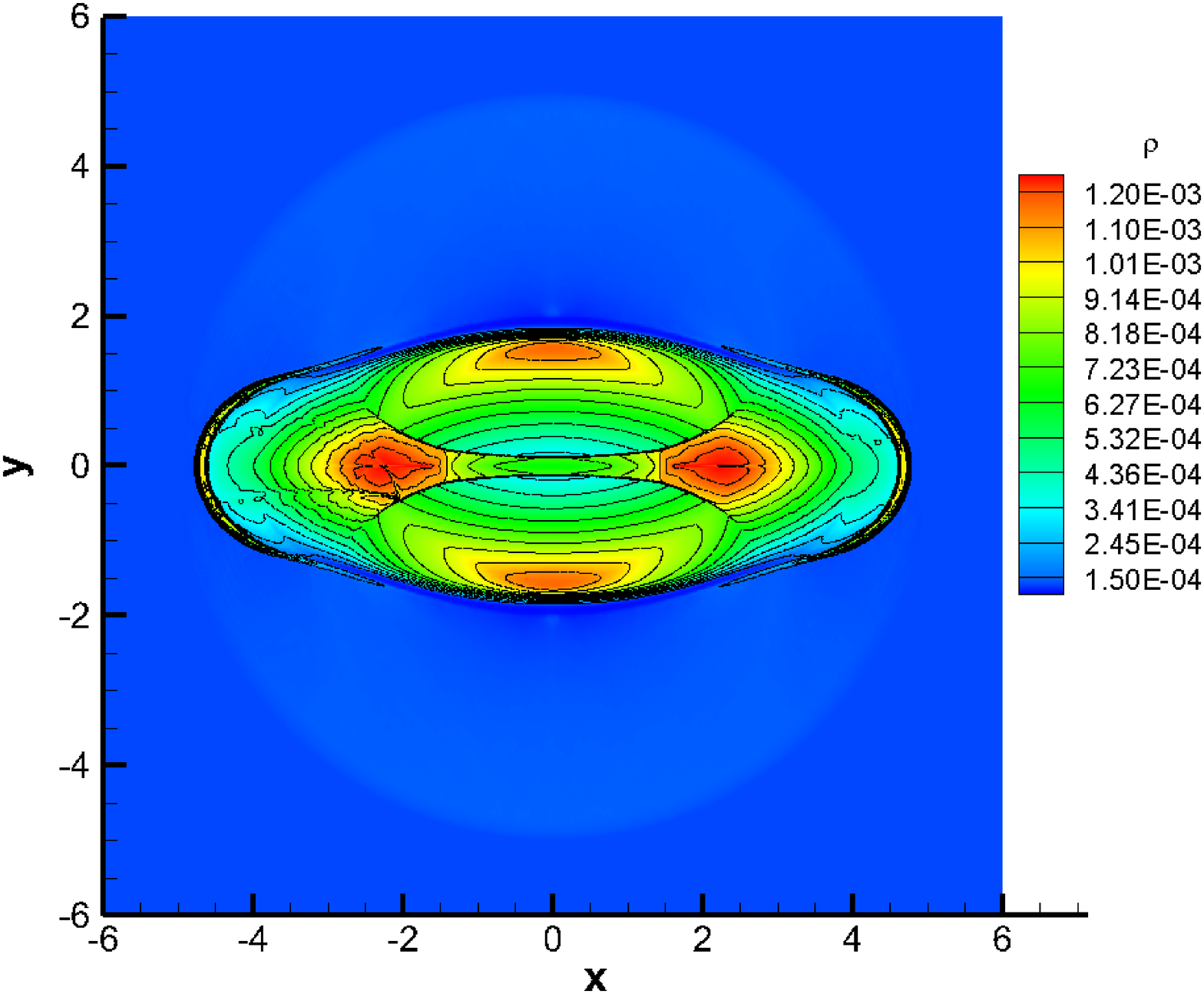}   &  
\includegraphics[width=0.45\textwidth]{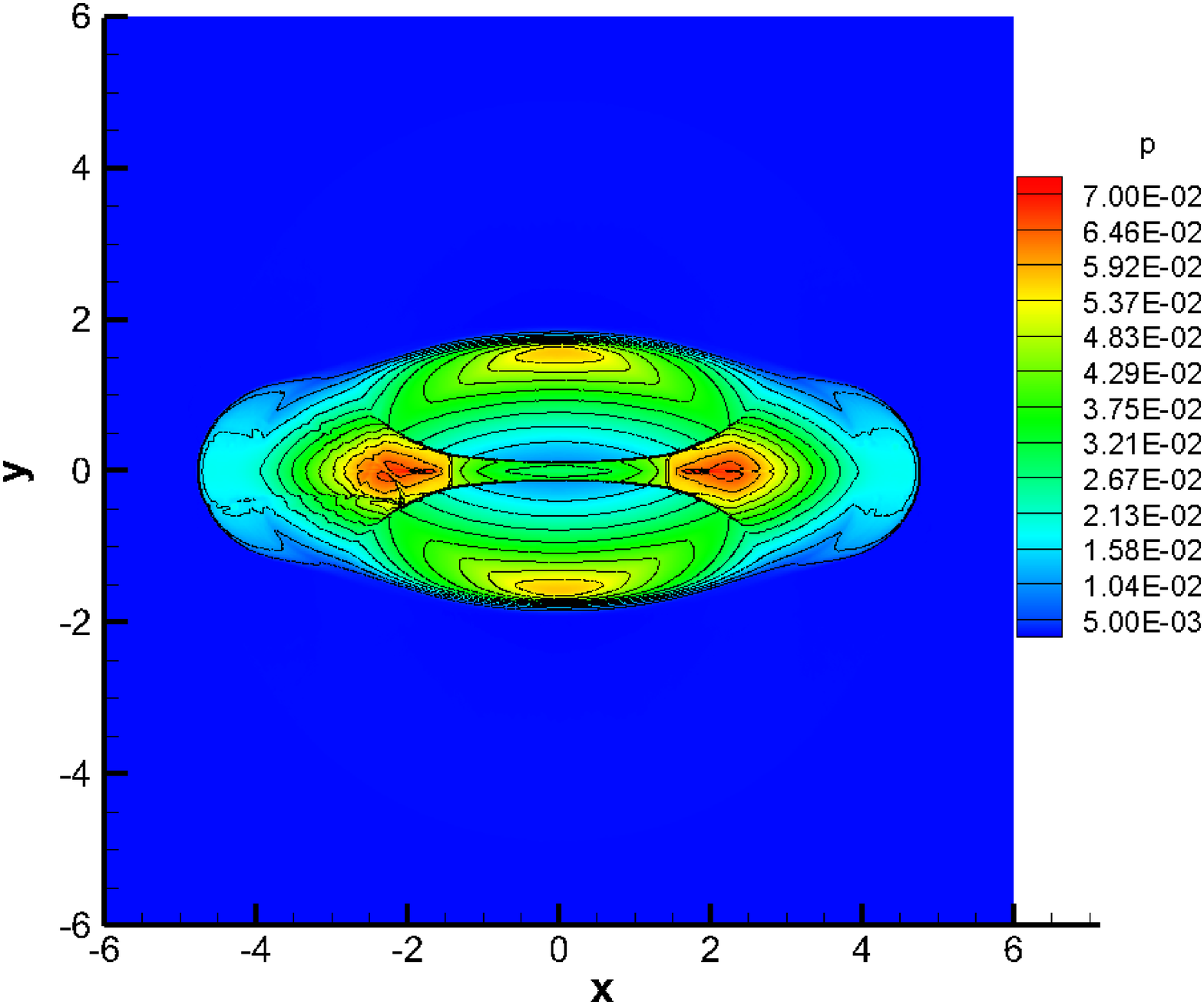}   \\  
\includegraphics[width=0.45\textwidth]{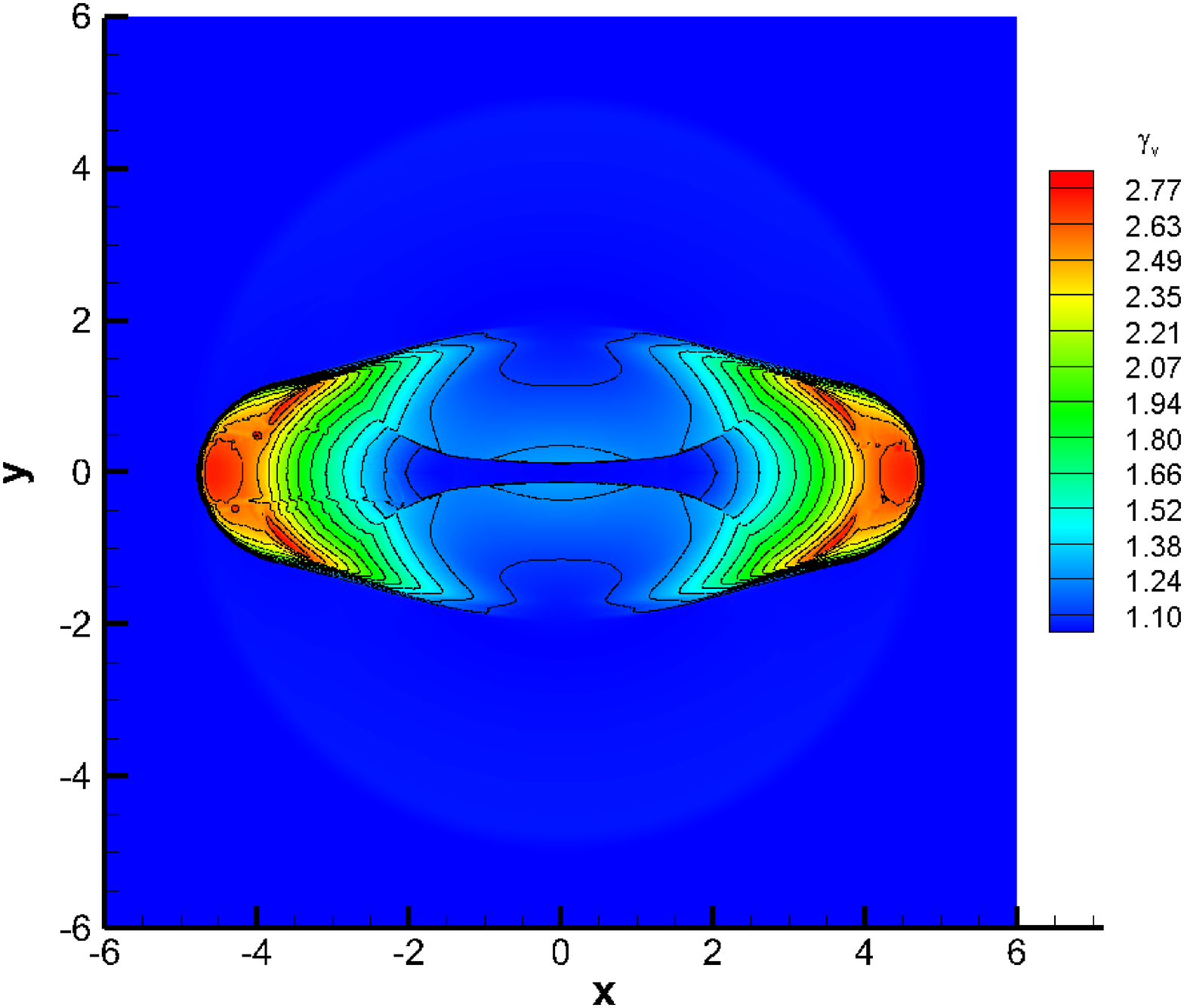} & 
\includegraphics[width=0.45\textwidth]{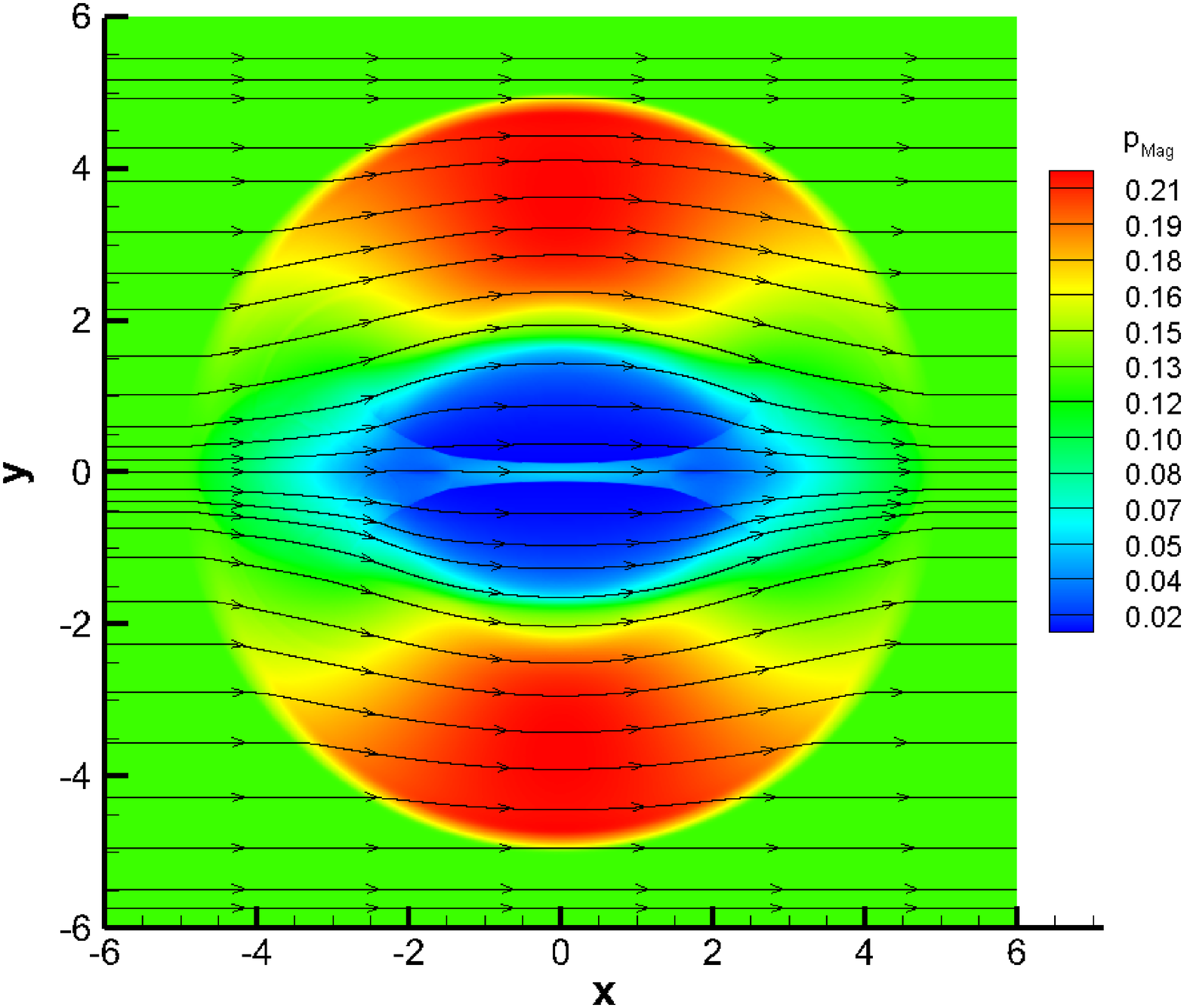} \\ 
\includegraphics[width=0.43\textwidth]{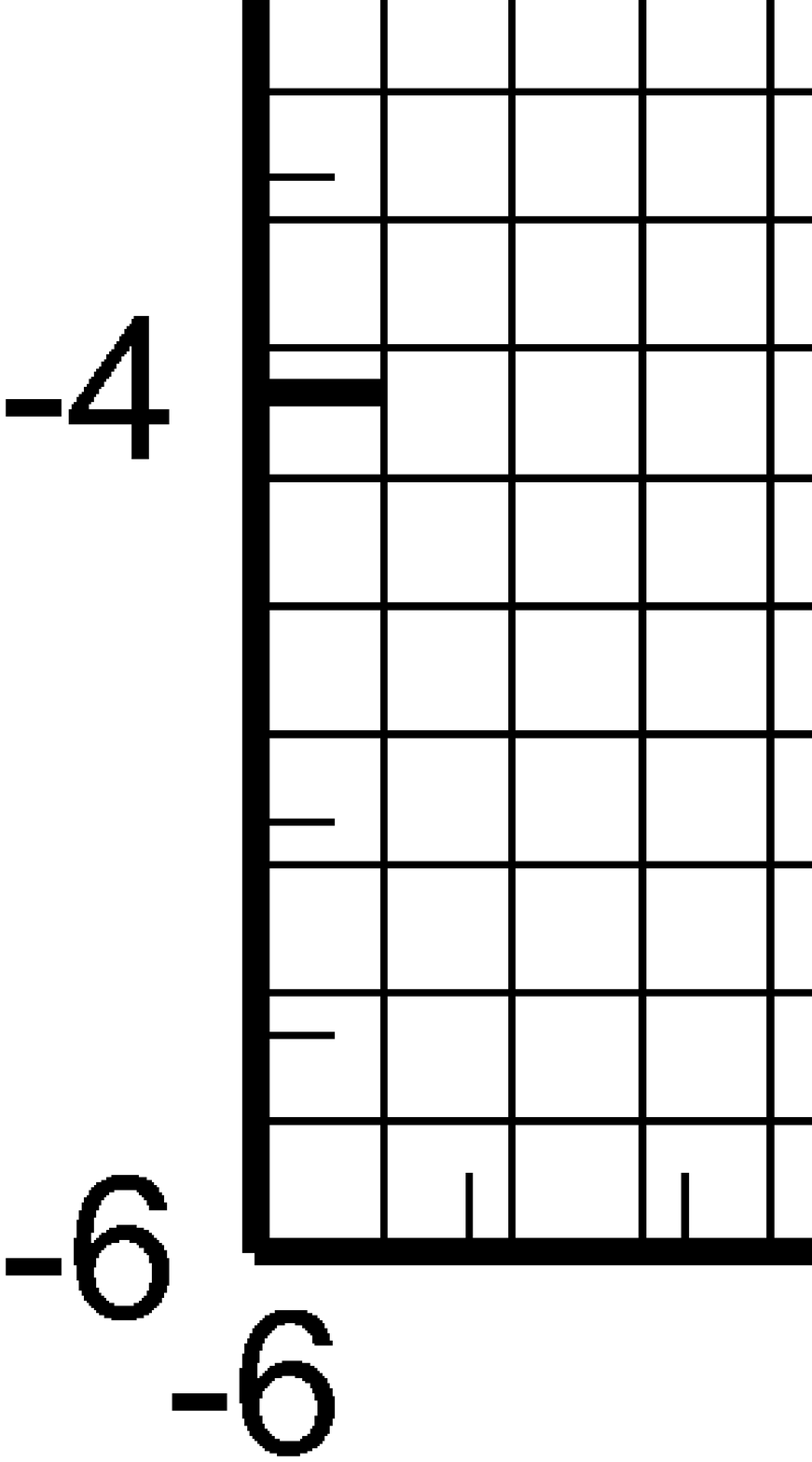}     &  
\includegraphics[width=0.43\textwidth]{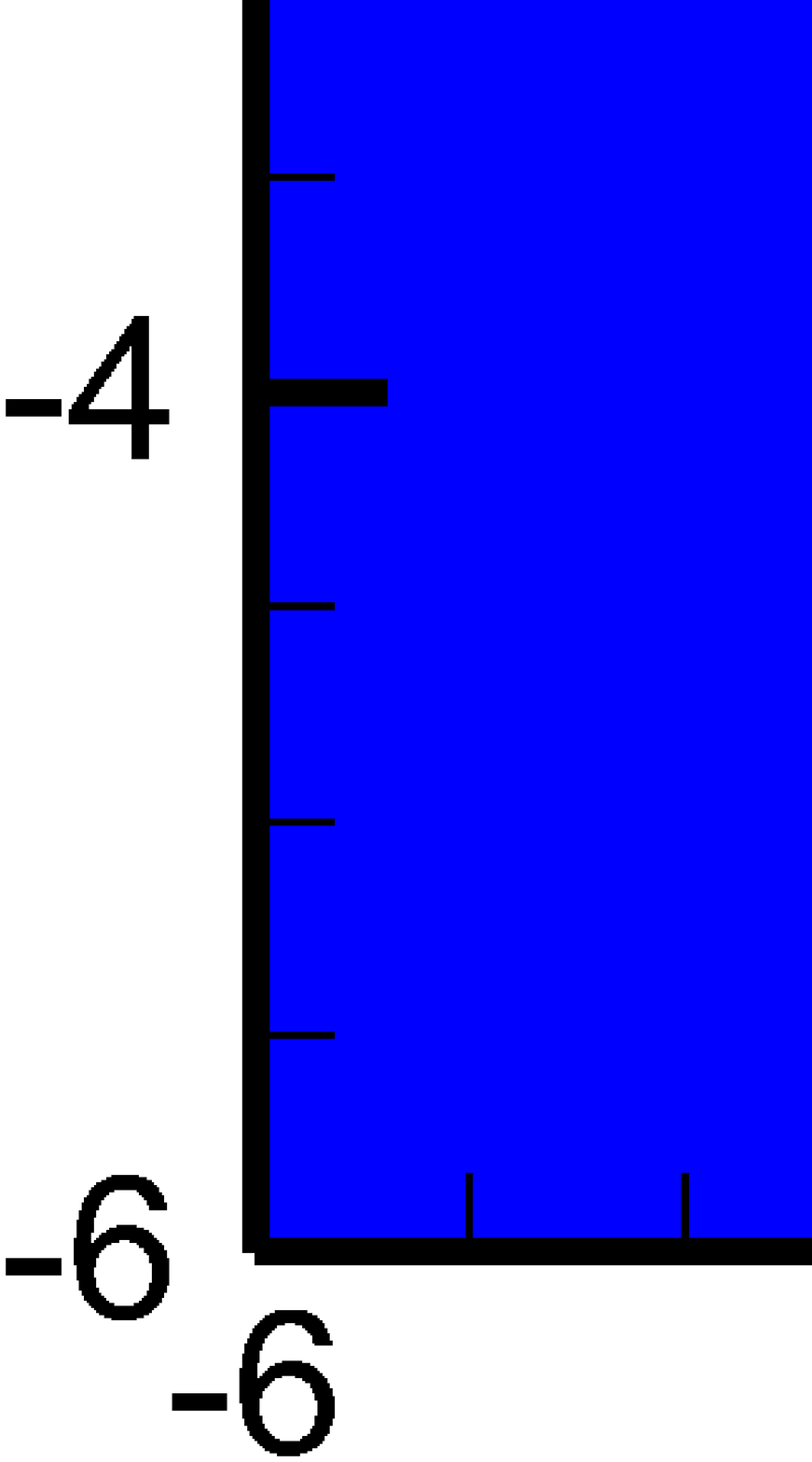} 
\end{tabular} 
\caption{Solution of the RMHD blast wave with $B_x=0.5$ at time $t=4.0$, obtained with the ADER-DG $\mathbb{P}_3$ scheme 
supplemented with the \aposteriori second order TVD subcell limiter. 
Top panels: rest-mass density (left) and thermal pressure (right). Central panels: Lorentz factor (left) and magnetic pressure (right), with magnetic field lines reported.
Bottom panels: AMR grid (left) and limiter map (right) with troubled cells marked in red and regular unlimited cells marked in blue. 
}
\label{fig:RMHDBlast-B05}
\end{center}
\end{figure*}
As a second two dimensional academic test we have considered the cylindrical expansion of a blast wave in a plasma with an initially uniform magnetic field.  This is notoriously a 
severe test, which became canonical after \cite{Komissarov1999}, and it has been solved by several authors, including \cite{Leismann2005,DelZanna2007,DumbserZanotti}. The initial conditions are
prescribed by assuming that, within a radius $R=1.0$, the rest-mass density and the pressure are $\rho=0.01$ and $p=1$, while outside the cylinder $\rho=10^{-4}$ and $p=5\times10^{-4}$. 
Like in \cite{Komissarov1999} and in \cite{DelZanna2007}, the inner and outer values are joined
through a smooth ramp function between $r=0.8$ and $r=1$, to avoid a sharp discontinuity in the initial conditions.
The plasma is initially at rest and subject to a constant magnetic field along the $x$-direction. In our tests we have considered two different magnetizations, the first one
with $B_x=0.1$, corresponding to the intermediate value chosen by \cite{Komissarov1999},
and the second one with $B_x=0.5$.
We have solved this problem over the computational domain 
$\Omega = [-6,6]\times[-6,6]$, with $40\times40$ elements on the coarsest refinement level,  $\mathfrak{r}=3$ and $\ell_\text{max}=2$. 
We have used the Rusanov Riemann solver with the $\mathbb{P}_3$ version of the ADER-DG scheme. Also for this test, 
a  robust second-order TVD scheme has been used on the subgrid where the limiter is activated. 
The results for $B_x=0.1$ are shown in Fig.~\ref{fig:RMHDBlast-B01}, which
reports the rest-mass density, the thermal pressure, the Lorentz factor and 
the magnetic pressure at time $t=4.0$. The wavepattern of the configuration at this time is composed by two main waves, an
external fast shock and a reverse shock, the former being almost circular, the latter being somewhat elliptic. The magnetic field is essentially confined between them,  
while the inner region is almost devoid of magnetization. We have detected a maximum Lorentz factor $W_{\rm max}\approx 4.3$ along the $x$ axis just on the back of the reversed shock.
The two bottom panels show the AMR grid and the map of the limiter, which is activated along the two main shock fronts.
In Fig.~\ref{fig:RMHDBlast-B05}, on the other hand, we have reported the results obtained 
for $B_x=0.5$, again at $t=4.0$.
In this case, the external circular fast shock, which is visible in the rest-mass density and in the magnetic pressure, is very weak, while the magnetic confinement of the plasma is increased. The maximum Lorentz factor detected in this case is $W_{\rm max}\approx 2.8$.

\begin{figure*}
\begin{center}
\begin{tabular}{lcr}
\includegraphics[width=0.33\textwidth]{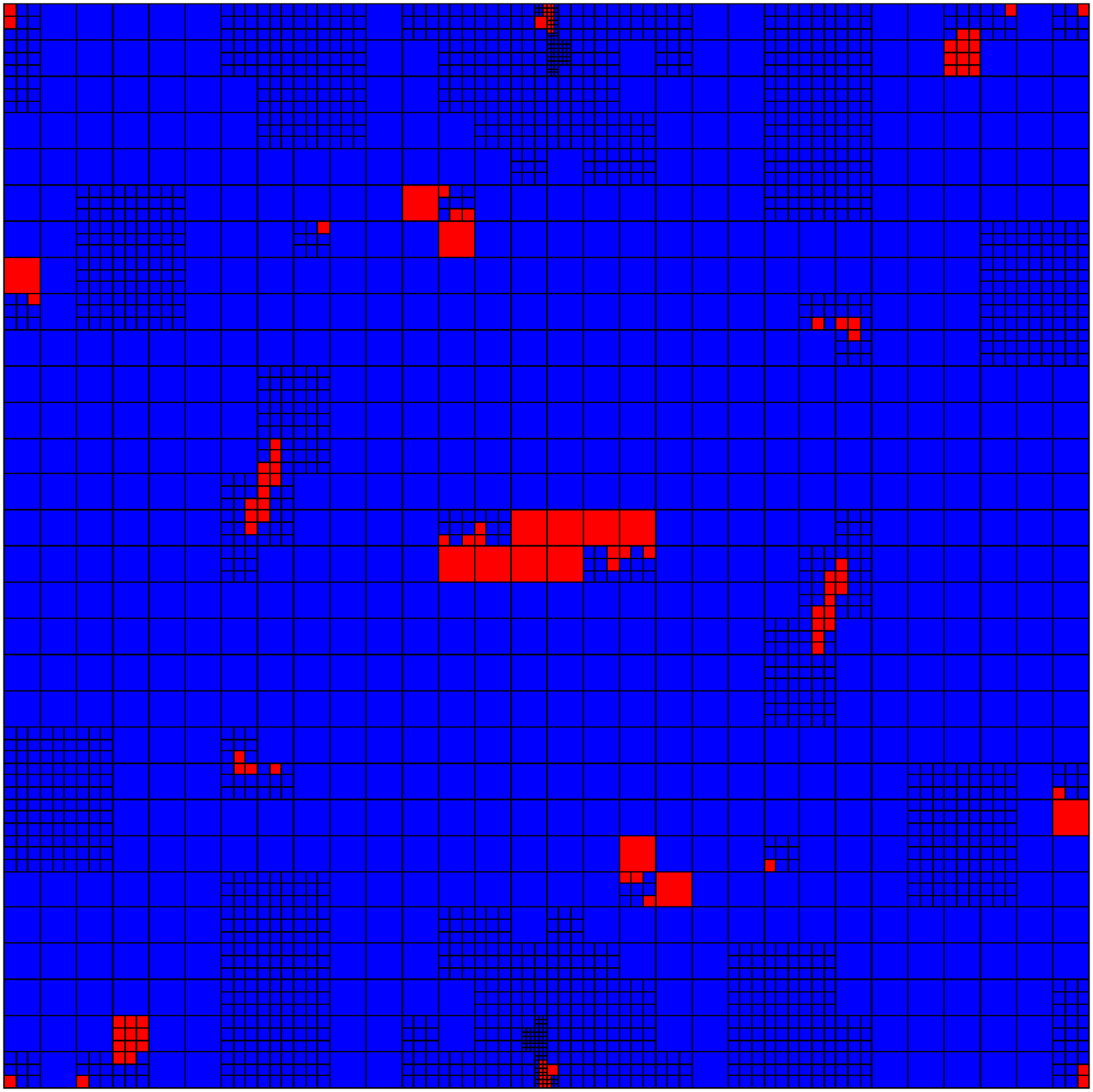}     &  
\includegraphics[width=0.33\textwidth]{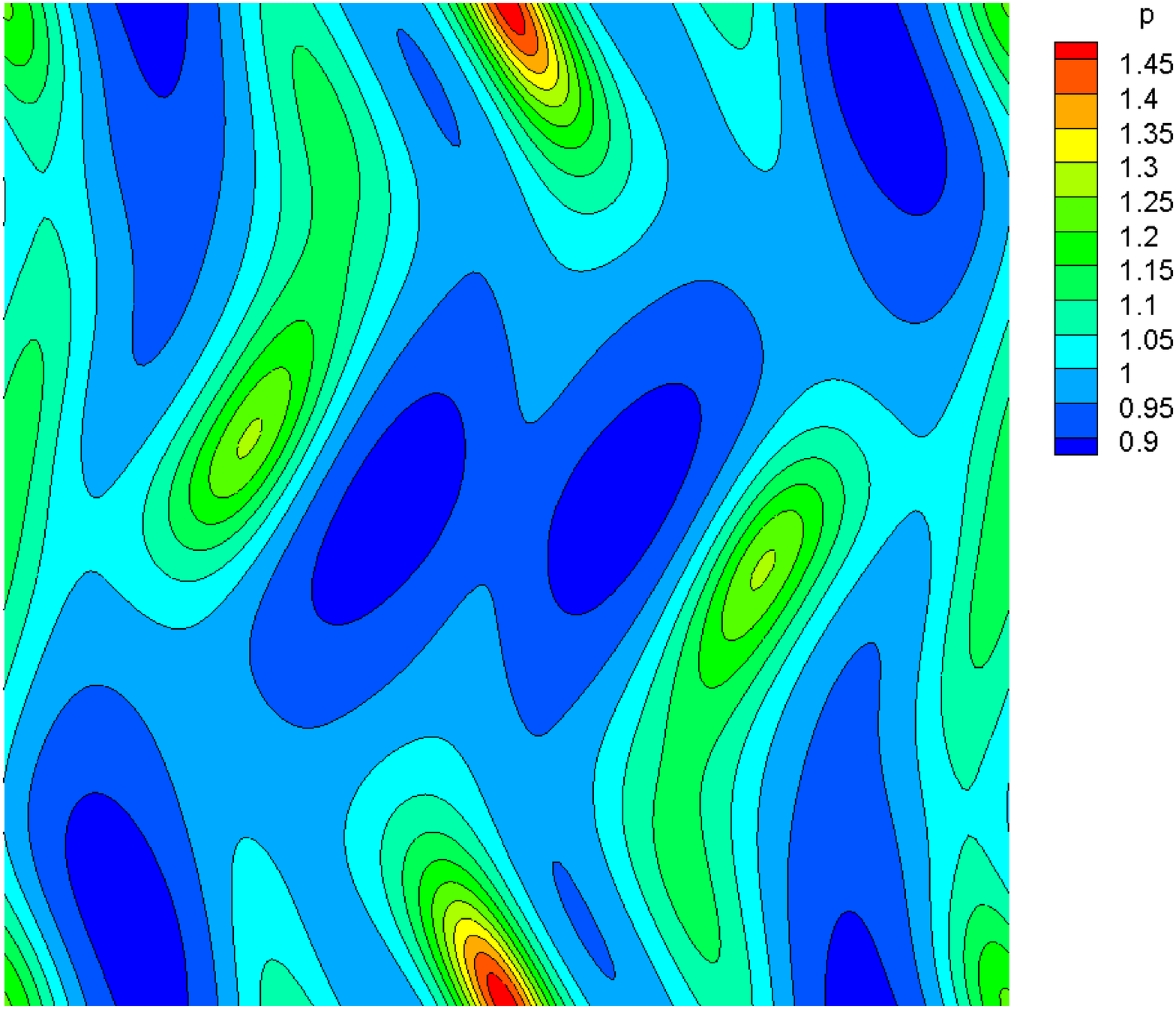}						 &  
\includegraphics[width=0.33\textwidth]{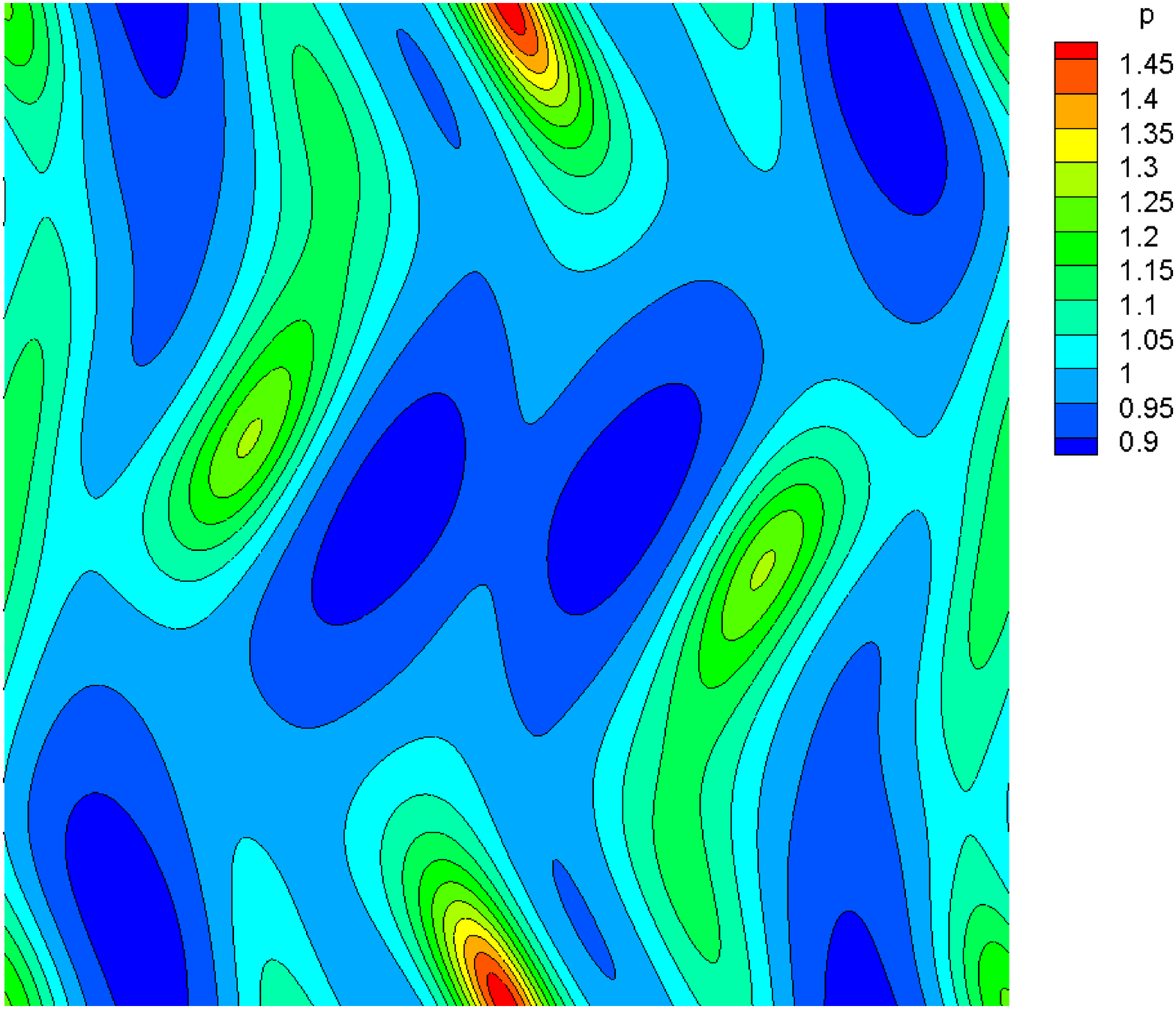}\\
\includegraphics[width=0.33\textwidth]{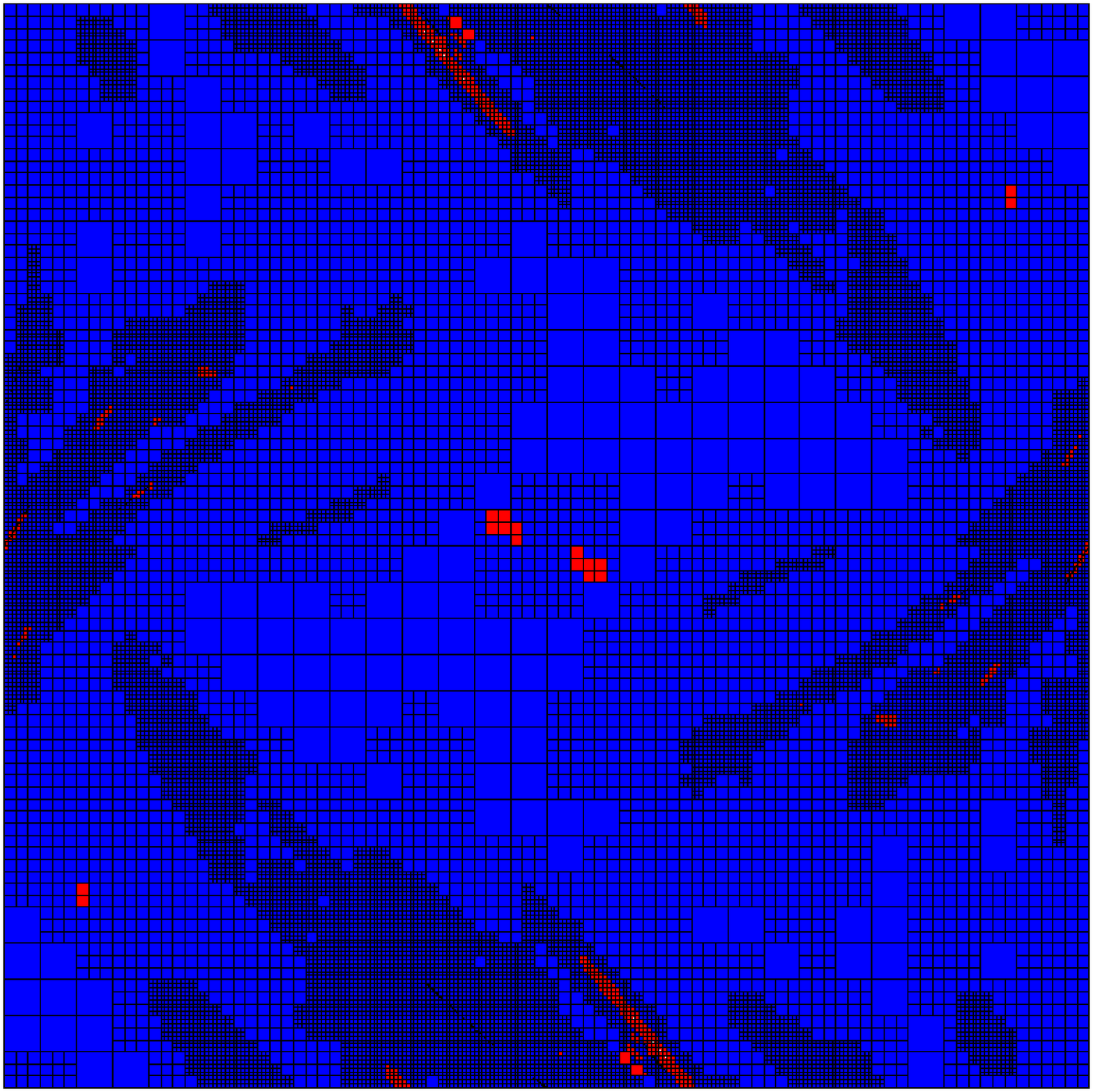}     &  
\includegraphics[width=0.33\textwidth]{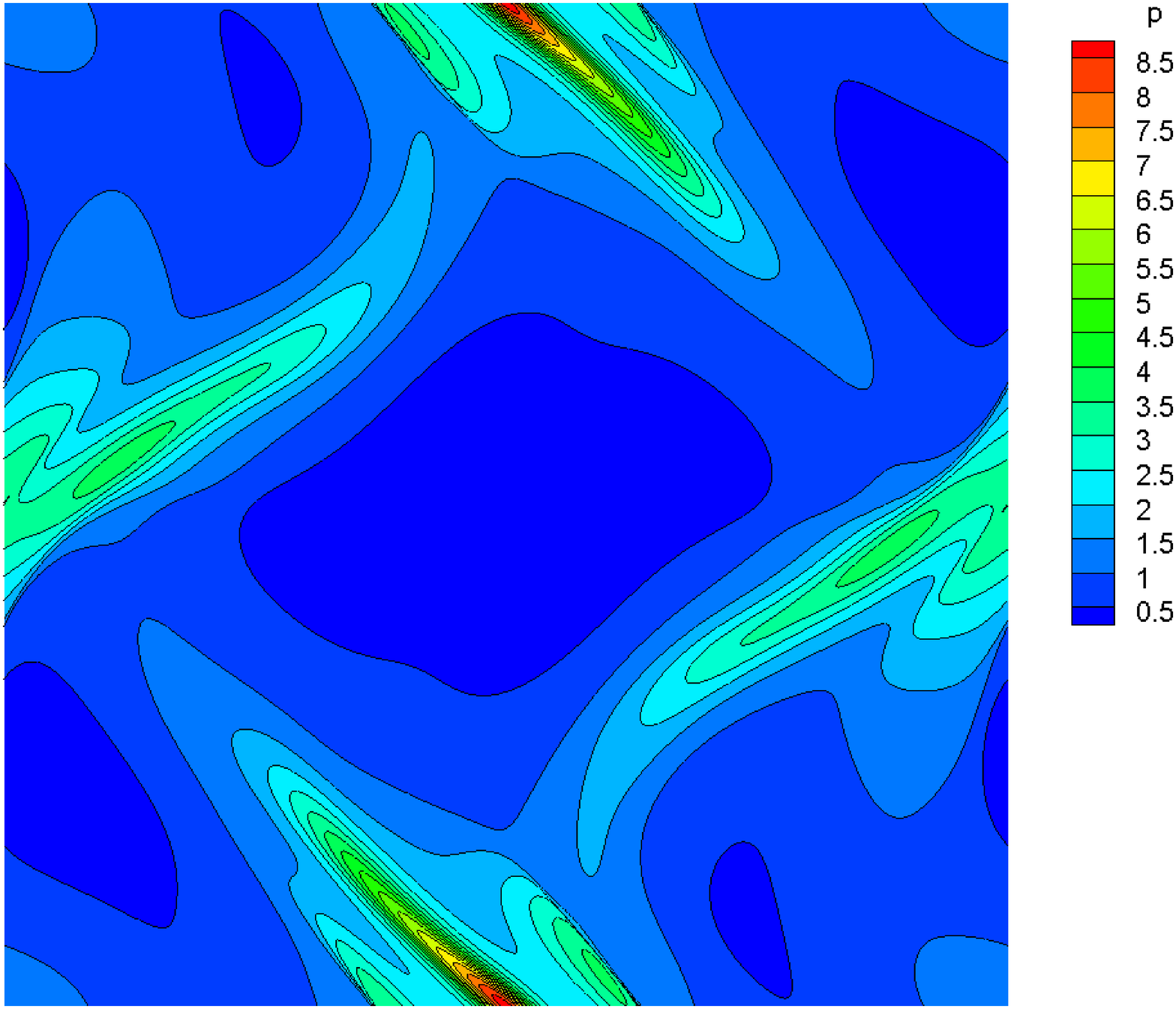}						 &  
\includegraphics[width=0.33\textwidth]{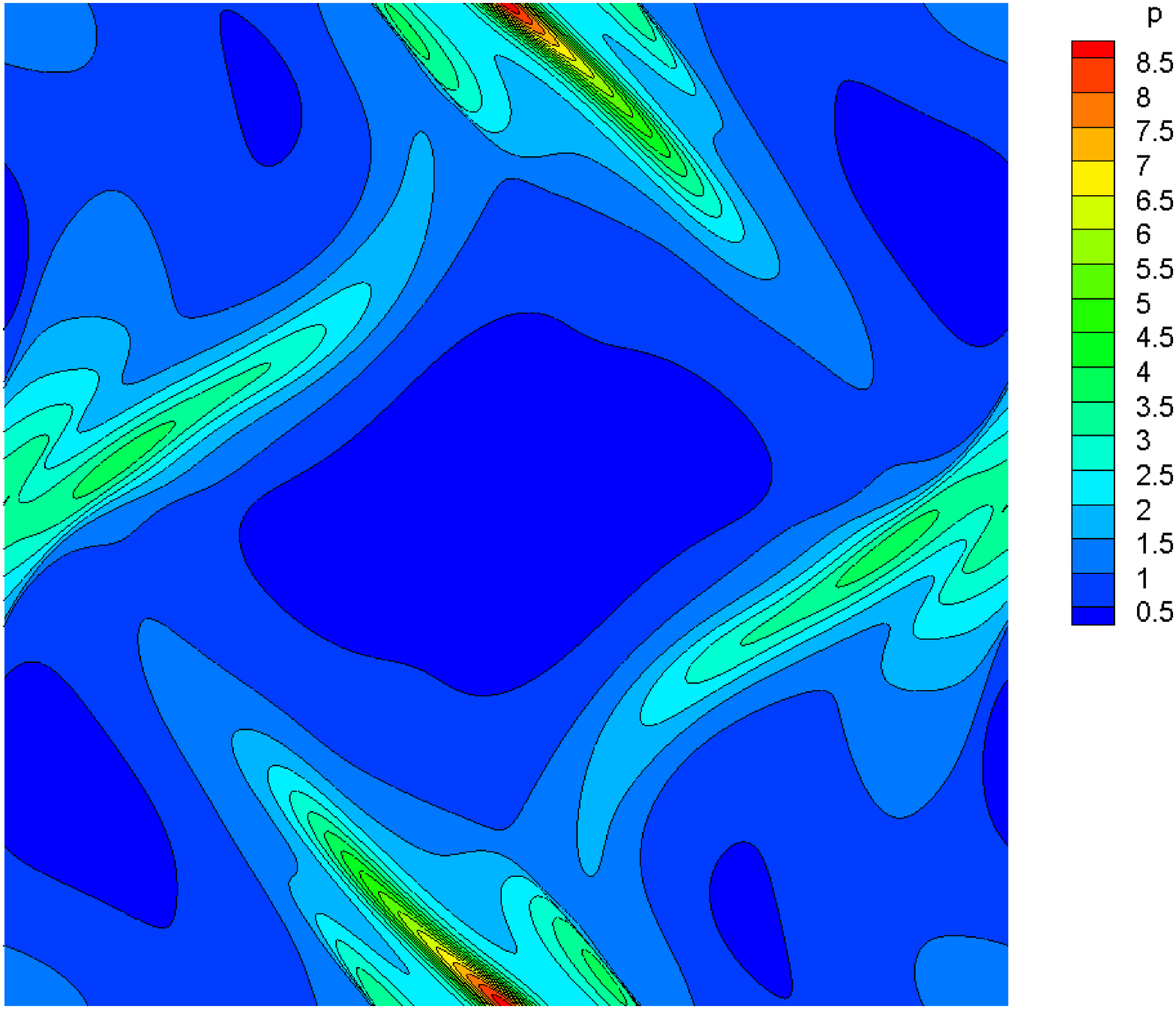}\\
\includegraphics[width=0.33\textwidth]{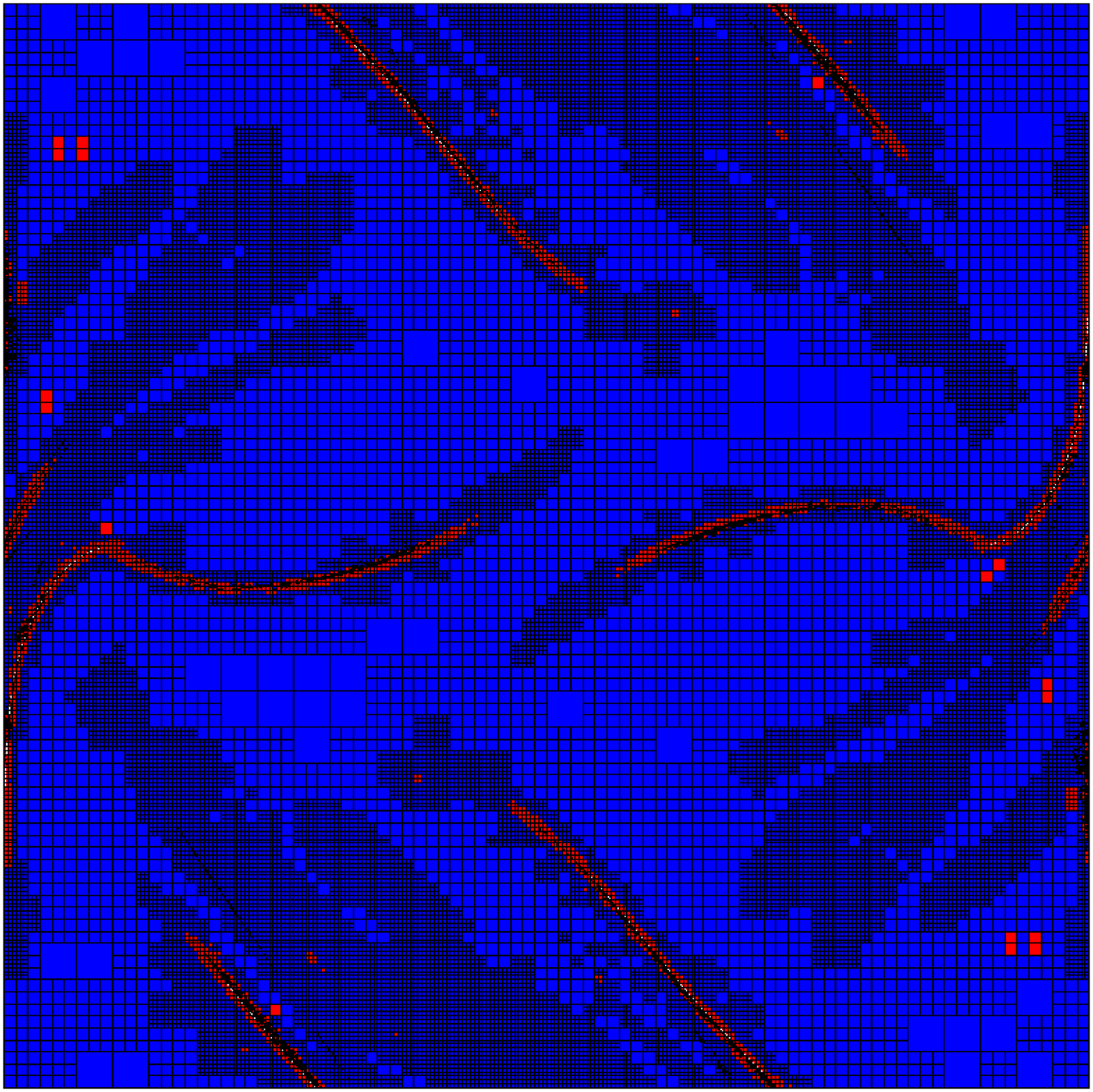}     &  
\includegraphics[width=0.33\textwidth]{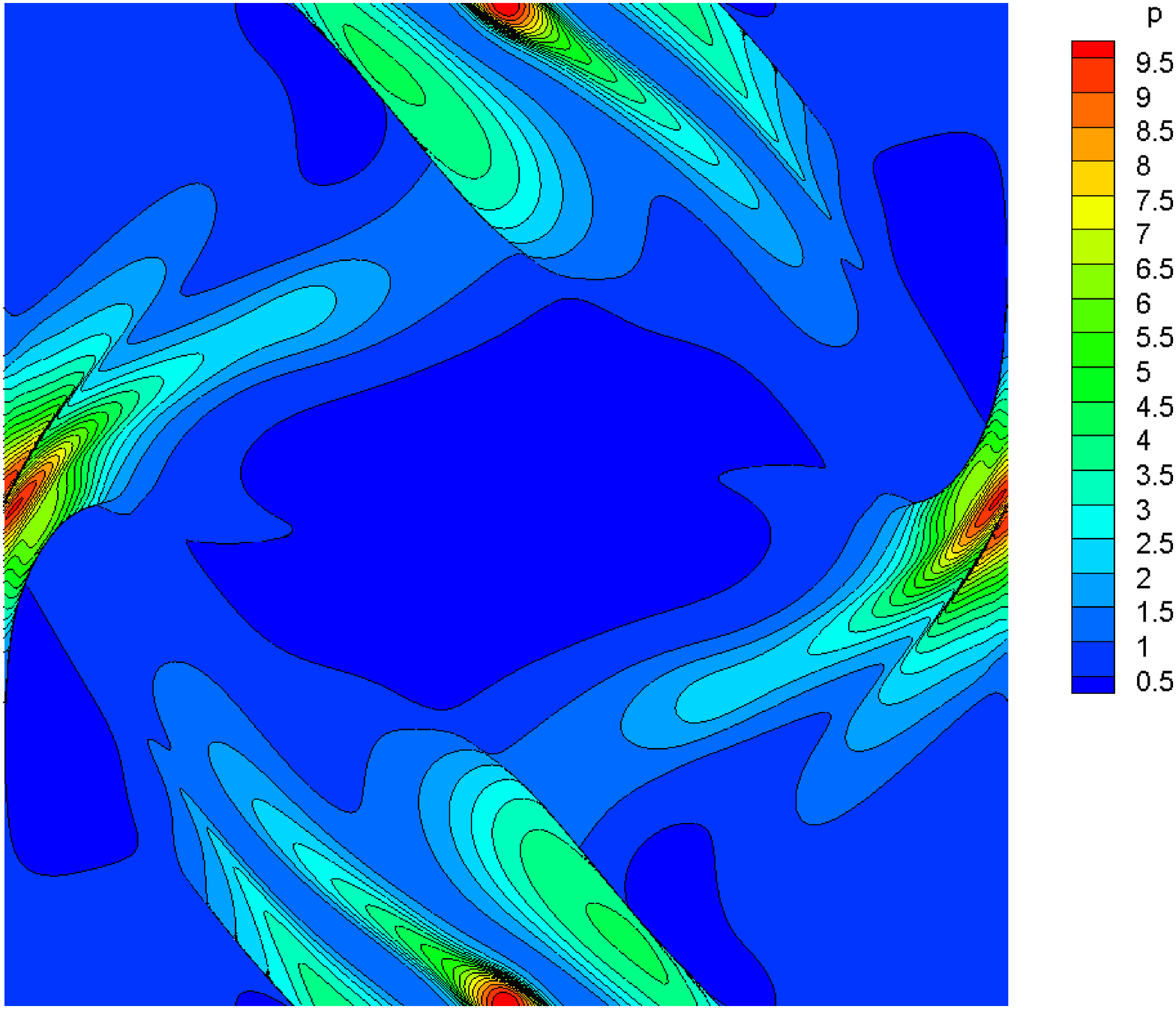}						 &  
\includegraphics[width=0.33\textwidth]{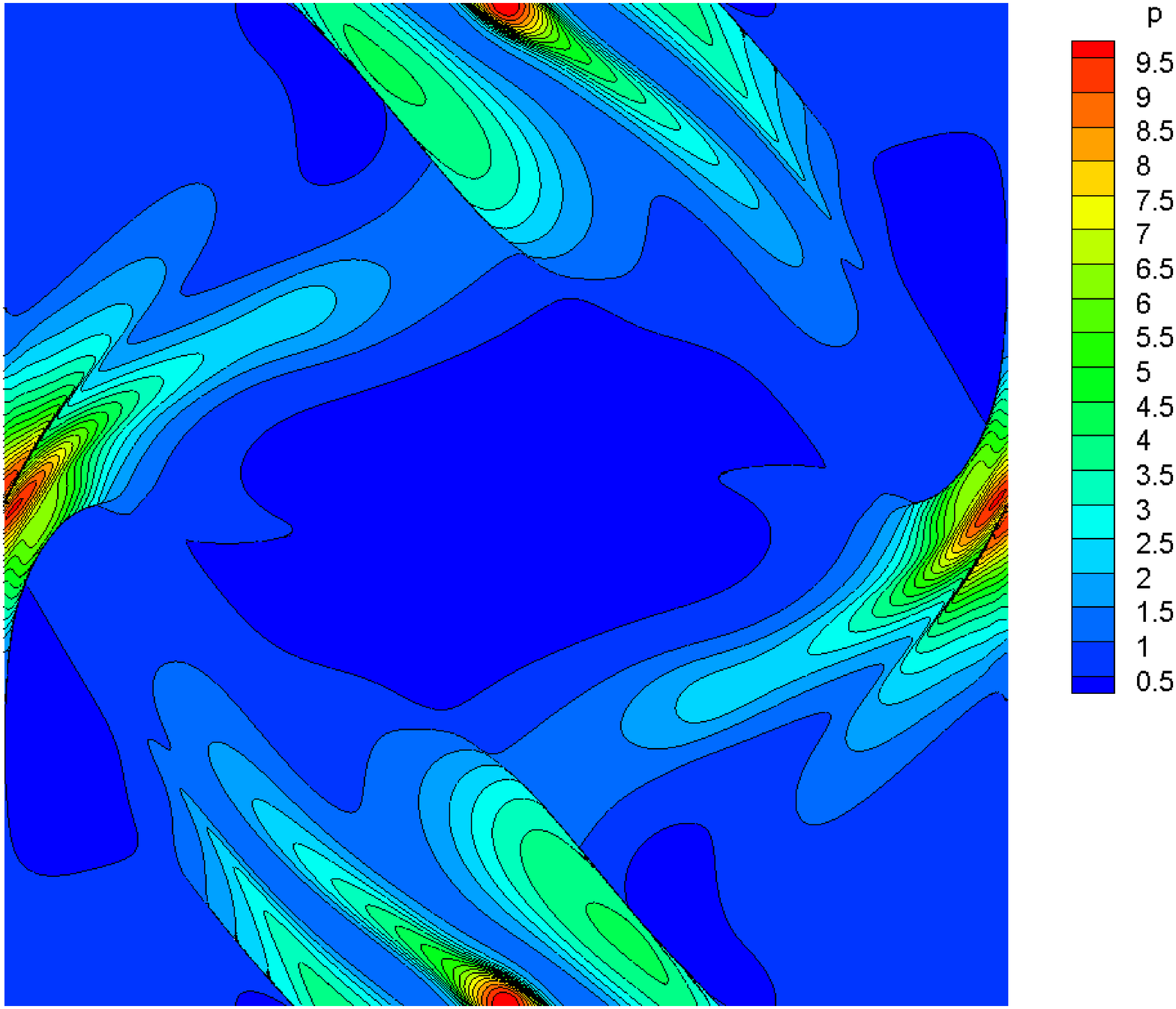}\\
\includegraphics[width=0.33\textwidth]{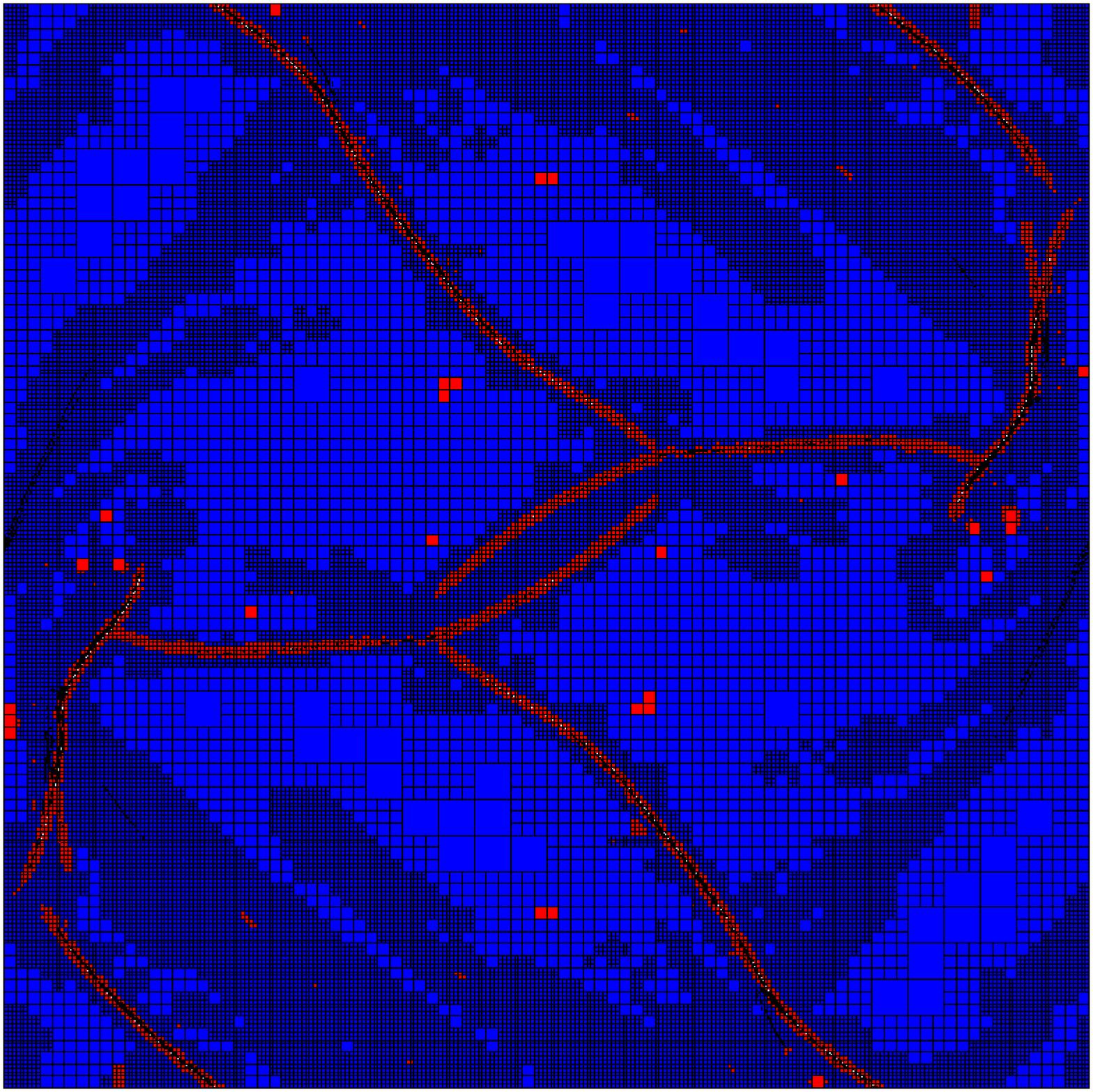}     &  
\includegraphics[width=0.33\textwidth]{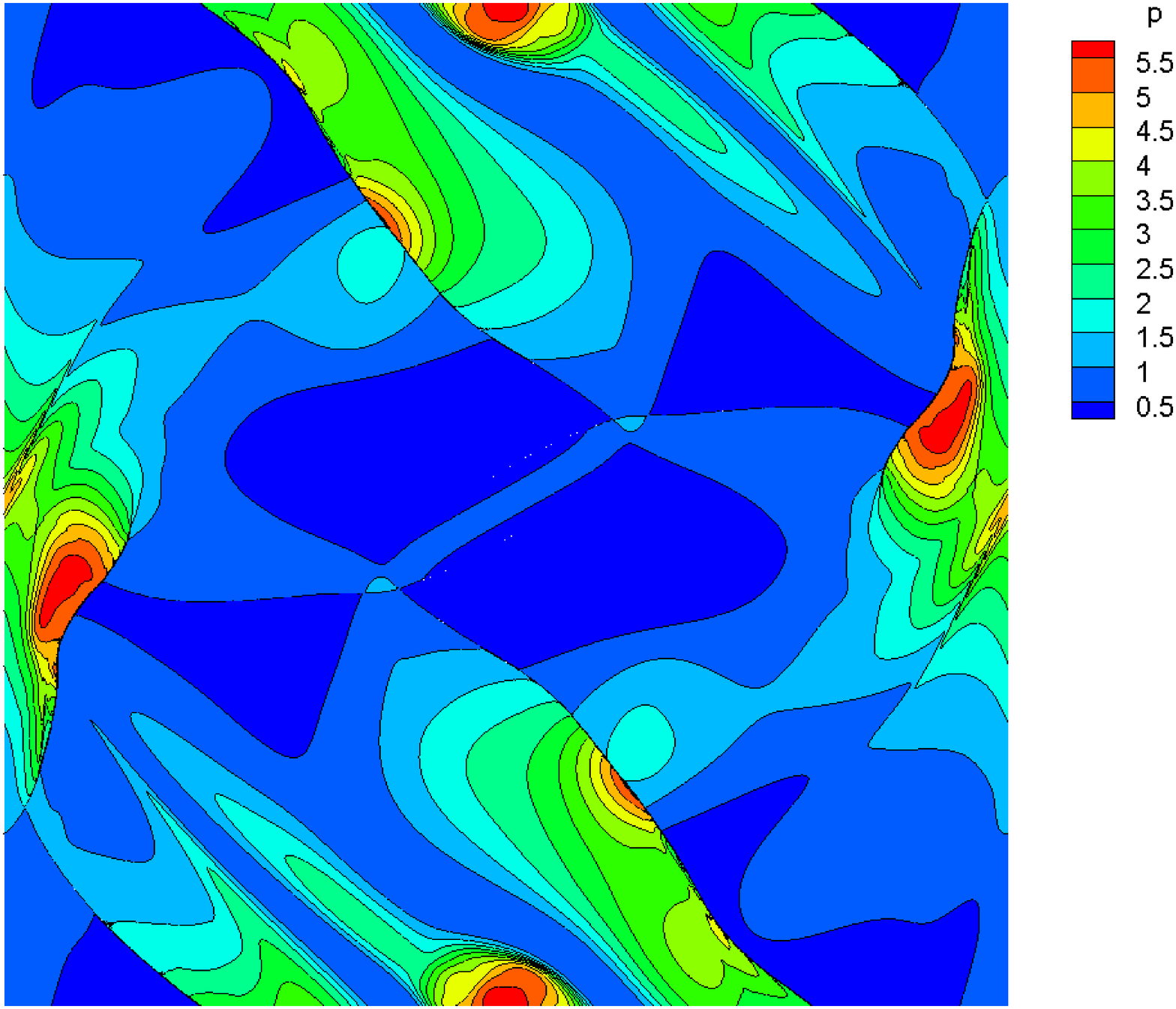}						 &  
\includegraphics[width=0.33\textwidth]{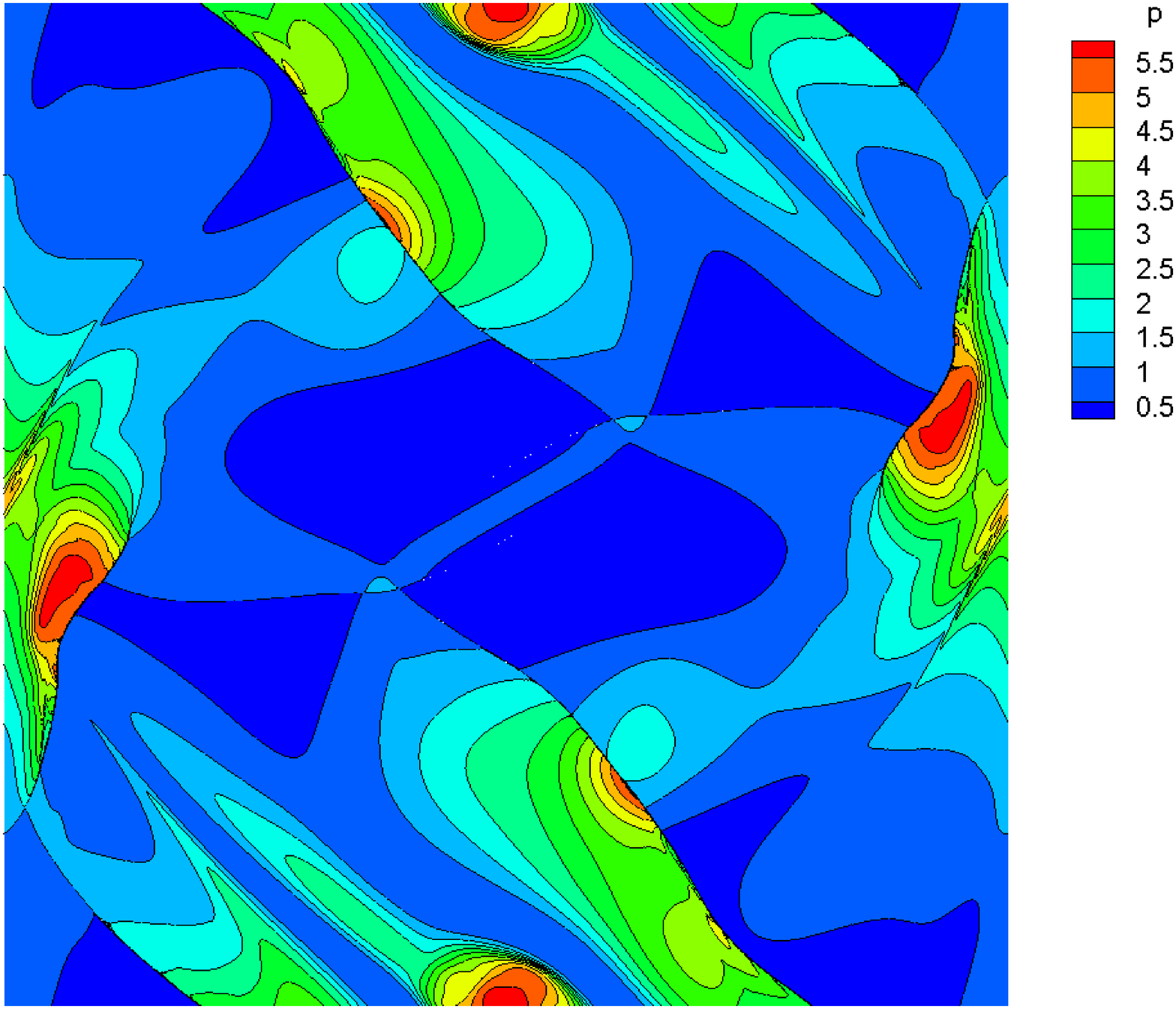}\\
\end{tabular} 
\caption{RMHD Orszag-Tang vortex problem at times $t=0.5$, $t=2.0$, $t=3.0$, $t=4.0$, from top to bottom, obtained through the ADER-DG-$\mathbb{P}_5$ scheme
supplemented 
with the third order \aposteriori ADER-WENO subcell limiter. Left panels: AMR-grid, troubled cells (red) and unlimited cells (blue). Central panels: $\mathbb{P}_5$-solution obtained on the AMR grid. Right panels:  $\mathbb{P}_5$-solution obtained on the fine uniform grid corresponding to the finest AMR grid level.}
\label{fig:OrszagTang}
\end{center}
\end{figure*}

%
\subsection{Orszag-Tang vortex system} 

Next, we have chosen the relativistic version of the well known Orszag-Tang vortex problem, proposed by \cite{OrszagTang}, and later considered by \cite{PiconeDahlburg} and \cite{DahlburgPicone}. The resistive case of this relativistic MHD problem has been investigated by \cite{DumbserZanotti}.
The initial conditions are given by
\begin{equation}
\left( \rho, u, v, w, p, B_x ,B_y,B_z \right) = \left(  1 , - \frac{3}{4\sqrt{2}}\sin\left(y\right), \frac{3}{4\sqrt{2}}\sin \left(x \right), 0, 1, - \sin\left(y\right), \sin \left(2x \right), 0 \right),
\label{eq:OrszagTang_ic}
\end{equation}
while the adiabatic index is $\gamma=4/3$. The equations are discretized over the computational domain $\Omega = [0,2\pi]\times[0,2\pi]$, with $30\times30$ elements on the coarsest refinement level at the initial state. Periodic boundary conditions are imposed at the borders
and the Rusanov Riemann solver is adopted.
The relevant AMR parameters are $\mathfrak{r}=3$ and $\ell_\text{max}=2$. 
We note that the maximally refined AMR mesh corresponds to
a uniform grid formed of $270\times270=72,900$ elements. Moreover, 
the $\mathbb{P}_5$ version of the ADER-DG scheme that we have adopted uses $6$ degrees of freedom per spatial dimension, 
amounting to a total resolution of $2,624,400$ spatial degrees of freedom.  
The computed solution for the rest-mass density is shown in the central column of Fig.~\ref{fig:OrszagTang}, at times $t=0.5,2.0,3.0,4.0$ respectively.
For comparison, the panels on the right column show the results of a 
simulation performed over the maximally refined uniform mesh, which can be used as a reference solution. 
Clearly, an excellent agreement between the AMR results and this  reference solution is obtained. 
As before, this test confirms the ability of the proposed method for solving complex two dimensional problems and, by showing the critical cells which
required the activation of the limiter, it provides an immediate visual sketch of the most delicate regions over the computational domain. 
\begin{figure*}
\begin{center}
\begin{tabular}{lcccr}
\includegraphics[width=0.21\textwidth]{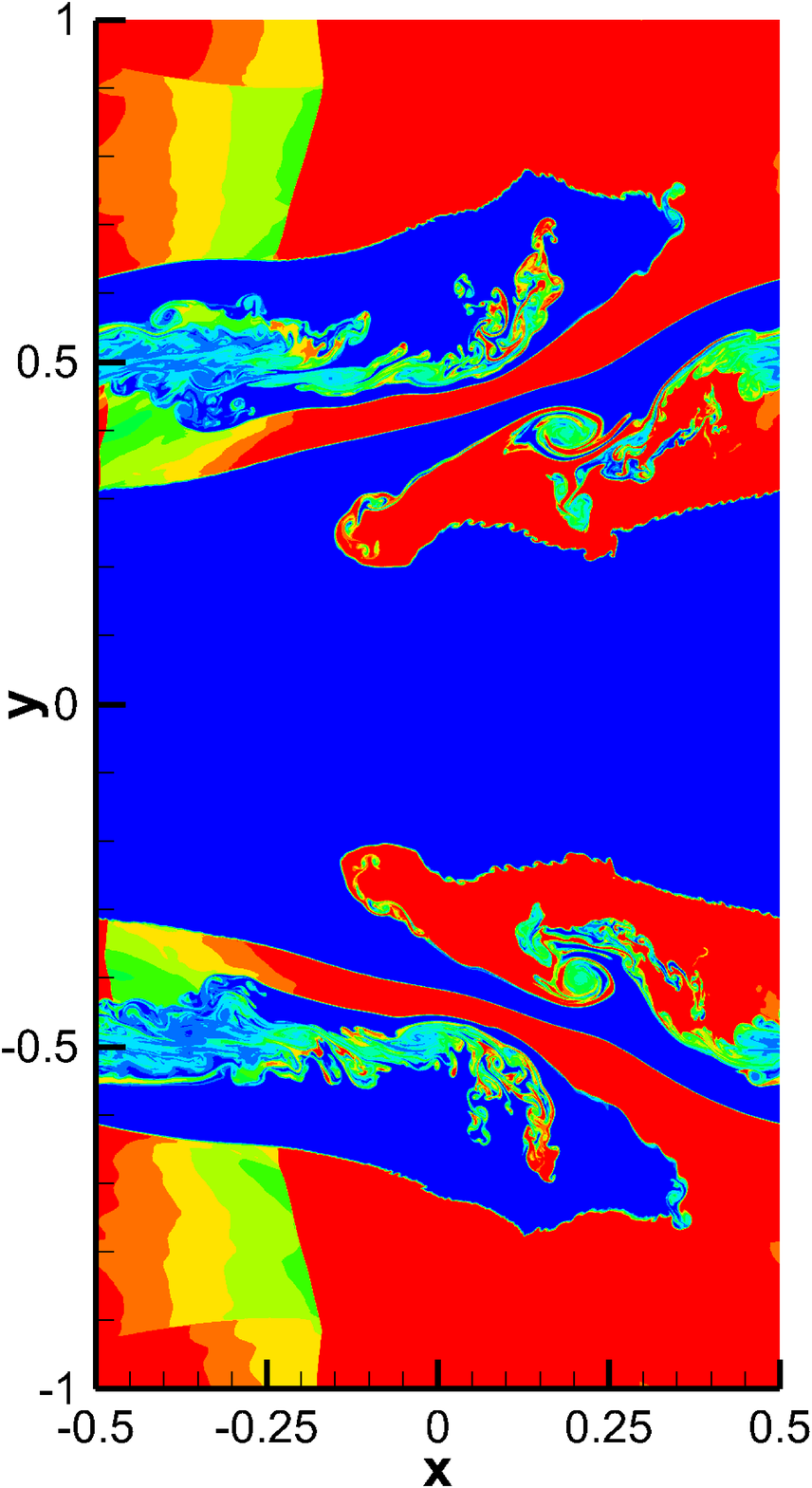}  &  
\includegraphics[width=0.21\textwidth]{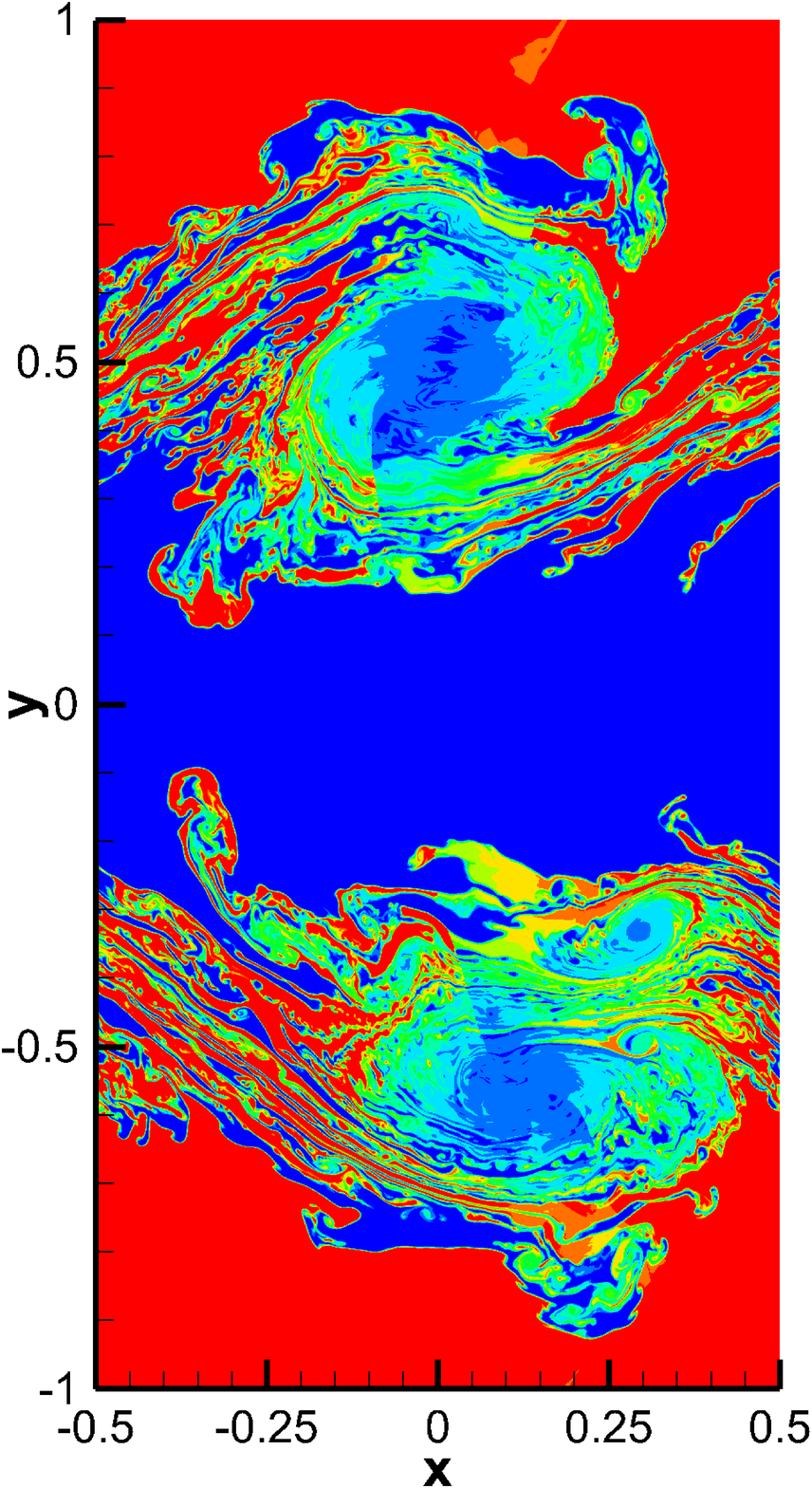}	 &  
\includegraphics[width=0.21\textwidth]{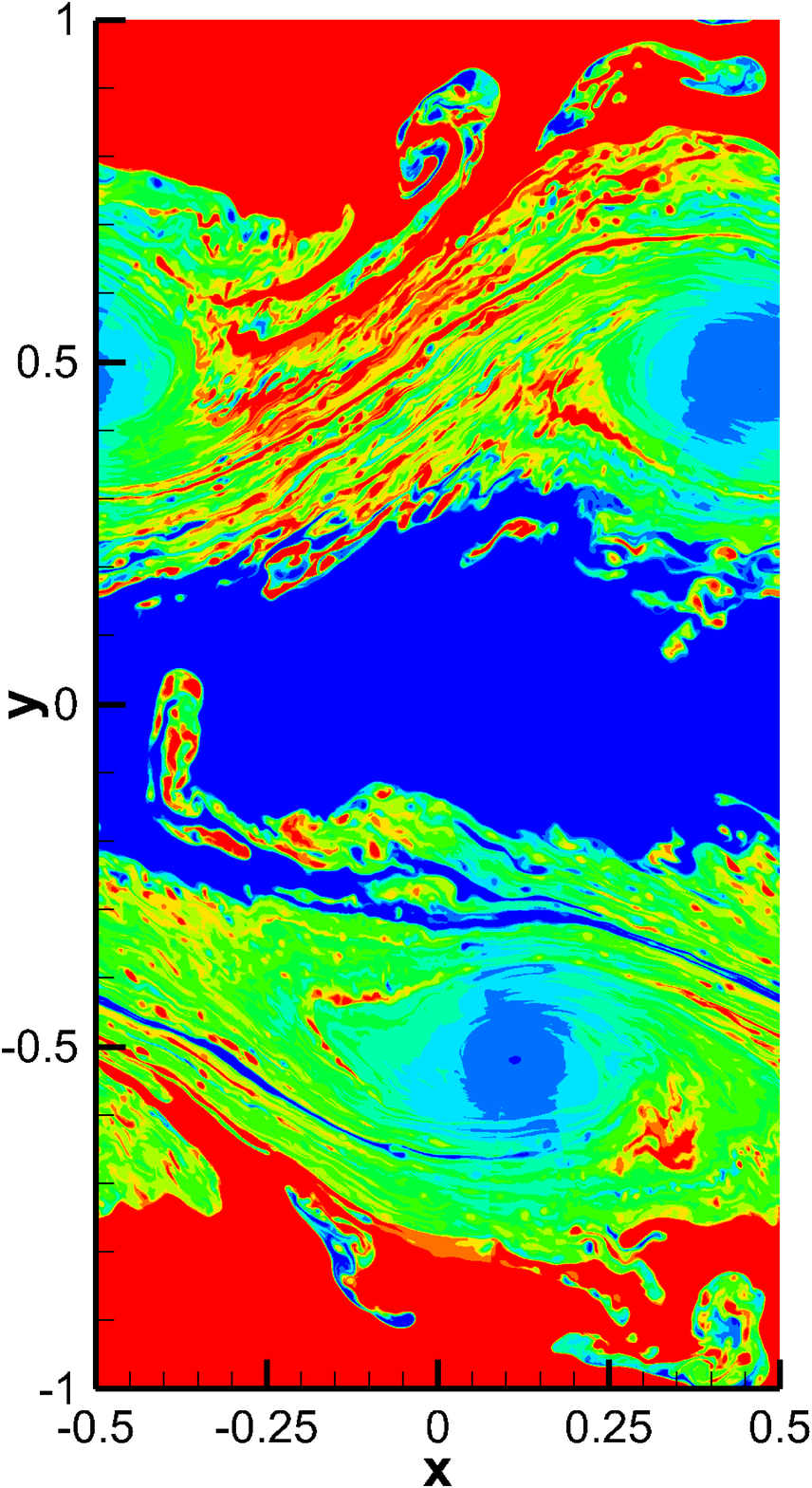}	 &  
\includegraphics[width=0.21\textwidth]{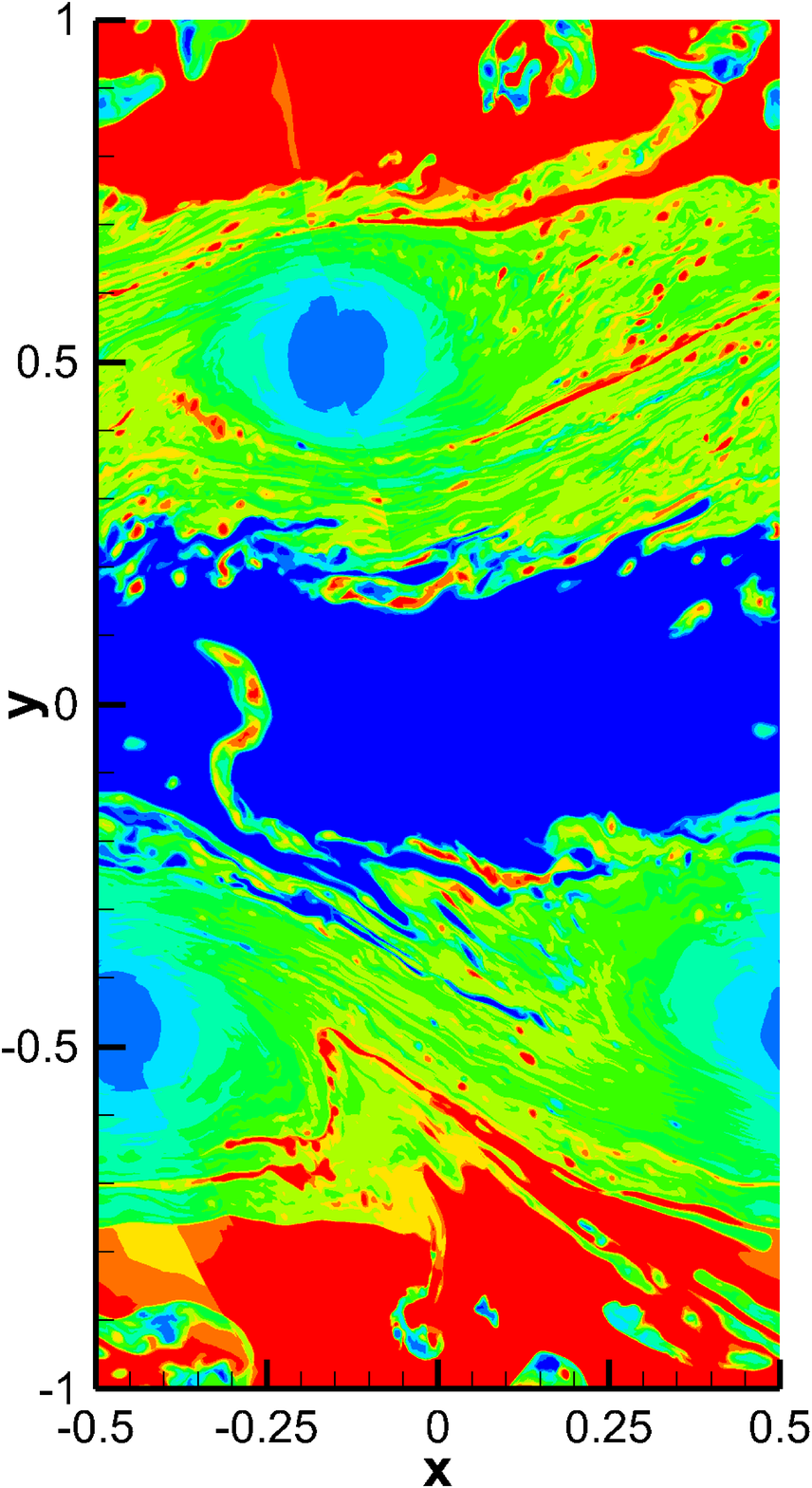}	 &  
\raisebox{15mm}{\includegraphics[width=0.06\textwidth]{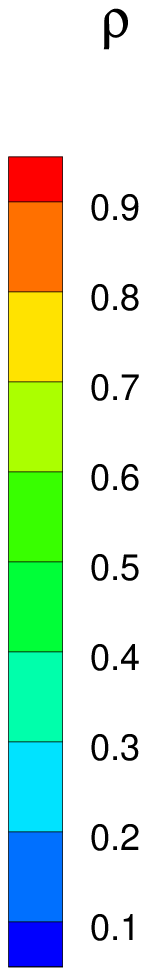}}\\				
\includegraphics[width=0.21\textwidth]{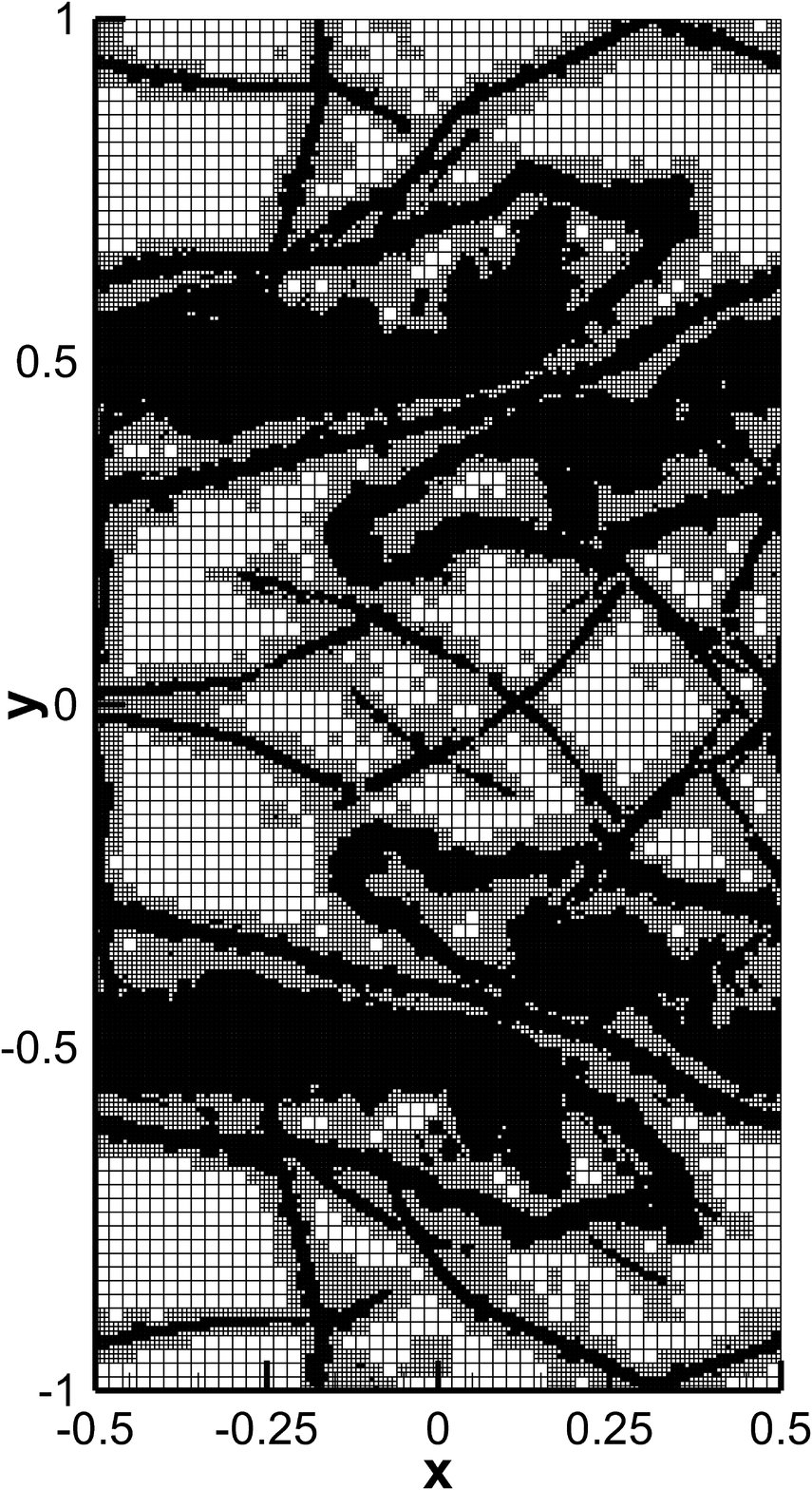}  &  
\includegraphics[width=0.21\textwidth]{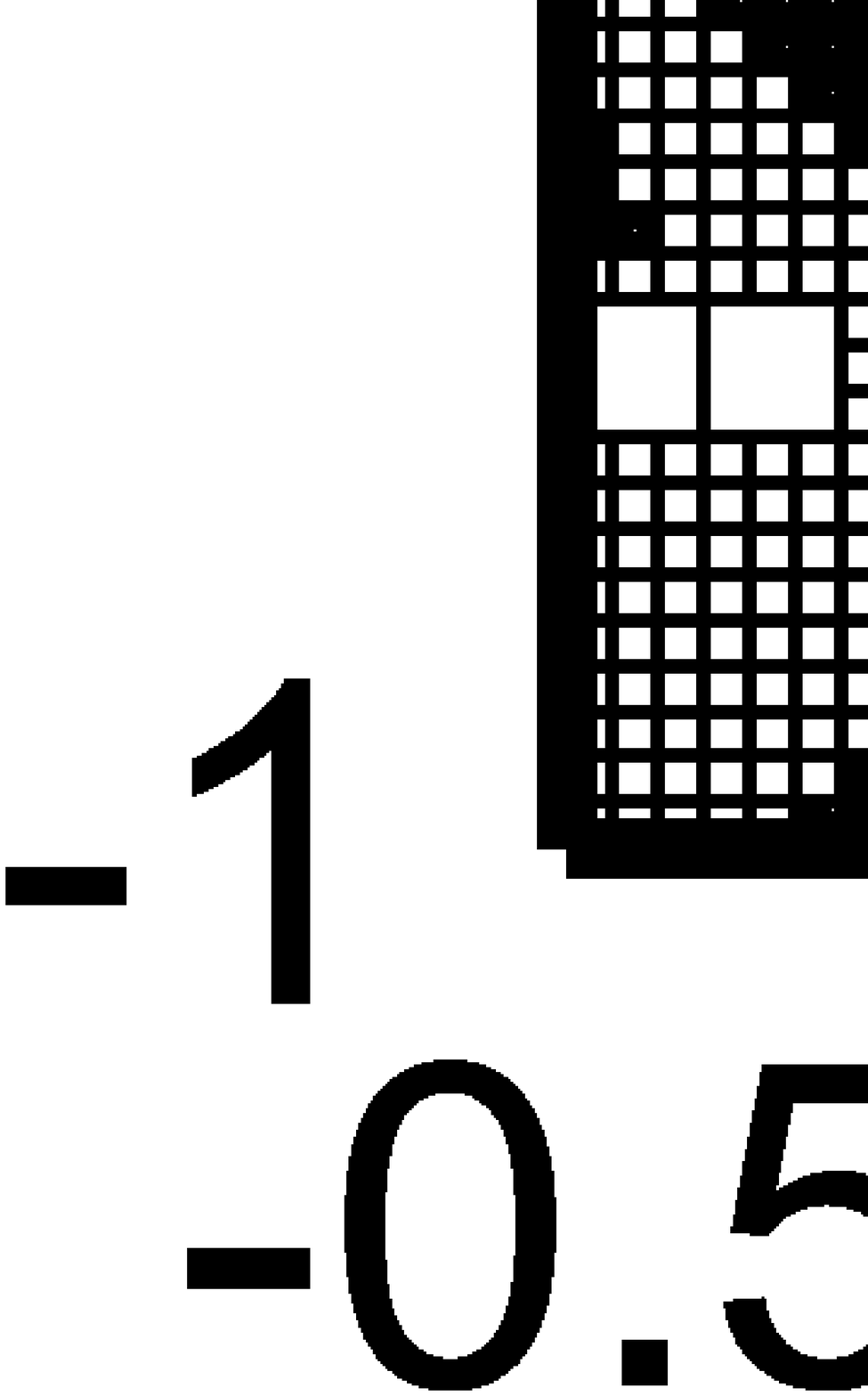}	 &  
\includegraphics[width=0.21\textwidth]{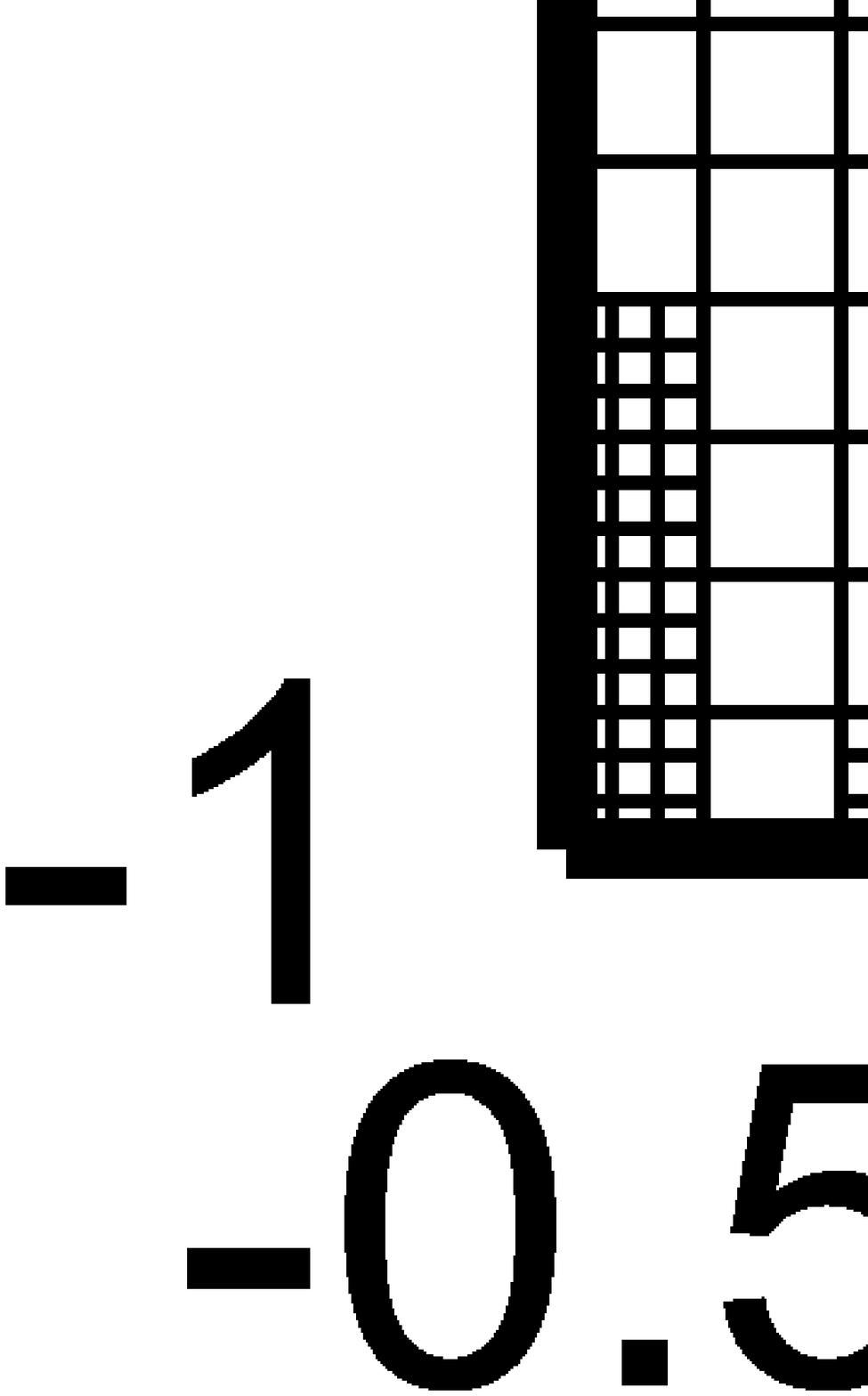}	 &  
\includegraphics[width=0.21\textwidth]{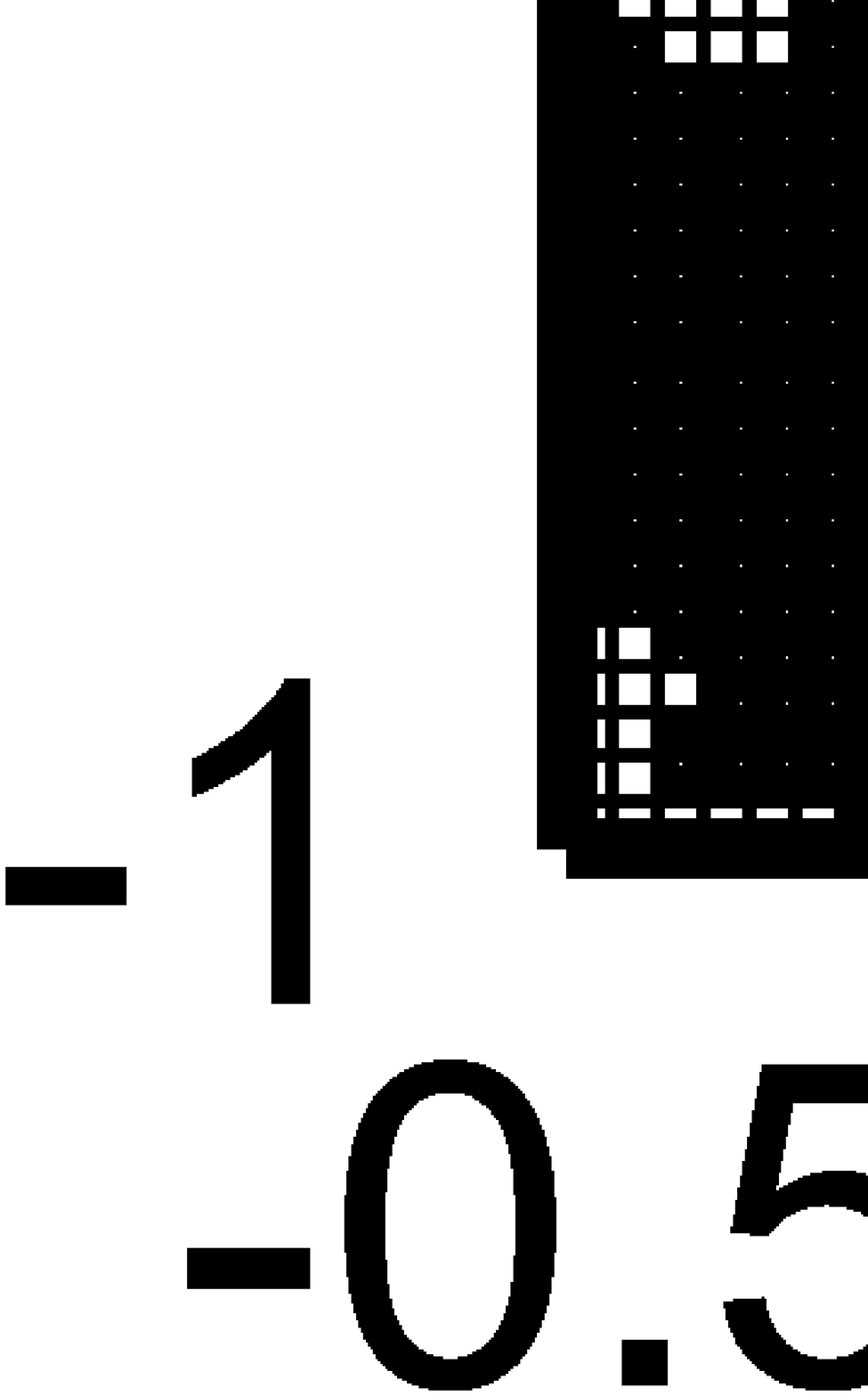}	 &  
\raisebox{15mm}{\includegraphics[width=0.06\textwidth]{./RMHD_KH_legend_nichts.eps}}\\				
\includegraphics[width=0.21\textwidth]{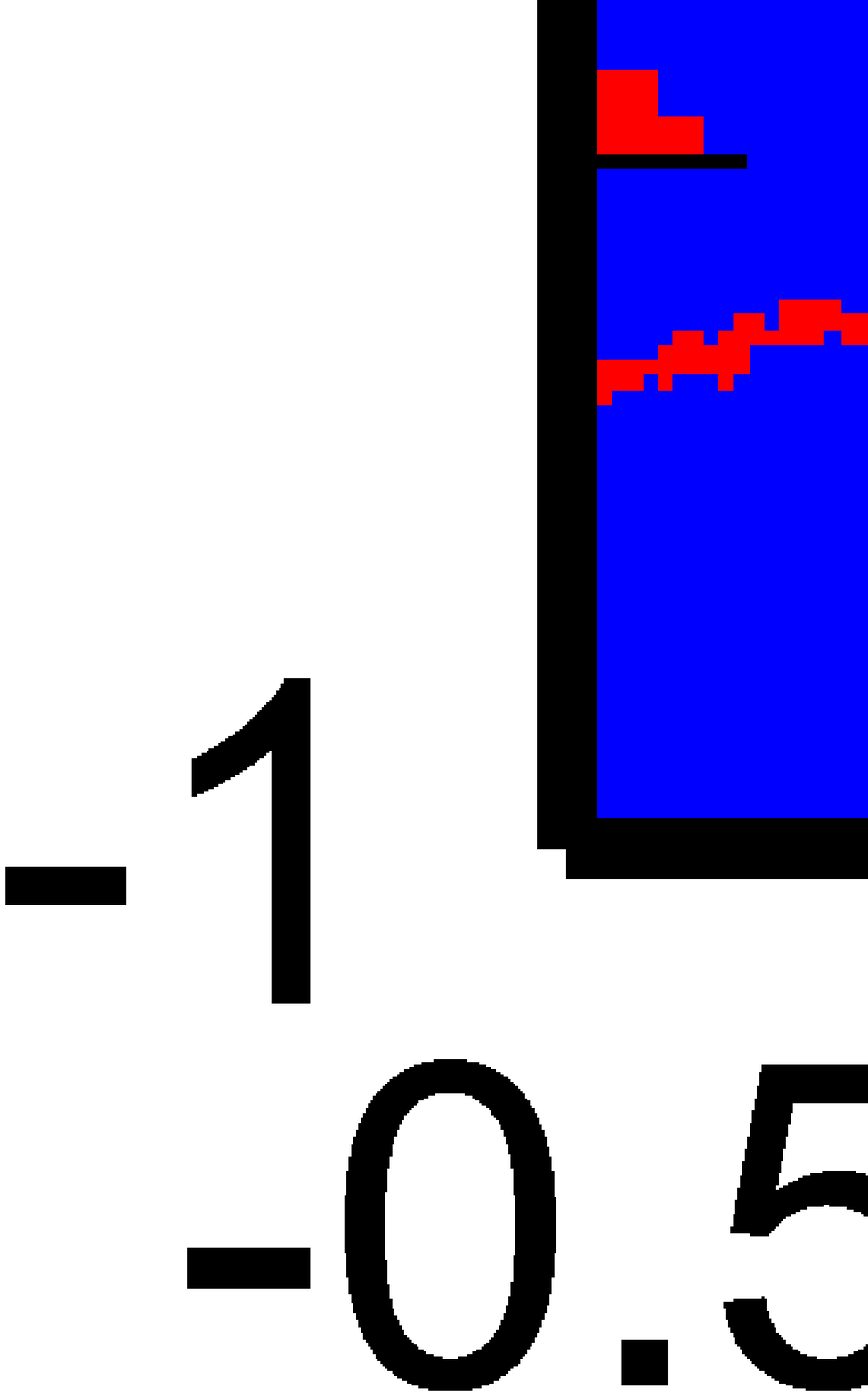}  &  
\includegraphics[width=0.21\textwidth]{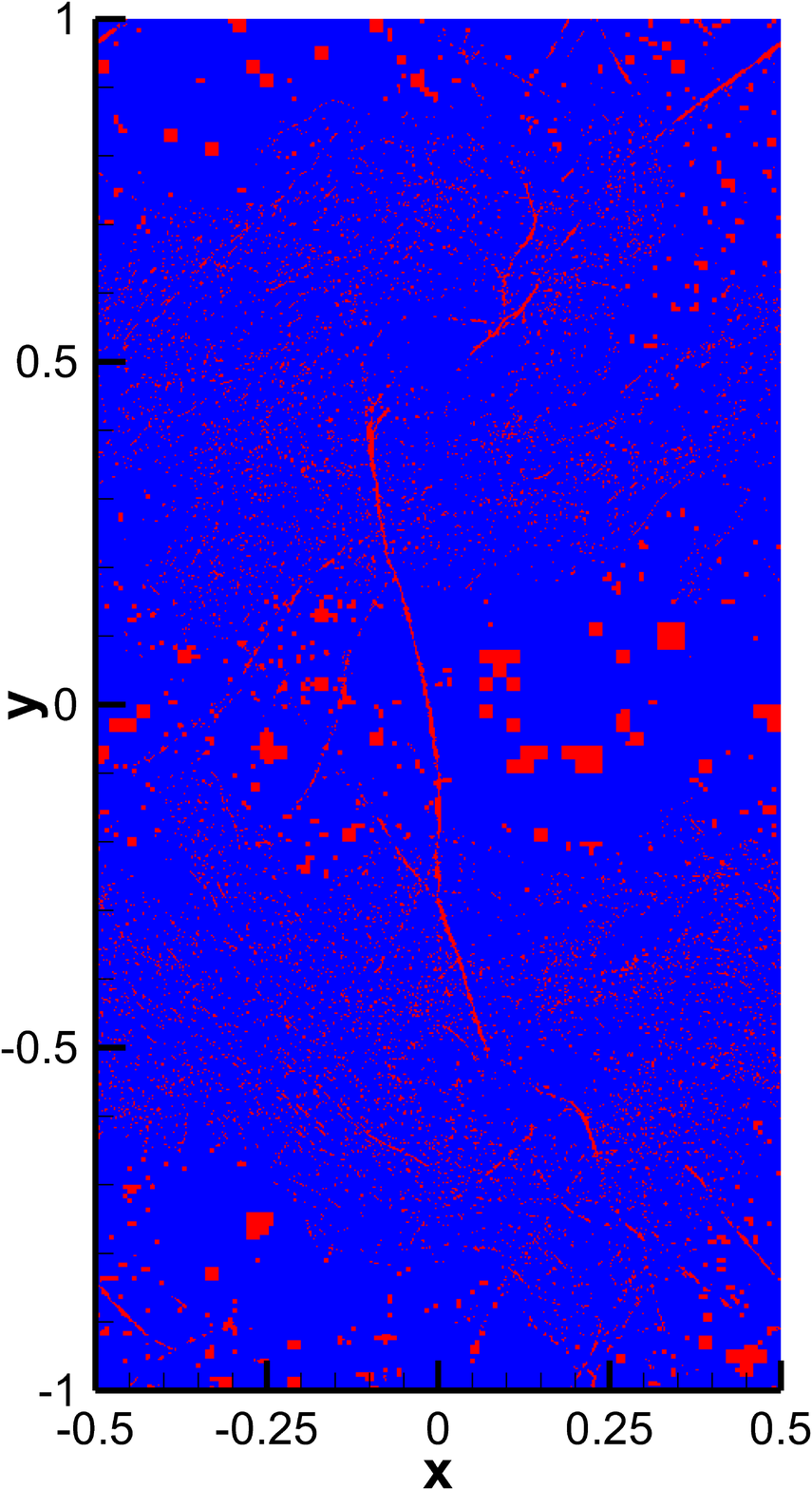}	 &  
\includegraphics[width=0.21\textwidth]{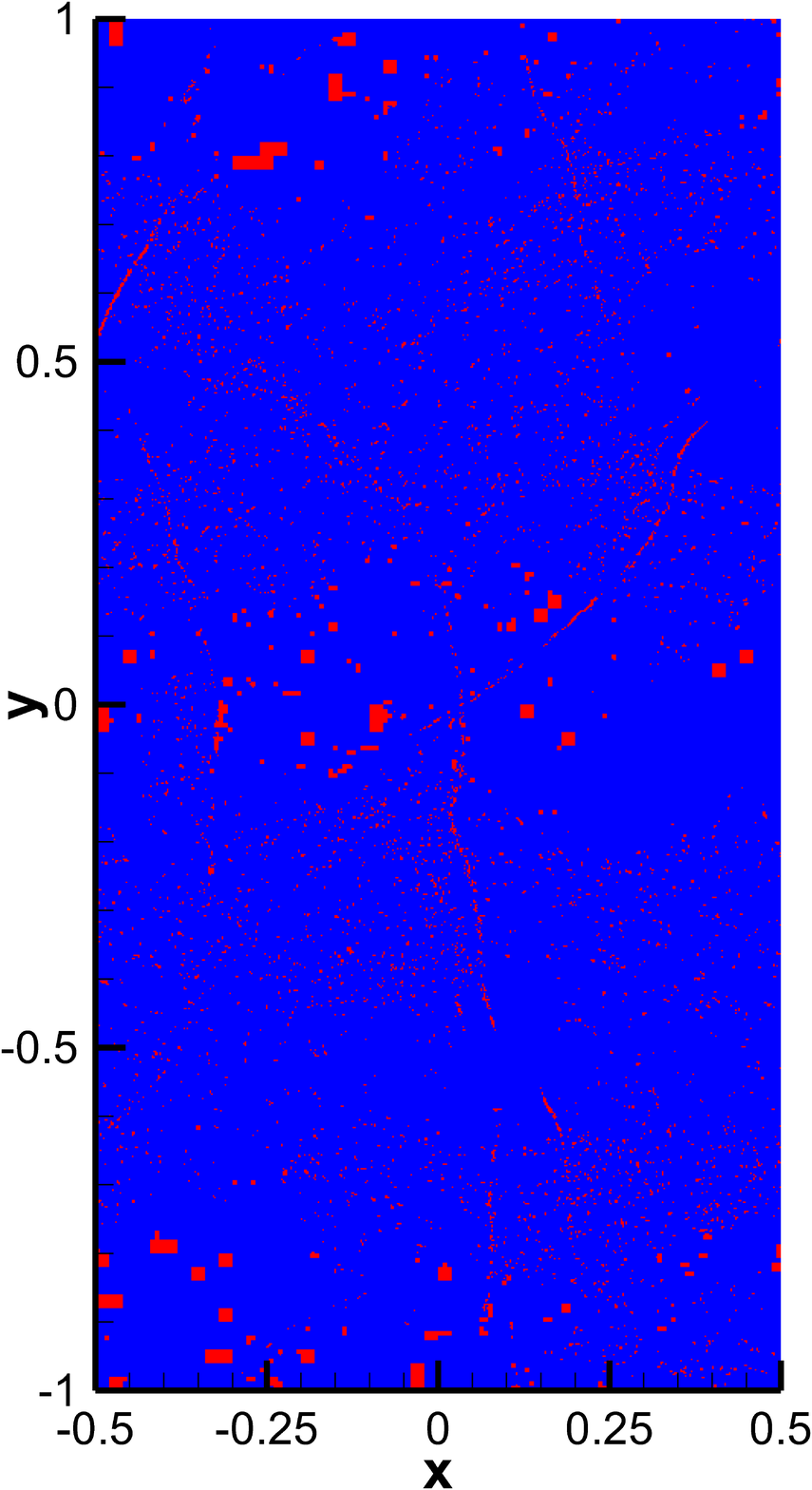}	 &  
\includegraphics[width=0.21\textwidth]{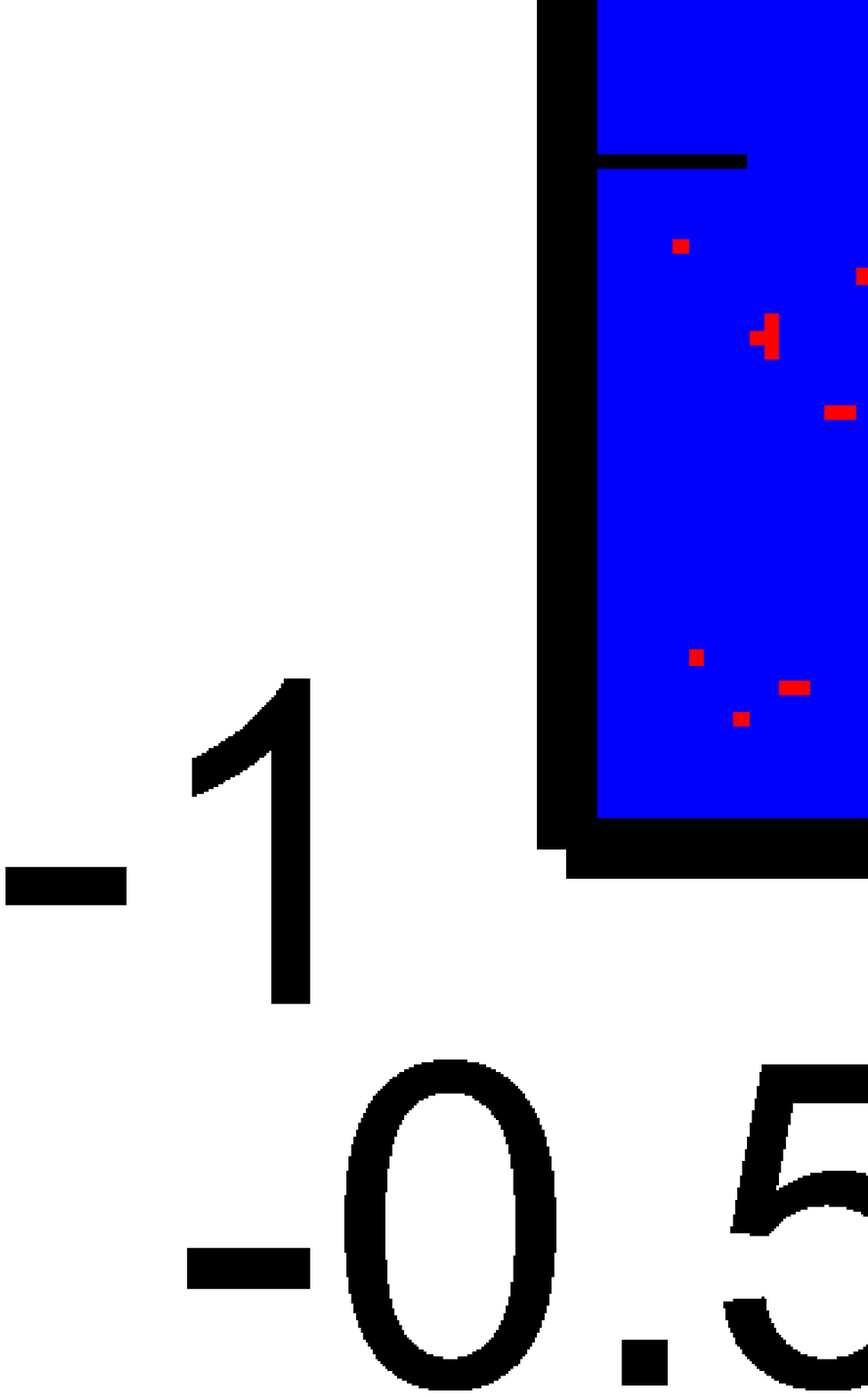}	 &  
\raisebox{15mm}{\includegraphics[width=0.06\textwidth]{./RMHD_KH_legend_nichts.eps}}\\				
\end{tabular} 
\caption{RMHD Kelvin--Helmholtz instability at times $t=5.0$, $t=10.0$, $t=20.0$, $t=30.0$ from left to right, obtained through the ADER-DG-$\mathbb{P}_3$ scheme
supplemented with the second order \aposteriori ADER-TVD subcell limiter. The computed solution of density (top), AMR grid (center) and limiter map (bottom) are shown.} 
\label{fig:RMHD-KH}
\end{center}
\end{figure*}

\begin{figure*}
\begin{center}
\begin{tabular}{c}
\includegraphics[width=0.4\textwidth]{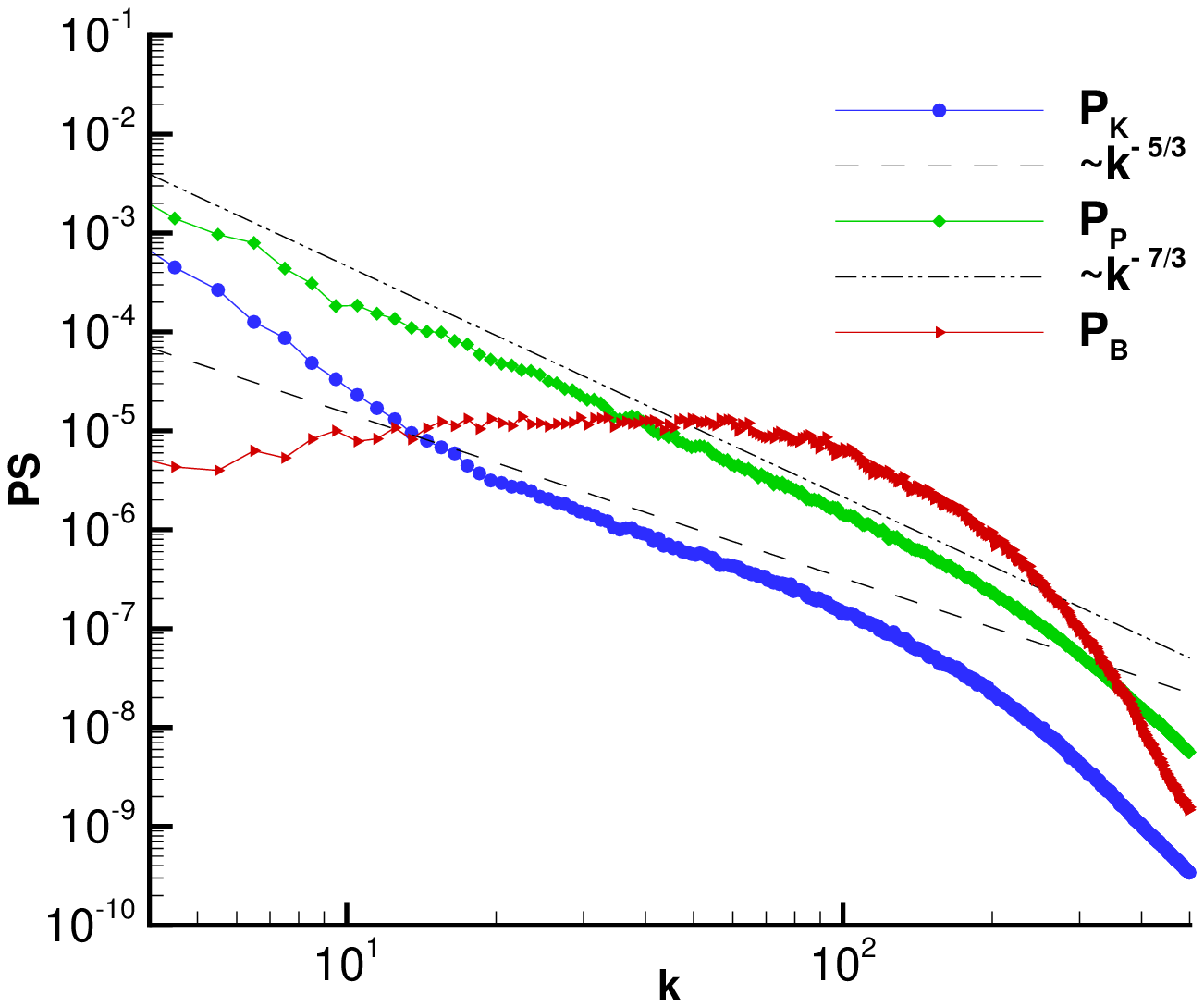}    			
\end{tabular} 
\caption{Power spectra for the  RMHD Kelvin--Helmholtz instability at time $t=30.0$ obtained through the ADER-DG-$\mathbb{P}_3$ scheme.
}
\label{fig:RMHD-KH-spectrum}
\end{center}
\end{figure*}

\section{The RMHD Kelvin--Helmholtz instability}
\label{sec:KH}
%
A two-dimensional test that is not only academic but may be relevant to explain the observed phenomenology of extended radio-jets [see \citet{Marti03} and references therein], we consider 
the Kelvin--Helmholtz (KH) instability with an initially uniform magnetic field.
Following the works of \citet{Mignone2009}, \citet{Beckwith2011} and \citet{Radice2012a}, we choose the initial conditions as
\begin{equation}\label{KHI-vx}
  v_x = \left\{\begin{array}{ll}
  v_s \tanh{[(y-0.5)/a]} & \quad y > 0\,, \\
 \noalign{\medskip}
 -v_s \tanh{[(y+0.5)/a]}  & \quad y \leq 0    \,, \\
 \noalign{\medskip}
 \end{array}\right.
\end{equation}
where $v_s=0.5$ is the velocity of the shear layer and $a=0.01$ is its characteristic size. Although not necessary in principle, it is convenient to introduce a small
transverse velocity to trigger the instability, hence fixing
\begin{equation}\label{KHI-vy}
  v_y = \left\{\begin{array}{ll}
  \eta_0 v_s \sin{(2\pi x)} \exp{[-(y-0.5)^2/\sigma]} & \quad y > 0\,, \\
 \noalign{\medskip}
 -\eta_0 v_s \sin{(2\pi x)} \exp{[-(y+0.5)^2/\sigma]}  & \quad y \leq 0    \,, \\
 \noalign{\medskip}
 \end{array}\right.
\end{equation}
where $\eta_0=0.1$ and $\sigma=0.1$. Finally, the rest-mass density is chosen as
\begin{equation}\label{KHI-rho}
  \rho = \left\{\begin{array}{ll}
  \rho_0 + \rho_1 \tanh{[(y-0.5)/a]} & \quad y > 0\,, \\
 \noalign{\medskip}
 \rho_0 - \rho_1 \tanh{[(y+0.5)/a]}  & \quad y \leq 0    \,, \\
 \noalign{\medskip}
 \end{array}\right.
\end{equation}
with $\rho_0=0.505$ and $\rho_1=0.495$.  The adiabatic index is $\gamma=4/3$, the pressure is $p=1$ everywhere, and we add a weak uniform magnetic field along the $x-$ direction, namely $B_x=0.001$.
The simulations are run with the ADER-DG-$\mathbb{P}_3$ scheme
over the computational domain 
$\Omega = [-0.5,0.5]\times[-1,1]$, using $50\times100$ elements on the coarsest refinement level at the initial state. 
Periodic boundary conditions are imposed along all borders
and the Rusanov Riemann solver is adopted. AMR is activated with $\mathfrak{r}=3$ and $\ell_\text{max}=2$. In this simulation 
the solution on the subgrid has been evolved through a second order TVD scheme, which turned out to be more robust than the usual third order WENO method.
Fig.~\ref{fig:RMHD-KH} shows the rest-mass density field at various times, up to $t=30$, and the corresponding development of the KH instability.
Since no physical viscosity or resistivity is present, it is very difficult to judge about the physical nature of the tiny structures, especially secondary instabilities, which are produced during the evolution, and which have been shown to 
depend sensibly on the order of accuracy of the scheme and on the Riemann solver used \citep{Beckwith2011,Radice2012a,Zanotti2015}. As the instability proceeds, the transition to a turbulent state
occurs. Although our final time is not large enough to allow for a fully developed turbulent state, and although this paper  is not devoted to a detailed study of relativistic MHD turbulence (see instead the works by \cite{Zhang09,Zrake2012,Garrison2015}),
we have nevertheless computed the power spectra of a few relevant quantities to confirm that the transition to turbulence is indeed taking place. Fig.~\ref{fig:RMHD-KH-spectrum}, in particular, shows the 
power spectra of the velocity field, of the pressure field and of the magnetic field, which have been computed according to 
\begin{equation}\label{eq:spectrum}
  P_v(k) = \frac{1}{2} \int_{|\boldsymbol{k}|=k} |\hat{v}(\boldsymbol{k})|^2\,
    d \boldsymbol{k}\,,~~~~~~
		P_p(k) = \int_{|\boldsymbol{k}|=k} |\hat{p}(\boldsymbol{k})|^2\,
    d \boldsymbol{k}\,,~~~~~~
		P_B(k) =  \int_{|\boldsymbol{k}|=k} |\hat{B}(\boldsymbol{k})|^2\,
    d \boldsymbol{k}
		\,,
\end{equation}
where ${\bf k}$ is the wave-number, while $\hat{v}(\boldsymbol{k})$, $\hat{B}(\boldsymbol{k})$, $\hat{p}(\boldsymbol{k})$ are the two-dimensional Fourier-transforms of $\boldsymbol{v}$, $p$ and $\boldsymbol{B}$, respectively. For $k\approx [20,70]$, in the so-called {\em inertial range} where the dynamics of the turbulence is not affected by large scale energy inputs nor by dissipation, we approximately recover Kolmogorov's trends, namely $P_v(k)\propto k^{-5/3}$ and $P_p(k)\propto k^{-7/3}$ \citep{Biskamp2008}. The power spectrum of the 
magnetic field is instead in qualitative agreement with results obtained by \cite{Zhang09} for fully turbulent configurations. A dedicated analysis to astrophysical RMHD turbulence
will be presented in a separate work.

\section{Discussion and conclusions}
\label{sec:Conclusions}
%
We have proposed a novel approach 
for the numerical solution of the special relativistic magnetohydrodynamics equations,
which is based on  the discontinuous Galerkin finite element method, but with some crucial modifications. Pure DG schemes, in fact, cannot
avoid the appearance of oscillations when 
discontinuities form in the solution. The common practice in these cases is to resort to either artificial viscosity, filtering, or to an \apriori finite-volume-type 
limiting of the higher order moments of the DG scheme, thus spoiling the subcell resolution capabilities of the DG scheme.  
On the contrary, following the recent works by \cite{Dumbser2014} and \cite{Zanotti2015c}, it is possible to verify \aposteriori the validity of the candidate
solution provided by the pure DG scheme by applying a relaxed form of the discrete maximum principle and by checking the numerical solution for physical validity, i.e.
for positivity, subluminal velocities and for possible failures in the the conversion from the conservative to the primitive variables. 
For those cells that violate any of the two criteria, we scatter the DG polynomials, computed at the previous (safe) time step, 
onto a set of $N_s=2N+1$ subcells along each spatial direction. After that, a traditional and more robust WENO finite volume scheme, or an even more robust TVD scheme, is applied at the 
\textit{subcell level}, thus \textit{recomputing} the solution in the troubled cells. In this way, our special \aposteriori subcell limiter of the DG method is intrinsically based on 
the governing partial differential equations, while standard DG limiters act independently of the governing PDE. 
The new, and thereby safe, subcell averages 
are subsequently gathered back into high order cell-centered DG polynomials on the main grid through a subgrid 
reconstruction operator. The \aposteriori limiter that we have used can also be regarded as a discontinuity detector. This is clear, for instance, after looking at the bottom right panel of Fig.~\ref{fig:RMHDrotor} and of Fig.~\ref{fig:RMHDBlast-B01}, where the activation of the limiter occurs where strong discontinuities are present.
In this respect, our \aposteriori correction may resemble the logic behind \apriori limiters based on {\em shock detectors}, several forms of which have been introduced, both in the Newtonian and in the relativistic framework. However, the \aposteriori approach that we have developed has at least
two evident advantages over traditional {\em shock detectors}. The first one is that it is 
both very simple and it applies unmodified for general systems of equations, while 
shock detectors become more and more elaborate as the complexity of the equations increases.\footnote{See \cite{Zanotti2010} for a shock detector valid in general relativistic hydrodynamics.}
The second advantage is that the \aposteriori approach can capture all kind of discontinuous solutions, including contact discontinuities, which are invisible to shock detectors but often represent a
serious challenge to the numerical scheme.

We also stress that both the DG scheme on the main grid and the WENO (or TVD) finite volume scheme on the subgrid are implemented with 
the local space–time discontinuous Galerkin predictor of \cite{DumbserEnauxToro}, thus providing a high order one-step ADER scheme in time, 
with no need for the Runge--Kutta time discretization that is typically used in the so-called method of lines. 
Finally, adaptive mesh 
refinement (AMR) is used, implying that the two AMR operations of projection and averaging need to involve the subcell averages of the solution on the sub-grids. 
For particularly challenging tests and applications, we have found convenient to resort to traditional second-order TVD schemes on the subgrid, thus achieving, at least formally, the same
robustness of those schemes.

The new ADER-DG-AMR scheme, which has been validated over stringent academic tests, can contribute significantly to the numerical modeling of high energy
astrophysics systems, such as extragalactic jets, gamma-ray bursts and magnetospheres of neutron stars. Work is already in progress to extend this approach 
to the full general relativistic regime. A particularly attracting field of application for the new method is represented by the study of relativistic turbulence, that we have
considered here in a simplified and preliminary two-dimensional set up in which the transition to turbulence is induced by 
the Kelvin--Helmholtz instability. We plan to investigate this problem in the future by means of three-dimensional simulations in which the high order capabilities of DG schemes are fully exploited.

\section*{Acknowledgments}

The research presented in this paper was financed by the European Research Council (ERC) under the
European Union's Seventh Framework Programme (FP7/2007-2013) with the research project \textit{STiMulUs}, 
ERC Grant agreement no. 278267. 

The authors are also very grateful for the subsequent financial support of the present research, already granted 
by the European Commission under the H2020-FETHPC-2014 programme with the research project \textit{ExaHyPE}, 
grant agreement no. 671698.

We are grateful to Bruno Giacomazzo and Luciano Rezzolla for providing the numerical code for the exact solution of the Riemann problem in RMHD.
We would also like to acknowledge PRACE for awarding access to the SuperMUC 
supercomputer based in Munich, Germany at the Leibniz Rechenzentrum (LRZ), 
and ISCRA, for awarding access to the FERMI supercomputer based in Casalecchio (Italy).

\bibliographystyle{mn2e}
\bibliography{references}

\bsp

\label{lastpage}

\end{document}